\documentclass{article}

\usepackage{arxiv}

\usepackage[utf8]{inputenc} 
\usepackage[T1]{fontenc}    
\usepackage{hyperref}       
\usepackage{url}            
\usepackage{booktabs}       
\usepackage{amsfonts}       
\usepackage{nicefrac}       
\usepackage{microtype}      
\usepackage{lipsum}
\usepackage{bm}
\usepackage{savesym}
\savesymbol{widering}
\usepackage{amsmath}
\usepackage{yhmath,scalerel}
\usepackage{amssymb}
\usepackage{subfigure}
\pdfoutput=1

\title{Modelling of Interfacial Flows Based on An Explicit Volume Diffusion Concept}

\author{
  B. Wang$^*$, \,\,  M. J. Cleary, \,\,  A. R. Masri
    \\
  School of Aerospace, Mechanical and Mechatronic Engineering\\
  The University of Sydney\\
  Sydney 2006, NSW, Australia \\
  \texttt{bosen.wang@sydney.edu.au} \\
}

\begin{document}
\maketitle

\begin{abstract}
A novel volume of fluid model (VoF) called explicit volume diffusion (EVD) is developed for the simulation of interfacial flows, including those with turbulence and primary spray atomisation. The EVD model is derived by volume averaging the VoF equations over a physically-defined length scale. This introduces unclosed sub-volume flux, sub-volume stress and volume averaged surface tension force. Sub-volume fluctuations arise due to both turbulent motions and other interface dynamics which can, in general, occur in both laminar and turbulent flows. Both of these types of fluctuations are attenuated by the volume averaging process. The sub-volume flux is closed by a gradient diffusion model and involves an explicit volume diffusion coefficient that is linked to the physical length scale. The sub-volume stress closure introduces an explicit volume viscosity {augmented} by turbulent viscosity {in turbulent flows}. The volume averaged surface tension force closure is based on fractal properties of wrinkled sub-volume interfaces. {These} closures are evaluated through \textit{a priori} analysis of resolved flow simulations for a series of two-dimensional laminar and three-dimensional turbulent interfacial shear flows. Subsequently, full EVD simulations are validated for these shear flows and for a laboratory airblast spray jet. Numerical convergence is demonstrated by keeping the physical length scale constant while reducing the numerical grid size so that numerical diffusion diminishes and becomes overwhelmed by the explicit volume diffusion. A sensitivity analysis is also conducted for variations in the physical length scale, which is compared to the boundary layer thickness on the light fluid side of the interface.
\end{abstract}


\section{Introduction}\label{sec:intro}
\noindent Spray {flows are common} in many engineering applications including agricultural spraying, medical and pharmaceutical technologies, food production, additive manufacturing and {various} combustion engines. {Spray atomisation is an interfacial physical process} which is the result of competing inertial, viscous and surface tension (or capillary) forces. The relative magnitudes of these forces are quantified by the dimensionless Weber number, $We$, which is the ratio of inertial to surface tension forces and the Ohnesorge number, $Oh$, which is the ratio of viscous to surface tension forces.
{In addition to the Weber number, the Reynolds numbers of the two phases and their momentum ratio classify the breakup regimes \cite{chigier1992morphological,lasheras2000liquid}.}
There are primary and secondary stages of atomisation, and each has a number of regimes which produce quite different downstream droplet morphology. The primary stage is characterised by growth of interfacial instabilities such as shear-driven Kelvin-Helmholtz instabilities that result in protrusions in the interface that may subsequently lead to secondary acceleration-driven Rayleigh-Taylor instabilities. This is followed by the disintegration of the continuous liquid stream into discrete elements that may be long ligaments, irregular-shaped blobs or spherical droplets \cite{marmottant2004spray}. In the secondary atomisation stage, these discrete liquid elements break further into smaller {fragments} and become increasingly spherical. The larger the value of $We$ the more catastrophic the disintegration and the smaller the size of the {resulting} droplets. The viscosity of the liquid tends to dampen instabilities at sufficiently high values of $Oh$ and the transition between breakup regimes occurs at higher values of $We$. Increasing the momentum ratio can also lead to finer spray with smooth drop size distributions \cite{hardalupas1996coaxial}. Atomising fragments spans a wide range of length scales ranging from the micron size droplets in the downstream spray cloud to the millimetre scale in the continuous liquid core. The interface itself is a sharp discontinuity. Turbulence imposes an additional spectrum of large and small scale motions which further induce interfacial instabilities.
\\\\
\noindent The complexity and multiscale nature of atomisation make both experimental and computational studies of atomisation very challenging. Detailed reviews are provided by \cite{lasheras2000liquid,dumouchel2008experimental,gorokhovski2008modeling} and only a few highlights are mentioned here. {Chigier \textit{et al.} \cite{chigier1992morphological} and Lasheras \textit{et al.} \cite{lasheras2000liquid} used high-speed visualization techniques to identify breakup regimes based on Reynolds number, Weber number and momentum ratio.} Marmottant and Villermaux \cite{marmottant2004spray} established correlations for the lengths of interfacial longitudinal and transverse waves in airblast spray jets and these were confirmed through direct {measurements} in high Weber number cases by Singh \textit{et al.} \cite{singh2020instability}. Matas \textit{et al.} \cite{matas2011experimental} and Fuster \textit{et al.} \cite{fuster2013instability} investigated primary instabilities in planar liquid sheets and demonstrated the accuracy of analytical linear instability theory \cite{huerre1990local,otto2013viscous}. Advanced optical techniques have enabled the measurement of liquid core length \cite{leroux2007experimental}, the size and shape of ligaments and droplets \cite{parker2006two} and the statistical spatial distributions of the liquid volume fraction downstream of atomisers \cite{singh2020volume}. 
\\\\
Computational fluid dynamics (CFD) approaches for atomisation tend to differ depending on whether the focus is on primary atomisation or secondary atomisation and subsequent droplet dispersion. The latter typically adopt hybrid Eulerian-Lagrangian schemes in which the droplets evolve as point particles \cite{crowe1977particle,miller1998evaluation}. Computational models for primary atomisation are usually based on the single-fluid paradigm in which the liquid-gas mixture is treated as a continuum with common velocity and thermodynamic properties at each spatial location. The main single-fluid approaches are the volume-of-fluid (VoF) method \cite{hirt1981volume}, the level set (LS) method \cite{sussman1994level} and the diffuse-interface (DI)  method \cite{anderson1998diffuse}. Each approach {has been assessed} for a range of different multiphase flows \cite{gopala2008volume,popinet2009accurate,loureiro2020primary,olsson2005conservative,desjardins2008accurate,sussman2000coupled,chiu2011conservative,sun2007sharp}. Comprehensive comparisons {of the performance of these three methods} are presented in \cite{desjardins2016performance,bilger2017evaluation,mirjalili2019comparison}. The LS method and some DI methods \cite{allen1976mechanisms} do not intrinsically conserve mass and additional numerical corrections are required \cite{olsson2005conservative,chiu2011conservative}. An earlier DI method by Cahn and Hilliard \cite{cahn1958free} has intrinsic mass conservations, however, discretisation of a required fourth-order derivative term is quite challenging. The VoF method, which is the focus of the present work, is {mass conserving} but it excessively smears the sharp interface through grid dependent numerical diffusion \cite{jasak1995interface,ubbink1997numerical}. {Accurate solutions using VoF} generally require very well resolved grids and high computational cost making it most suitable for fundamental studies of canonical flow configurations \cite{fulgosi2003direct,duret2013improving,hasslberger2019flow}.
\\\\
\noindent Large eddy simulation (LES), in which the large scale motions are resolved and the small dissipative and interfacial scales are modelled, have a more acceptable computational cost. However, LES of two-phase interfacial flows is still at a very early stage of development 
and to our knowledge the research community has not yet reached a consensus about the governing equations or sub-grid closures \footnote{Here and in the remainder of this paper we focus on the single-fluid paradigm for two phase flows and note that LES for droplet dispersion using Eulerian-Lagrangian formulations is more advanced, e.g. \cite{crowe1977particle,khan2018two}.}. Many LES of interfacial flows simply adopt concepts and closures that were developed initially for single phase LES. For example, standard turbulent viscosity and turbulent diffusivity models are widely used for the sub-grid terms in the filtered velocity and liquid volume fraction equations \cite{de2004large,chesnel2011large,navarro2014large,ketterl2019large,gartner2020numerical}. {However,} the inherent assumptions that scale similarity and the energy cascade are uninterrupted across the interface are highly questionable. {Attempts} have been made to model sub-grid interface dynamics in a more physically realistic way. Liovic and Lakehal \cite{liovic2012subgrid} and Herrmann \cite{herrmann2013sub} addressed sub-grid surface tension modelling based on curvature estimates at the grid scale and an auxiliary level set approach, respectively. Hecht \cite{hecht2016simulation} and Anez \textit{et al.} \cite{anez2019eulerian} suggested a hybrid strategy that aims to sharpen the interface through the application of a compression velocity along continuous regions of the interface \cite{jasak1995interface} and to use turbulent diffusion to account for sub-grid liquid dispersion in the regions with discrete ligaments and droplets. 
The artificial compression term aims to compensate for numerical diffusion {which can be dominant due to the absence of volume fraction molecular diffusion}. The addition of the turbulent diffusivity, even if it is physically unjustified near the interface, may improve the convergence behaviour to an extent but this diminishes as the grid/filter scale is refined. A criterion based on the local interface curvature was proposed to determine where the interface is continuous \cite{hecht2016simulation} but this too has been found to be highly grid dependent \cite{canu2020curvature}. {It should be noted that good numerical convergence studies in the LES of interfacial flows are very scarce.}
\\\\
\noindent The purpose of the present contribution is to formulate a new volume-of-fluid type computational model which treats interfacial dynamics in a physically realistic way{, converges} with refinement of the grid{, and} is not unduly affected by numerical diffusion. The concept is based on volume averaging of the instantaneous VoF equations over a physically-defined length scale. The volume averaged volume fraction is resolved by the grid and interface topology, which is a sub-volume artefact, is modelled through closures for the sub-volume flux and sub-volume stress. These closures are obtained through the introduction of a volume diffusivity and thus the model is called the explicit volume diffusion (EVD) model. This volume diffusivity is finite only in the vicinity of the interface and, importantly, is dependent on the physical and explicitly defined length scale and independent of the numerical grid. The physically defined volume diffusion is employed to account for a "diffusion" between the two immiscible fluids otherwise the Bachelor scale of volume fraction is infinitely small and numerical diffusion dominates the interface topology. A closure for the volume averaged surface tension force is also formulated. The volume averaging is relevant to laminar and turbulent flows. {Where} turbulence is present, additional turbulent viscosity and turbulent diffusivity appear but the latter does not act on the interface region. The sub-volume closures are validated initially based on an \textit{a priori} analysis of resolved flow simulations (RFS) of a series of 2D laminar and 3D turbulent interfacial shear flows. Further validation for the shear layer cases is performed through numerical solution of the EVD transport equations while varying the volume length scale and grid resolution to test accuracy and demonstrate numerical convergence. Suggestions are given for selecting the volume length scale based on the interfacial boundary layer thickness. Finally the EVD model is validated against existing experimental data for a turbulent airblast ethanol spray jet.

\section{Formulation of the explicit volume diffusion model}\label{sec:EVDform}
\noindent In this section the EVD model equations are derived. We consider the case of two immiscible fluids, one denoted as the heavy fluid (typically a liquid) and the other as the light fluid (typically a gas), which are quantified by volume fractions. The instantaneous Navier-Stokes and volume fraction transport equations are averaged over a volume with a physically defined length scale. Density weighted (Favre) averaging is used, although the conventional (Reynolds) average is of interest and a conversion is made for the purposes of observation. {Volume} averaging introduces unclosed terms involving the sub-volume fluctuations of the velocity and volume fractions and closure models are developed below.

\subsection{Volume averaged transport equations}

\noindent The VoF concept is based on the idea that two immiscible fluids can be modelled as a single fluid having singular values of thermodynamics, transport and flow properties. The density, $\rho$, and dynamic viscosity, $\mu$, at each point in space and time are
\begin{equation}
\rho=\alpha\rho^h + \left(1-\alpha\right)\rho^l,
\end{equation}
\begin{equation}
\mu=\alpha\rho^h\nu^h + \left(1-\alpha\right)\rho^l\nu^l.
\end{equation}
The superscripts $h$ and $l$ denote the heavy and the light fluids, respectively, and $\nu$ is the kinematic viscosity. The volume fraction of the heavy fluid {is $\alpha$} and the volume fraction of the light fluid is $1 - \alpha$. The single fluid transport equations for continuity, momentum and volume fraction are 
\begin{equation}\label{eq:instRho}
\frac{\partial \rho}{\partial t}+\frac{\partial \rho u_{i}}{\partial x_i} = 0,
\end{equation}
\begin{equation}\label{eq:instU}
\frac{\partial \rho u_{i}}{\partial t}+\frac{\partial \rho u_i u_j}{\partial x_j}  = -\frac{\partial p}{\partial x_i} + \frac{\partial \tau_{ij}}{\partial x_j} + F_{s,i} + \rho g_i,
\end{equation}
\begin{equation} \label{eq:instAlpha}
\frac{\partial \rho \alpha}{\partial t}+\frac{\partial \rho u_{i} \alpha }{\partial x_i} = 0.
\end{equation}
Here, $u$, $p$ and $g$ denote velocity, pressure and gravitational acceleration, respectively, $t$ is time and $x_i$ is the spatial coordinate in the $i$th direction. We select this version of the VoF equations (e.g., \cite{burluka2001development}) to obtain an evident sub-volume flux term by Favre volume averaging and thus to introduce an explicit volume diffusion model even though some other equivalent formulations exist. {The viscous stress, $\tau_{ij}$,} reads as
\begin{equation}
\tau_{ij}=\mu\left[2S_{ij}-\frac{2}{3}\frac{\partial u_k}{\partial x_k}\delta_{ij}\right],
\end{equation}
where $S_{ij}$ is the strain rate tensor,
\begin{equation}
S_{ij}=\frac{1}{2}\left(\frac{\partial u_i}{\partial x_j}+\frac{\partial u_j}{\partial x_i}\right),
\end{equation}
where $\delta$ is a Kronecker delta function. $F_{s,i}$ is the surface tension force and the details of its modelling are delayed to a later section.  Since the fluids are immiscible and phase change is not considered here, the real interface is sharp and the volume fractions in the interior of the heavy and light fluids are $\alpha=1$ and $\alpha = 0$, respectively. Due to numerical diffusion associated with discretisation of Eq.~(\ref{eq:instAlpha}), numerical solutions of the above equations inevitably produce locations where $0 < \alpha < 1$ such that the interface is an artificial continuum. Computational values of $\alpha$ and the other flow and thermodynamics properties represent numerical averages within the grid cells and they are highly dependent on the grid resolution \cite{schonfeld2003simulation,ndinisa2005computational}. With greater grid refinement the interface topology becomes more detailed and, in general, the resolved flow dynamics change.
\\\\
\noindent We can average the above transport equations over physically defined volumes, $V$, that are independent of the grid, in an attempt to replace the grid dependent continuum interface with a grid independent continuum interface with characteristic length scale $l_v \sim V^{1/3}$. Within a fixed volume centred at point $\bm{x_0}$, the volume averaged value of an arbitrary function $\phi$ is given by
\begin{equation}\label{eq:phixt2}
\widehat{\phi} = \widehat{\phi(\bm{x_0},t)} = \frac{1}{V} \int_V \phi(\bm{x'},t) d \bm{x}',
\end{equation}
where $\bm{x'}$ is a local coordinate centred at $\bm{x_0}$ and the integral is performed over all coordinate directions. $\widehat{\phi}$ represents a conventional (or Reynolds) volume average. Since the volumes can encompass both fluids with vastly different densities it is convenient for the derivation of subsequent sub-volume closures to introduce density weighted (or Favre) volume averaging,
\begin{equation}\label{eq:FavreDef}
\wideparen{\phi} = \wideparen{\phi(\bm{x_0},t)} = \frac{\widehat{\rho \phi}}{\widehat{\rho}}.
\end{equation}
At the sub-volume scale, the fluctuations relative to the Reynolds and Favre volume averaged quantities are
\begin{equation}\label{eq:phiFlucReynolds}
\phi^{\backprime} = \phi^{\backprime}(\bm{x_0},t) = \phi - \widehat{\phi},
\end{equation}
and
\begin{equation}\label{eq:phiFlucFavre}
\phi^{\backprime\backprime}  = \phi^{\backprime\backprime}(\bm{x_0},t) = \phi - \wideparen{\phi},
\end{equation}
respectively. The use of fixed volumes centred at $\bm{x_0}$ implies that
\begin{equation}
\widehat{\widehat{\phi}} = \widehat{\phi}, \,\, \widehat{\phi^{\backprime}} = 0,
\end{equation}
\begin{equation}
\wideparen{\wideparen{\phi}} = \wideparen{\phi}, \,\, \wideparen{\phi^{\backprime\backprime}} = 0,
\end{equation}
which is convenient in the subsequent manipulations as it avoids unclosed cross interactions between volume averaged fields and sub-volume fluctuations. The definition given by Eq.~(\ref{eq:phixt2}) is mathematically similar to a top-hat spatial filtering operation that is commonly used in LES of turbulent flows. The conceptual difference is that the volume averaging is not specifically linked to the filtering of turbulent frequencies above a certain cut-off value. Rather the volume averaging attenuates fluctuations that result from interfacial dynamics that can occur in both laminar and turbulent flows. Similar volume averaging concepts that can be applied to multiphase flow with discontinuity due to the existence of interface can be found in \cite{drew2006theory,zhang1994averaged}.
Note that the hat and paren (curved overline) notations $\widehat{\phi}$ and $\wideparen{\phi}$ are distinct from the symbols traditionally used for Reynolds and Favre statistical averaging and filtering in turbulent flows, viz. $\overline{\phi}$ and $\widetilde{\phi}$. Also, the fluctuations are denoted here by backward sloping primes whereas forward sloping primes are generally used for turbulent notations. 
\\\\
\noindent Volume averaging Eqs~(\ref{eq:instRho})~-~(\ref{eq:instAlpha}) gives
\begin{equation}\label{eq:avRho}
\frac{\partial \widehat{\rho}}{\partial t}+\frac{\partial \widehat{\rho} \wideparen{u_{i}}}{\partial x_i} = 0,
\end{equation}
\begin{equation}\label{eq:avU}
\frac{\partial \widehat{\rho} \wideparen{u_i}}{\partial t}+\frac{\partial \widehat{\rho} \wideparen{u_i} \wideparen{u_j}}{\partial x_j} = -\frac{\partial \widehat{p}}{\partial x_i} + \frac{\partial \widehat{\tau_{ij}}}{\partial x_j} - \frac{\partial \tau_{ij}^{va}}{\partial x_j} + \widehat{F_{s,i}} + \widehat{\rho} g_i,
\end{equation}
\begin{equation}\label{eq:avAlpha}
\frac{\partial \widehat{\rho} \wideparen{\alpha} }{\partial t}+\frac{\partial \widehat{\rho} \wideparen{u_i} \wideparen{\alpha}}{\partial x_i} = - \frac{\partial J_{\alpha,i}^{va}}{\partial x_i}.
\end{equation}
The volume averaged single fluid density and viscosity are
\begin{equation}
\widehat{\rho}=\widehat{\alpha}\rho^h + \left(1-\widehat{\alpha}\right)\rho^l,
\end{equation}
and
\begin{equation}
\widehat{\mu}=\widehat{\alpha}\rho^h\nu^h + \left(1-\widehat{\alpha}\right)\rho^l\nu^l.
\end{equation}
Reynolds volume averages of terms involving $\rho$ in Eqs~(\ref{eq:avRho})~-~(\ref{eq:avAlpha}) have been replaced with Favre volume averages through the application of Eq.~(\ref{eq:FavreDef}) and this eliminates unclosed forces arising from sub-volume variations of density and dynamic viscosity \cite{labourasse2007towards}. The volume averaged viscous stress is expressed as
\begin{equation}
\widehat{\tau_{ij}}=\widehat{\mu}\left[2\wideparen{S_{ij}}-\frac{2}{3}\frac{\partial\wideparen{u_k}}{\partial x_k}\delta_{ij}\right]
\end{equation}
where $\wideparen{S_{ij}}$ is the volume averaged strain rate tensor, viz.
\begin{equation}
\wideparen{S_{ij}}=\frac{1}{2}\left(\frac{\partial \wideparen{u_i}}{\partial x_j}+\frac{\partial \wideparen{u_j}}{\partial x_i}\right).
\end{equation}
Through relations for the conservation of mass of the heavy fluid, the Favre volume averaged solutions to the above equations are easily converted to {the physically meaningful Reynolds averaged forms that are of physical interest through}
\begin{equation}\label{eq:rhoAlpha}
\widehat{\alpha} = \frac{\widehat{\rho}\wideparen{\alpha}}{\rho^l}.
\end{equation}
\\\\
\noindent Volume averaging of the transport equations introduced new terms for the sub-volume flux,
\begin{equation}
J_{\alpha,i}^{va} = \widehat{\rho} \wideparen{\alpha^{\backprime\backprime} u_i^{\backprime\backprime}}
\end{equation}
and the sub-volume stress,
\begin{equation}
\tau_{ij}^{va} = \widehat{\rho} \wideparen{u_i^{\backprime\backprime} u_j^{\backprime\backprime}}.
\end{equation}
Closure models for these along with a model for the volume averaged surface tension force, $\widehat{F_{s,i}}$, are presented in the following three subsections.

\subsection{A closure for sub-volume flux}\label{sec:modelFlux}

\noindent Since the interface between the two fluids is sharp, the volume fraction field at the sub-volume scale has a bimodal probability density function (PDF) and this fact can be used to find a closure for the sub-volume flux. A similar approach has been used for modelling the unclosed sub-filter flux and stress for turbulent premixed flames based on a thin flame assumption \cite{bray1985unified}. The PDF of volume fraction is denoted by $P_\alpha(A)$ where $A$ is the sample space variable for the random variable $\alpha$. Since $P_\alpha(A)$ is bimodal and bounded between zero and one, the normalisation constraint is simplified as
\begin{equation}
\int_{-\infty}^{\infty} P_\alpha(A)dA = P_\alpha(A=1) + P_\alpha(A=0)=1.
\end{equation}
Following on from this we can define the joint PDF of velocity and volume fraction that satisfies
\begin{equation}
\int_{-\infty}^{\infty}\int_{-\infty}^{\infty}P_{\boldsymbol{u},\alpha}\left(\boldsymbol{U},A\right) d\boldsymbol{U} dA  = \int_{-\infty}^{\infty}P_{\boldsymbol{u},\alpha}\left(\boldsymbol{U},A=1\right) d\boldsymbol{U} + \int_{-\infty}^{\infty}P_{\boldsymbol{u},\alpha}\left(\boldsymbol{U},A=0\right) d\boldsymbol{U} = 1,
\end{equation}
where $\boldsymbol{U}$ is the sample space variable for the random velocity variable $\boldsymbol{u}$. Next we introduce the conditional PDF of $\boldsymbol{u}$ subject to the condition that $\alpha=A$ which is denoted as $P_{\boldsymbol{u}\mid \alpha}\left(\boldsymbol{U} \mid \alpha=A\right)$ and abbreviated as $P_{\boldsymbol{u}\mid \alpha}\left(\boldsymbol{U} \mid A\right)$ and defined by
\begin{equation}\label{eq:Pua1}
P_{\boldsymbol{u}\mid \alpha}\left(\boldsymbol{U} \mid A\right) = \frac{P_{\boldsymbol{u},\alpha}\left(\boldsymbol{U},A\right)}{P_\alpha(A)}.
\end{equation}
Since $P_\alpha(A)$ is bimodal, $P_{\boldsymbol{u}\mid \alpha}\left(\boldsymbol{U} \mid A\right)$ takes finite values only at $A=1$ and $A=0$ and therefore its normalisation constraint is simplified to
\begin{equation}
\int_{-\infty}^{\infty}P_{\boldsymbol{u}\mid \alpha}\left(\boldsymbol{U} \mid A=1\right) d\boldsymbol{U} = \int_{-\infty}^{\infty}P_{\boldsymbol{u}\mid \alpha}\left(\boldsymbol{U} \mid A=0\right) d\boldsymbol{U} = 1.
\end{equation}
Through these conditional PDF definitions, the (unconditional) volume averaged velocity can be expressed as
\begin{equation}\label{eq:phiCond}
\wideparen{\boldsymbol{u}} = \wideparen{\alpha}\wideparen{\boldsymbol{u}}^h + \left(1-\wideparen{\alpha}\right)\wideparen{\boldsymbol{u}}^l,
\end{equation}
where
\begin{equation}\label{eq:phih}
\wideparen{\boldsymbol{u}}^h= \int_{-\infty}^{\infty}\boldsymbol{U} P_{\boldsymbol{u}\mid \alpha}\left(\boldsymbol{U} \mid A=1\right) d\boldsymbol{U}
\end{equation}
and
\begin{equation}\label{eq:phil}
\wideparen{\boldsymbol{u}}^l= \int_{-\infty}^{\infty}\boldsymbol{U} P_{\boldsymbol{u}\mid \alpha}\left(\boldsymbol{U} \mid A=0\right) d\boldsymbol{U}.
\end{equation}
are the conditional mean velocities in the heavy and light fluids, respectively. Within the pure fluids where there are no density variations, the Favre and Reynolds volume averages are equivalent so that $\wideparen{\boldsymbol{u}}^h = \widehat{\boldsymbol{u}}^h$ and $\wideparen{\boldsymbol{u}}^l = \widehat{\boldsymbol{u}}^l$.
\\\\
Using an expression similar to Eq.~(\ref{eq:phiCond}), the sub-volume correlation of the fluctuations of volume fraction and velocity in direction $i$ can be expanded as
\begin{equation}\label{eq:alphaPrimeuPrime1}
\wideparen{\alpha^{\backprime\backprime}u_i^{\backprime\backprime}} = \wideparen{\alpha}\wideparen{\alpha^{\backprime\backprime} u_i^{\backprime\backprime}}^h + \left(1-\wideparen{\alpha}\right)\wideparen{\alpha^{\backprime\backprime} u_i^{\backprime\backprime}}^l.
\end{equation}
The terms $\wideparen{\alpha^{\backprime\backprime} u_i^{\backprime\backprime}}^h$ and $\wideparen{\alpha^{\backprime\backprime} u_i^{\backprime\backprime}}^l$ can be derived in a similar routine as shown by Eqs~(\ref{eq:phih}) and (\ref{eq:phil}). With an additional associate of Eq.~(\ref{eq:phiFlucFavre}), Eq.~(\ref{eq:alphaPrimeuPrime1}) can be derived as
\begin{equation}\label{eq:alphaPrimeuPrime2}
\begin{split}
\wideparen{\alpha^{\backprime\backprime}u_i^{\backprime\backprime}} & = \wideparen{\alpha}\int_{-\infty}^{\infty}\left(1-\wideparen{\alpha}\right)\left(u_i-\wideparen{u_i}\right)P_{u_i \mid \alpha}(U_i \mid A=1) dU_i \\ & + \left(1-\wideparen{\alpha}\right)\int_{-\infty}^{\infty}\left(0-\wideparen{\alpha}\right)\left(u_i-\wideparen{u_i}\right)P_{u_i \mid \alpha}(U_i \mid A=0) dU_i.
\end{split}
\end{equation}
Multiplying by density gives a new expression for the sub-volume flux,
\begin{equation}\label{eq:alphaU}
J_{\alpha,i}^{va} = \widehat{\rho}\wideparen{\alpha^{\backprime\backprime}u_i^{\backprime\backprime}} = \widehat{\rho}\wideparen{\alpha}\left(1-\wideparen{\alpha}\right)\left(\wideparen{u_i}^h-\wideparen{u_i}^l\right),
\end{equation}
and introduces the unclosed term $\wideparen{u_i}^h-\wideparen{u_i}^l$ which is called the mean drift velocity. Equation\,(\ref{eq:alphaU}) reveals that sub-volume inhomogeneity of both the volume fraction and the drift velocity lead to sub-volume flux. These inhomogeneities are caused by interface dynamics which occur in laminar as well as turbulent flows.
\\\\
To close Eq.~(\ref{eq:alphaU}), a gradient diffusion model is proposed,
\begin{equation}\label{eq:svFlux}
\begin{split}
J_{\alpha,i}^{va} = -\widehat{\rho}D_V\frac{\partial \wideparen{\alpha}}{\partial x_i},
\end{split}
\end{equation}
where $D_V$ is called the explicit volume diffusion coefficient that scales as
\begin{equation}\label{eq:Dv1}
\begin{split}
D_V \sim \left|\frac{Q}{E}\right|^{3/2}\wideparen{\alpha}\left(1-\wideparen{\alpha}\right)\left|\wideparen{\boldsymbol{u}}^h-\wideparen{\boldsymbol{u}}^l\right|\frac{1}{\left|\nabla \wideparen{\alpha}\right|},
\end{split}
\end{equation}
which implies that the mean drift velocity in direction $i$ is modelled using the magnitude of the mean drift velocity vector (which still needs to be closed) and the normalised volume fraction gradient in direction $i$,
\begin{equation}
\wideparen{u_i}^h-\wideparen{u_i}^l = -\left|\frac{Q}{E}\right|^{3/2}\left|\wideparen{\boldsymbol{u}}^h-\wideparen{\boldsymbol{u}}^l\right|\frac{1}{\left|\nabla \wideparen{\alpha}\right|} \frac{\partial \wideparen{\alpha}}{\partial x_i}.
\end{equation}
Here, $\left|\frac{Q}{E}\right|$ is a coherent structure function  \cite{hunt1988eddies,kobayashi2005subgrid} with $Q$ being the well-known second invariant of the velocity gradient tensor,
\begin{equation}
Q = \frac{1}{2}\left({\wideparen{W_{ij}}}{\wideparen{W_{ij}}}-{\wideparen{S_{ij}}}{\wideparen{S_{ij}}}\right),
\end{equation}
and $E$ being the magnitude of a velocity gradient tensor,\begin{equation}
E=\frac{1}{2}\left({\wideparen{W_{ij}}}{\wideparen{W_{ij}}}+{\wideparen{S_{ij}}}{\wideparen{S_{ij}}}\right).
\end{equation}
$\wideparen{W_{ij}}$ is the vorticity tensor given by
\begin{equation}
\wideparen{W_{ij}}=\frac{1}{2}\left(\frac{\partial \wideparen{u_i}}{\partial x_j}-\frac{\partial \wideparen{u_j}}{\partial x_i}\right).
\end{equation}
High values of $\left|{Q}/{E}\right|$ characterise regions of high vorticity such as in homogeneous or strong shear turbulence, while low values characterise regions of low vorticity which are typical in laminar or mildly turbulent flows. It is introduced in the present model to give the correct limiting behaviour of the interface. Prior to the appearance of interfacial instabilities, the curvature of the interface approaches zero. There, $\left|{Q}/{E}\right| \to 0$ and as a result $D_V \to 0$ ensuring that there is not any spurious interfacial flux at the sub-volume scale and no excessive volume diffusion. This effect is similar to that of the damping function for sub-grid turbulent viscosity proposed in \cite{fulgosi2003direct,reboux2006large}. As instabilities generate interfacial curvature $\left|{Q}/{E}\right|$ and hence $D_V$ increase smoothly to thicken the volume averaged interface.
\\\\
{As derived in detail below,} the magnitude of the mean drift velocity {is found} to scale with the root mean square (rms) of sub-volume velocity fluctuations, $\left|u_V^{\backprime\backprime}\right|=\sqrt{\wideparen{{\boldsymbol{u}^{\backprime\backprime}}^2}}$, and these are obtained by application of the sub-volume conditional PDF concept,
\begin{equation}\label{eq:urpime0}
\begin{split}
\wideparen{{\boldsymbol{u}^{\backprime\backprime}}^2} 
& = \wideparen{\alpha}{\wideparen{{\boldsymbol{u}^{\backprime\backprime}}^2}}^h + \left(1-\wideparen{\alpha}\right){\wideparen{{\boldsymbol{u}^{\backprime\backprime}}^2}}^l \\
& = \wideparen{\alpha}\int_{-\infty}^{\infty}\left(\boldsymbol{u}-\wideparen{\boldsymbol{u}}\right)^2 P_{\boldsymbol{u} \mid \alpha}(\boldsymbol{U} \mid A=1) d\boldsymbol{U} \\ & + \left(1-\wideparen{\alpha}\right)\int_{-\infty}^{\infty}\left(\boldsymbol{u}-\wideparen{\boldsymbol{u}}\right)^2 P_{\boldsymbol{u} \mid \alpha}(\boldsymbol{U} \mid A=0) d\boldsymbol{U} \\
& = \wideparen{\alpha}\left(1-\wideparen{\alpha}\right)\left(\wideparen{\boldsymbol{u}}^h-\wideparen{\boldsymbol{u}}^l\right)^2 + \wideparen{\alpha}\wideparen{{\left(\boldsymbol{u}^{\backprime\backprime,h}\right)}^2} + \left(1-\wideparen{\alpha}\right)\wideparen{{\left(\boldsymbol{u}^{\backprime\backprime,l}\right)}^2}. \\
\end{split}
\end{equation}
Here, $\boldsymbol{u}^{\backprime\backprime,h} = \boldsymbol{u}^h - \wideparen{\boldsymbol{u}}^h$ and $\boldsymbol{u}^{\backprime\backprime,l} = \boldsymbol{u}^l - \wideparen{\boldsymbol{u}}^l$ are conditional fluctuations in the pure heavy and light fluids streams, respectively, and $\wideparen{{\left(\boldsymbol{u}^{\backprime\backprime,h}\right)}^2}$ and $\wideparen{{\left(\boldsymbol{u}^{\backprime\backprime,l}\right)}^2}$ are the sub-volume variances. It follows that
\begin{equation}\label{eq:driftU}
\left|\wideparen{\boldsymbol{u}}^h-\wideparen{\boldsymbol{u}}^l\right| = \frac{\sqrt{\wideparen{{\boldsymbol{u}^{\backprime\backprime}}^2}-{\wideparen{\alpha}}\wideparen{{\left(\boldsymbol{u}^{\backprime\backprime,h}\right)}^2}-\left(1-\wideparen{\alpha}\right)\wideparen{{\left(\boldsymbol{u}^{\backprime\backprime,l}\right)}^2}}}{\sqrt{\wideparen{\alpha}\left(1-\wideparen{\alpha}\right)}}.
\end{equation}
Away from the interface within the pure heavy fluid stream $\wideparen{{\left(\boldsymbol{u}^{\backprime\backprime,h}\right)}^2} =  \wideparen{{\boldsymbol{u}^{\backprime\backprime}}^2}$ and within the pure light fluid stream $\wideparen{{\left(\boldsymbol{u}^{\backprime\backprime,l}\right)}^2} =  \wideparen{{\boldsymbol{u}^{\backprime\backprime}}^2}$. Therefore we have the two limiting relations,
\begin{equation}
\lim_{{\wideparen{\alpha}} \to 1} \left[\sqrt{{\wideparen{{\boldsymbol{u}^{\backprime\backprime}}^2}}-{\wideparen{\alpha}}{\wideparen{{\left(\boldsymbol{u}^{\backprime\backprime,h}\right)}^2}}
-\left(1-{\wideparen{\alpha}}\right)\wideparen{{\left(\boldsymbol{u}^{\backprime\backprime,l}\right)}^2}}\right] = 0, 
\end{equation}
and
\begin{equation}
\lim_{{\wideparen{\alpha}} \to 0} \left[\sqrt{{\wideparen{{\boldsymbol{u}^{\backprime\backprime}}^2}}-{\wideparen{\alpha}}{\wideparen{{\left(\boldsymbol{u}^{\backprime\backprime,h}\right)}^2}}
-\left(1-{\wideparen{\alpha}}\right)\wideparen{{\left(\boldsymbol{u}^{\backprime\backprime,l}\right)}^2}}\right] = 0,
\end{equation}
which imply that $\left|\wideparen{\boldsymbol{u}}^h-\wideparen{\boldsymbol{u}}^l\right|$, and hence $D_V$, are non-zero only in the interface region. From empirical observation of interfacial regions of resolved flow simulations (see \textit{a priori} analysis in Section~\ref{sec:apriori}), we introduce the functional approximation
\begin{equation}\label{eq:approxUprime}
\sqrt{{\wideparen{{\boldsymbol{u}^{\backprime\backprime}}^2}}-{\wideparen{\alpha}}{\wideparen{{\left(\boldsymbol{u}^{\backprime\backprime,h}\right)}^2}}
-\left(1-{\wideparen{\alpha}}\right)\wideparen{{\left(\boldsymbol{u}^{\backprime\backprime,l}\right)}^2}} \sim \sqrt{{\wideparen{\alpha}}\left(1-{\wideparen{\alpha}}\right){\wideparen{{\boldsymbol{u}^{\backprime\backprime}}^2}}} \sim \sqrt{{\wideparen{\alpha}}\left(1-{\wideparen{\alpha}}\right)}\left|u_V^{\backprime\backprime}\right|,
\end{equation}
The sub-volume root mean square of velocity fluctuations, $\left|u_V^{\backprime\backprime}\right|$, is formulated as a Smagorinsky-Lilly type model,\cite{smagorinsky1963general,lilly1966representation},
\begin{equation}\label{eq:magUprime}
\begin{split}
\left|u_V^{\backprime\backprime}\right|&= C_{SL} l_V \sqrt{2{\wideparen{S_{ij}}}{\wideparen{S_{ij}}}},
\end{split}
\end{equation}
where $C_{SL}$ is a constant. This model takes inspiration from sub-filter turbulence closures but here it covers fluctuations due to both interface dynamics and turbulence (if it is present). Most importantly, the model for $\left|u_V^{\backprime\backprime}\right|$ is linked to the explicit volume length scale, $l_V$, and is strictly independent of the numerical grid scale. In the vicinity of a sharp interface, the sub-volume velocity fluctuations are highly anisotropic. This fact is not contradicted by Eq.~(\ref{eq:magUprime}) which implies only that the magnitude of the sub-volume velocity fluctuations scale with the volume averaged strain rate. This is physically plausible because both velocity fluctuations and strain rate are positively correlated with the drift velocity between the phases.
\\\\
Although $\left|\nabla \wideparen{\alpha}\right|$, which appears in the denominator of Eq.~(\ref{eq:Dv1}) for $D_V$, is known from the solution for $\wideparen{\alpha}$, it can lead to destablisation of the numerical solution as it approaches zero away from the interface. It is noted again that $D_V$ itself is also zero away from the interface. We therefore employ the approximation,
\begin{equation}\label{eq:magGradA}
\begin{split}
\left|\nabla \wideparen{\alpha}\right| \sim \frac{\sqrt{\wideparen{{\alpha^{\backprime\backprime}}^2}}}{l_V} \sim \frac{\sqrt{\wideparen{\alpha}\left(1-\wideparen{\alpha}\right)}}{l_V},
\end{split}
\end{equation}
where $\sqrt{\wideparen{{\alpha^{\backprime\backprime}}^2} }=\sqrt{\wideparen{\alpha}\left(1-\wideparen{\alpha}\right)}$ is a property of the bimodal sub-volume distribution of $\alpha$. 
\\\\
Substituting Eqs\,(\ref{eq:driftU}) and (\ref{eq:approxUprime})~-~(\ref{eq:magGradA}) into Eq.~(\ref{eq:Dv1}), the model for the explicit volume diffusion coefficient can finally be expressed as
\begin{equation}\label{eq:DV}
D_V = C_{\alpha u}\left|\frac{Q}{E}\right|^{3/2}l_V^2\sqrt{{\wideparen{\alpha}}\left(1-{\wideparen{\alpha}}\right)}\sqrt{2{\wideparen{S_{ij}}}{\wideparen{S_{ij}}}}.
\end{equation}
Note that the constant $C_{\alpha u}$ incorporates $C_{SL}$ in Eq.~(\ref{eq:magUprime}) but the latter is retained as it is used in modelling of other terms in the coming subsections. Suitable values for both constants are determined by \textit{a priori} analysis in Section~\ref{sec:apriori}.

\subsection{A closure for sub-volume stress}
\noindent Based on the Boussinesq viscosity assumption that was demonstrated for free shear flows with turbulent/non-turbulent interface \cite{wilcox1998turbulence}, the sub-volume stress here is assumed to be proportional to the strain rate through a volume viscosity, $\nu_V$,
\begin{equation}\label{eq:UiUj}
\begin{split}
\tau_{ij}^{va} = -2\widehat{\rho}\nu_V \wideparen{S_{ij}},
\end{split}
\end{equation}
where $\nu_V$ is a new term called the explicit volume viscosity. Where $0<\wideparen{\alpha}<1$ in the interface region, $\nu_V$ is approximated from the explicit volume diffusion coefficient via a Schmidt number,
\begin{equation}
\nu_V=D_VSc_V.
\end{equation}
Away from the interface, in the interior of the pure fluid streams where $\wideparen{\alpha} = 1$ or $\wideparen{\alpha} = 0$, sub-volume fluctuations of velocity are not zero, especially in turbulent flows. This is in contrast to sub-volume fluctuations of the volume fraction which are zero. Therefore the explicit volume viscosity away from the interface can be modelled by a conventional subgrid turbulence model and, here, the Smagorinsky model is used\cite{smagorinsky1963general},
\begin{equation}
\nu_V = \nu_t = C_s l_V^2 \sqrt{2{\wideparen{S_{ij}}\wideparen{S_{ij}}}},
\end{equation}
where the constant $C_s$ is taken as 0.01927. Other turbulence models are also possible. To account for the reduction in turbulent dissipation as the distance to the interface decreases, a smooth blending function is applied so that the effective turbulent viscosity is given by
\begin{equation}\label{eq:nueff}
\nu_t^{eff} = \Omega(\alpha)\nu_t,
\end{equation}
where
\begin{equation}
\Omega(\alpha) = 1-4\wideparen{\alpha}\left(1-\wideparen{\alpha}\right).
\end{equation}
The choice of this function takes inspiration from the the sensor function that is commonly used in modelling turbulent premixed flames where the flame is considered as an interface \cite{durand2007implementation}. Finally, the sub-volume stress is modelled as
\begin{equation}\label{eq:subStress}
\tau_{ij}^{va} = -2\widehat{\rho} \left(D_VSc_V+\nu_t^{eff}\right)\wideparen{S_{ij}}.
\end{equation}
The explicit volume Schmidt number, $Sc_V$, will be calibrated by an \textit{a priori} analysis in Section~\ref{sec:apriori}.

\subsection{A closure for volume averaged surface tension force}
\noindent The surface tension force in VoF formulations is commonly modelled by the continuum model of Brackbill et al. \cite{brackbill1992continuum}, and volume averaging this model gives
\begin{equation}
\widehat{F_{s,i}} = \sigma \widehat{\kappa n_{i} \delta_s},
\end{equation}
where $\sigma$ is the surface tension coefficient, $\kappa=-\nabla\cdot \boldsymbol{n} $ is the interface curvature found through the unit normal vector $\boldsymbol{n}=\frac{\nabla \alpha}{\left|\nabla \alpha\right|}$, and $\delta_s=\left|\nabla \alpha \right|$ is an interface detecting function. {This continuum surface tension model has been widely used and validated for LES and DNS of spray atomisation in macroscopic configurations \cite{shinjo2010simulation,chesnel2011large,navarro2014large,warncke2017experimental,hasslberger2019flow}. However, a closure accounting for the sub-grid fluctuations of surface tension force is a long-standing challenge which is the focus of the present study.} Integrating over an explicit volume and simplifying we have
\begin{equation}\label{eq:Fs0}
\begin{split}
\widehat{F_{s,i}} = \sigma \frac{1}{V}\int_{V} \kappa n_{i} \left|\nabla \alpha\right| d V.
\end{split}
\end{equation}
Following the $(n-1)$~-~dimensional Hausdorff approach for obtaining the integral of the modulus of a gradient \cite{maz2013sobolev}, the volume integral can be transformed to the surface integral over iso-surfaces of $\alpha$,
\begin{equation}\label{eq:Fs1}
\begin{split}
&\widehat{F_{s,i}} = \sigma \frac{1}{V}\int_0^1 d \alpha \int_{A_s} \kappa n_{i} d A = \sigma \frac{1}{V} \int_{A_s} \kappa n_{i} d A. \\ 
\end{split}
\end{equation}
The surface integral can be replaced by the surface averaged value of the integrand (denoted by ${\langle \cdot \rangle}_s$) multiplied by the interfacial surface area and with further approximation leads to
\begin{equation}\label{eq:AsKappa}
\int_{A_s} \kappa n_{i} d A = {\langle \kappa n_{i} \rangle}_s A_s \cong \widehat{\kappa}\widehat{n_{i}} A_s.
\end{equation}
It should be noted Eq.~(\ref{eq:AsKappa}) can be found in \cite{worner2001volume} but the derivation was not provided there. Here, $\widehat{\kappa} = -\nabla \cdot \widehat{\boldsymbol{n}}$ denotes the curvature of the volume averaged interface with $\widehat{\boldsymbol{n}}=\frac{\nabla \widehat{\alpha}}{\left|\nabla \widehat{\alpha}\right|}$. Once again applying the $(n-1)$~-~dimensional Hausdorff approach, the surface density is obtained as ${\widehat{\Sigma}} =\widehat{\left|\nabla\alpha\right|}$. The surface area of the interface is modelled using the surface density, {$\widehat{\Sigma}$}, as
\begin{equation}\label{eq:As}
{A_s} = {\widehat{\Sigma}} V = \widehat{\left|\nabla\alpha\right|} V.
\end{equation}
A dynamic wrinkling factor, $\Xi$, is now used for the ratio of the surface density to the resolved interface indicator function, $\left|\nabla\widehat{\alpha}\right|$ \cite{hawkes2001physical}, and combining it with a fractal interface model \cite{charlette2002power} gives,
\begin{equation}\label{eq:Xi}
\Xi = \frac{{\widehat{\Sigma}}}{\left|\nabla\widehat{\alpha}\right|} = \left( 1 + \frac{l_V}{l_\sigma} \right)^{D_f-2},
\end{equation}
where $l_\sigma$ is the interface cut-off length scale and $D_f$ is the fractal dimension of interface. The fractal dimension is a natural property of the interface. As a first attempt it is assumed to be constant and is estimated about 7/3 for naturally wrinkled interfaces \cite{hentschel1984relative}. Here, the cut-off length scale is defined as a scale at the interface where the surface tension force can balance the inertial force of the heavy fluid induced by the sub-volume velocity fluctuations. Therefore, a critical sub-volume Weber number is given by 
\begin{equation}
We_{h,c}=\frac{\rho_h \left|u_V^{\backprime\backprime}\right|^2 l_\sigma}{\sigma} \sim 1 .
\end{equation}
and thus
\begin{equation} \label{eq:lsigma}
l_\sigma \sim \frac{\sigma}{\rho_h \left|u^{\backprime\backprime}\right|^2}.
\end{equation}
Using Eqs.\,(\ref{eq:Fs1}-\ref{eq:lsigma}), the volume averaged surface tension force becomes
\begin{equation}\label{eq:Fs}
\begin{split}
\widehat{F_{s,i}} = C_{sf}\left( 1 + \frac{l_V \rho_h \left|u_V^{\backprime\backprime}\right|^2}{\sigma} \right)^{1/3} \sigma\widehat{\kappa}\widehat{n_{i}}\left|\nabla\widehat{\alpha}\right|.
\end{split}
\end{equation}
The modelling constant, $C_{sf}$, is determined in the \textit{a priori} analysis in Section~\ref{sec:apriori}.


\section{Numerical implementation}\label{sec:numericalImp}
\noindent The EVD model has been implemented in a new solver called \textit{evdFoam} that is incorporated into OpenFOAM \cite{weller1998tensorial}. The volume averaged transport equations are discretised on numerical grids and grid fields are solved directly, including $\widehat{\rho}^{\Delta}$, $\wideparen{\alpha}^{\Delta}$ and $\wideparen{u_i}^{\Delta}$ where $\Delta$ represents the length scale of the grid cell. A second mesh of length scale $l_V \geq \Delta$ is defined and integrating each of the grid quantities over the volume gives the corresponding volume integrated values such as $\widehat{\rho}^V$, $\wideparen{\alpha}^V$ and $\wideparen{u_i}^V$. They are used for computing the explicit volume diffusion coefficient. For example, the volume integrated volume fraction and velocity are given by
\begin{equation}
	\wideparen{\alpha}^{V} = \frac{\int_V \widehat{\rho}^{\Delta} \wideparen{\alpha}^{\Delta} d V}{\widehat{\rho}},
\end{equation}
\begin{equation}
	\wideparen{u_i}^{V} = \frac{\int_V \widehat{\rho}^{\Delta} \wideparen{u_i}^{\Delta} d V}{\widehat{\rho}}.
\end{equation}
They are taken as inputs for Eq.\,(\ref{eq:DV}) for the calculation of $D_V^{V}$. The values of volume diffusion on the grid, $D_V^\Delta$, that are used in the discretised transport equations are obtained from $D_V^{V}$ by linear interpolation. In the interface region $l_V$ is generally larger than $\Delta$ (discussion of this later) {such that the volume averaged interface region is resolved by the LES grid.} In the interior of each fluid it is convenient and logical to make the two scales equal.
\\\\
Some additional explanations are needed about the implementation of the volume averaged surface tension model given by Eq.\,(\ref{eq:Fs}). The magnitude of the integrated surface tension force on the explicit volume mesh, $\widehat{F_{s}}^V$ is firstly computed by using the volume integrated values of the corresponding parameters in Eq.\,(\ref{eq:Fs}), and then the magnitude of the force at each grid cell, $\widehat{F_{s}}^\Delta$, is obtained by linear interpolation from from $\widehat{F_{s}}^V$. To better capture the interface dynamics, the direction of the force is given directly from the grid resolved interface normal direction.
\\\\
\noindent The numerical schemes in \textit{evdFoam} are adapted from the previously validated standard OpenFOAM solver called \textit{interFoam} \cite{deshpande2012evaluating}. A forward Euler scheme is used for the discretisation of the unsteady time derivatives. The convection and diffusion terms are discretised through second-order central difference schemes, with the exception of the convection of volume fraction which is solved using the multidimensional universal limiter for explicit solution (MULES) algorithm \cite{weller2008new} that is based on the flux corrected transport (FCT) technique \cite{zalesak1979fully}. MULES guarantees boundedness of volume fraction fields for immiscible interfaces with large liquid/gas density ratios and {a limited mass flux is obtained as input for solving} the momentum equation. The volume averaging operation and the numerical implementations do not violate the intrinsic mass conservation of the VoF algorithm. This has been demonstrated by simulations of interfacial shear flows in Section~\ref{sec:shearCases} where the total volume fraction is shown to remain constant during the simulations. In the test cases investigated here, the maximum interface Courant number is $Co_i = 0.25$ but the numerical method integrates the volume fraction over four sub-steps such that it is effectively 0.0625. At the same time the maximum flow Courant number is $Co_f = 0.75$. Reducing both Courant numbers to $Co_i=Co_f=0.1$ does not affect the numerical solutions.

\section{Validation of EVD based on interfacial shear layer cases}\label{sec:shearCases}
\noindent A series of 2D laminar and 3D turbulent interfacial shear flows are constructed to validate the EVD model. For comparison {purposes, a database} is generated by resolved flow simulations (RFS) that are described in detail below. Initially this data is used in an \textit{a priori} analysis to demonstrate the accuracy of the closures for sub-volume velocity fluctuations, sub-volume flux, sub-volume stress and volume averaged surface tension force, and to provide estimates for the unknown model constants. Subsequently, EVD simulations are performed to validate the overall model and its numerical implementation.

\subsection{Resolved flow simulations and case setup}\label{sec:aprioriSetup}
\begin{figure}[h!]
 \centering
 \subfigure[]{\includegraphics[clip=true, trim=380 300 380 300, height=0.24\textwidth]{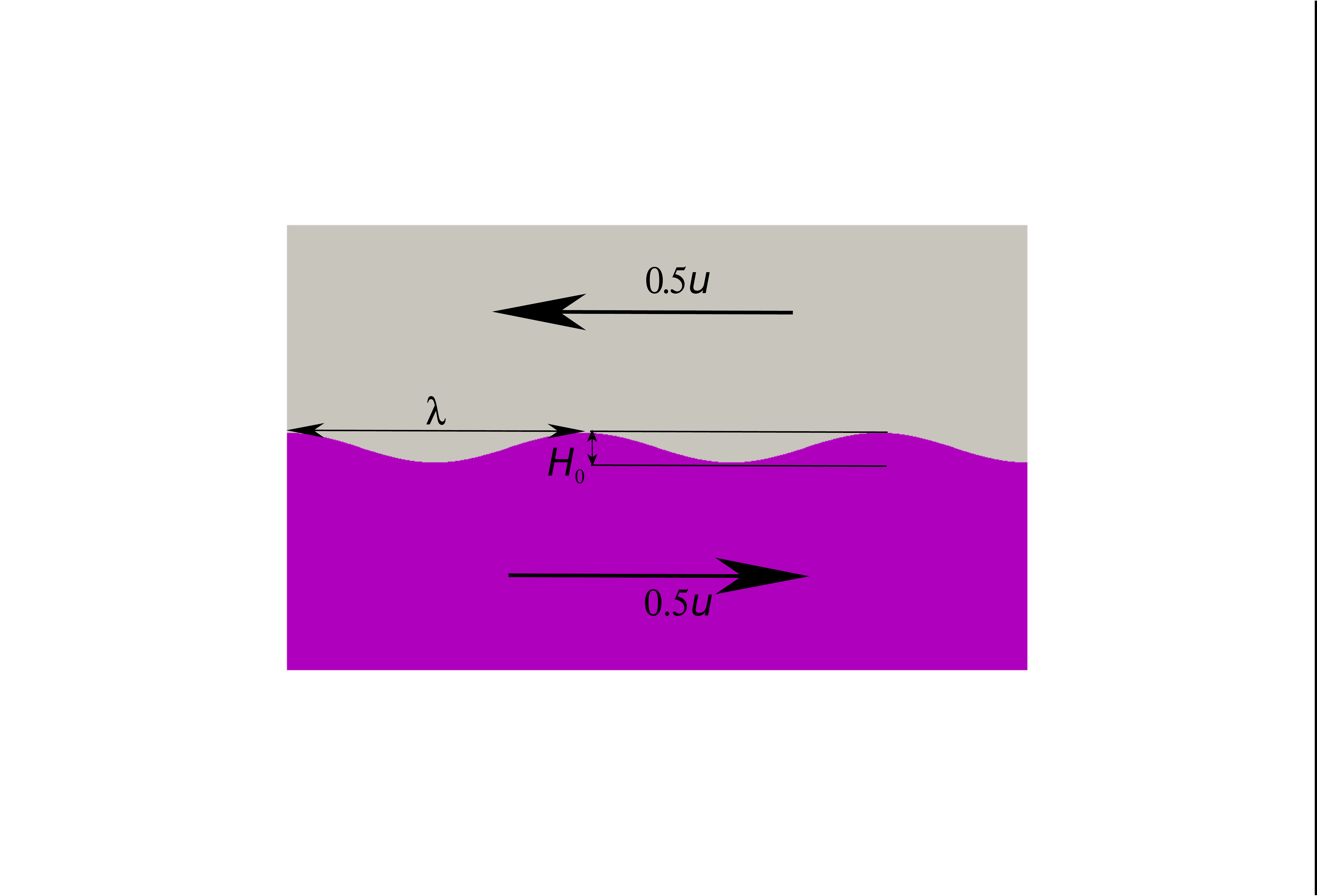}\label{fig:intialWave}}
 \subfigure[]{\includegraphics[clip=true, trim=380 300 380 300, height=0.24\textwidth]{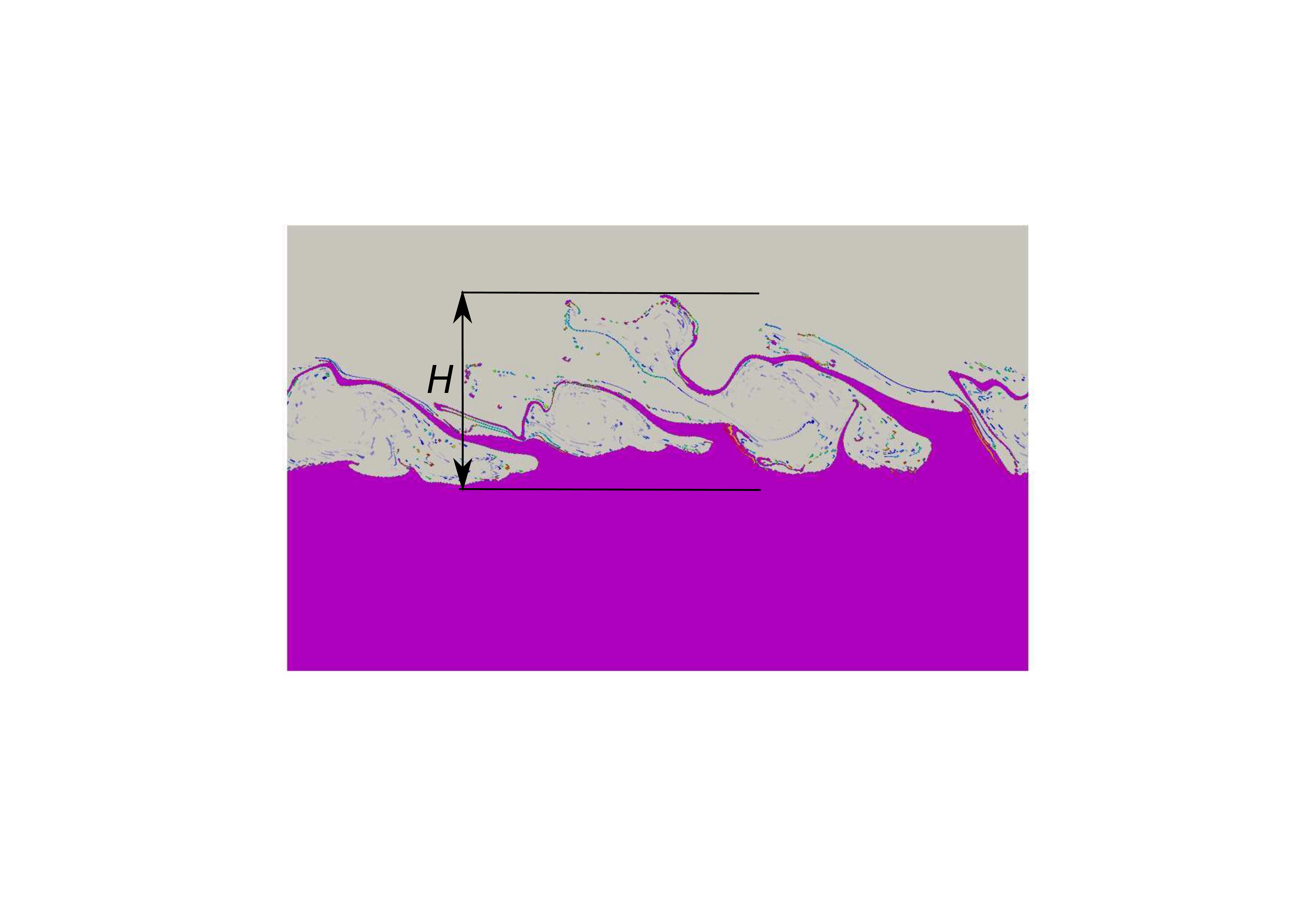}\label{fig:tWave}}
\caption{Flow configuration at initial conditions (left) and after the shear layer has developed (right). Images are for a RFS of C1.}\label{fig:picSetup}
\end{figure}

\begin{figure}
\footnotesize 
 \centering
 \includegraphics[clip=true, trim=0 0 0 0, height=0.3\textwidth]{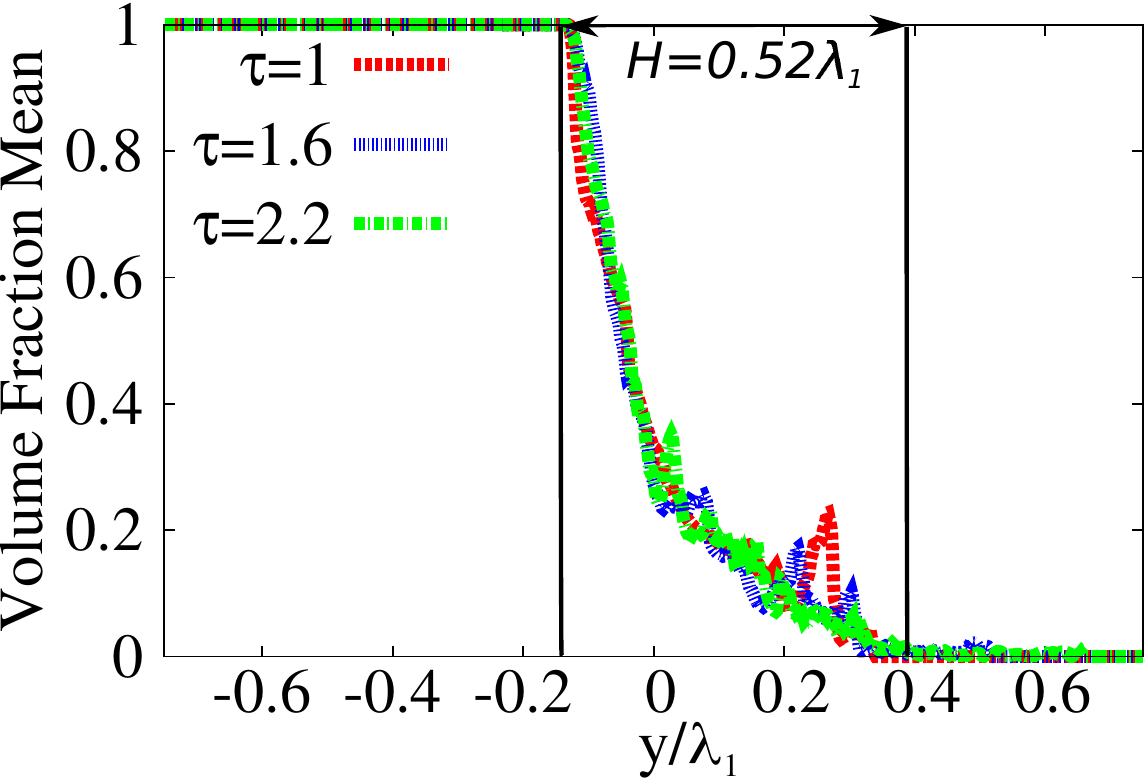}
\caption{Mean volume fraction in the transverse direction, $\overline{\alpha_y}$, at three instants of normalised time. The vertical lines correspond to $\overline{\alpha_y}=0.99$ and $\overline{\alpha_y}=0.01$, respectively, and the distance between them is equivalent to the outer length scale, $H$.}\label{fig:AlphaYC1delta}
\end{figure}

\noindent A high-fidelity RFS database is generated through the VoF method with the addition of an artificial compression term to counter the effects of numerical diffusion and sharpen the interface \cite{jasak1995interface,weller2008new}. This numerical model is part of OpenFOAM's \textit{interFoam} solver. A comprehensive assessment of this method is presented in \cite{deshpande2012evaluating} where, subject to sufficient resolution, it has been demonstrated to capture the physics of interfacial flows in similar flow configurations to those investigated here. We have separately validated the VoF artificial compression method (abbreviated as VoF-AC) for the canonical flow case of Popinet \cite{popinet1999front} but this {is} not shown here for brevity. As a further test of the accuracy of VoF-AC for RFS, the Appendix contains a comparison against the \textit{isoAdvector} method \cite{roenby2016computational} in which an iso-surface concept is used to reconstruct the unresolved interface inside grid cells as an alternative to artificial compression. An important advantage of using the VoF-AC approach to generate RFS data and for validations of EVD is that both solvers use the same numerical schemes thus eliminating those a potential source of difference. Note that the VoF-AC method is not intended for under-resolved simulations, such as LES or RANS, where turbulent fluctuations need to be accounted for. A recent study by Anez et al. \cite{anez2019eulerian} proposed a hybrid VoF-AC and turbulent diffusion method. The present EVD model overcomes the problems of numerical diffusion without the need for artificial compression (which is a numerical concept), and some turbulent simulations are presented below. 
\\\\
In VoF-AC the volume fraction transport equation is given by \cite{jasak1995interface,weller2008new}
\begin{equation}\label{eq:vofSR}
\frac{\partial \alpha}{\partial t}+\frac{\partial \alpha u_{i}}{\partial x_i} + \frac{\partial \alpha \left(1-\alpha\right) u_r} {\partial x_i} = 0,
\end{equation}
where
\begin{equation}\label{eq:vofUr}
u_r = C_r\left|u_i\right|\frac{\nabla \alpha}{\left|\nabla \alpha\right|}
\end{equation}
is a compression velocity and the model parameter $C_r = 1$ which is the standard value although values between 1 and 4 have been used elsewhere \cite{weller2008new}. In the VoF approach, values of $\alpha$ other than zero and one are generated only by numerical diffusion, and due to there not being any diffusion in a molecular sense, the numerical diffusion can be dominant and grid convergence can be elusive. The artificial compression term in Eq.~(\ref{eq:vofSR}) contains the weighting $\alpha \left(1-\alpha\right)$ which is non-zero only for $0 < \alpha < 1$.
\\\\
\noindent Five interfacial shear-layer cases characterised by different Reynolds numbers and different heavy and light fluid density ratios are investigated. The key parameters are listed in Table\,\ref{tab:set}. The initial conditions are configured as shown in Fig.\,\ref{fig:intialWave} for C1 with similar setups for the other
\begin{table}[h!]
  \begin{center}
\def~{\hphantom{0}}
  \begin{tabular}{lccccc}
  \toprule
    \textbf{Case}  & \textbf{Dimension}  & $\bf{Re_0}$ & $\bf{Re}$ & $\boldsymbol{\rho_h/\rho_l}$ \\[3pt]
    \midrule
    ~C1 & ~~~~2D~~~~  & ~320   & ~2368 & ~10\\
    ~C2 & ~~~~2D~~~~  & ~~32   & ~~156 & ~10\\    
    ~C3 & ~~~~2D~~~~  & ~640   & ~2144   & 100\\       
    ~C4 & ~~~~3D~~~~  & 2224   & 30570  & ~10\\
    ~C5 & ~~~~3D~~~~  & 2224   & 23437  & 100\\
    \bottomrule    
  \end{tabular}
  \caption{Interfacial shear-layer cases. $Re_0$ is the initial Reynolds number, $Re$ is the outer-scale Reynolds number and $\rho_h/\rho_l$ is the heavy to light fluid density ratio.}
  \label{tab:set}
  \end{center}
\end{table}
two-dimensional laminar cases, C2 and C3. The coordinate origin is at the centre of the domain. The location of the interface is initially given by $ y(x)=-0.05\lambda_1 \cos[(2 \pi / \lambda_1 )x-1.25 \lambda_1] $ where $\lambda_1=1$ is the initial wavelength. The peak to trough height is $H_0=0.1\lambda_1$. The domain spans from -1.25$\lambda_1$ to 1.25$\lambda_1$ in the axial direction and from -0.75$\lambda_1$ to 0.75$\lambda_1$ in the transverse direction. Similar configurations are found in numerous studies, e.g. \cite{tauber2002nonlinear,shirani2006turbulence,hoepffner2011self}. The computational domain for the 3D cases, C4 and C5, have the same configuration as the turbulent, single-phase mixing layer in \cite{hawkes2007scalar}. The interface is initialised as $y(x)=-0.139\lambda_2 \cos[(2 \pi / \lambda_2 )x-1.25 \lambda_2]$ with $\lambda_2 = 0.003456$. The domain spans from -1.25$\lambda_2$ to 1.25$\lambda_2$ in the axial direction and from -1.458$\lambda_2$ to 1.458$\lambda_2$ in the transverse direction. The additional cross-stream dimension spans from -0.833$\lambda_2$ to 0.833$\lambda_2$. The value of $H_0$ is set to be $0.278\lambda_2$ which is also the same as the height of the central jet in \cite{hawkes2007scalar}. For all cases, the heavy fluid below the interface is moving to the right with a velocity of 0.5$U$, and the light fluid above moves to the left at the same speed. The kinematic viscosity of both fluids is the same, $\nu = \nu_h = \nu_l$ and two different density ratios are considered with the density of the heavy fluid set to 1000~kg/m$^3$ in all cases. The surface tension coefficient is set to be constant with $\sigma = 0.07$~Nm$^{-1}$. The initial Reynolds number is defined as $Re_0=UH_0/\nu$ and is given in Table\,\ref{tab:set}. As the shear layer evolves, the outer length scale of the interfacial region, $H$, increases and eventually reaches an approximately stable value. A snapshot of the shear layer for C1 at a later time is shown in Fig.\,\ref{fig:tWave}. The corresponding outer-scale Reynolds number is $Re=UH/\nu$ \cite{rahmani2014effect}. The scale $H$ is evaluated as shown in Fig.\,\ref{fig:AlphaYC1delta} and is based on the profile of mean volume fraction in the transverse direction and spatially averaged in the other directions. It is defined as the length scale corresponding to the range of values $0.01<\overline{\alpha_y}<0.99$. As can be seen in the figure, $H$ is approximately the same at three different instants in time; namely, $\tau=1$,1.6 and 2.2, where $\tau=t/t_f = t /(L_x/U)$ is the time normalised by the characteristic domain flow through time where $L_x$ is the domain length in the streamwise direction. Past work \cite{rahmani2014effect,breidenthal1981structure,koochesfahani1986mixing} suggests that three-dimensional turbulent motions arise for $Re$ between 3000 and 6500. Therefore, in the present study the 2D simulations denoted as C1~-~C3 are classified as laminar flows and, and C4 and C5 are 3D turbulent cases. In C1 we set $U = 10$~m/s.
Then, in C2 the Reynolds number is reduced by increasing $\nu$ by a factor of 10 while for C3 we set $U=20$~m/s to increase $Re_0$ relative to C1 while $Re$ at later times is approximately the same. For C4 and C5 we set $U=72.5$~m/s while the density ratio varies by a factor of 10 and $Re$ differs.  
\\\\
\noindent For the laminar cases C1~-~C3, the RFS have a grid resolution of $2000\times 1200$ in the streamwise and transverse directions, respectively. The thickness of the boundary layer on the light fluid side of the interface is resolved by at least 45 cells (see Section \,\ref{sec:hypoLv}). The fidelity is very high compared to some previous simulations \cite{desjardins2013direct,agbaglah2017numerical} which suggest that as few as two cells in the boundary layer is sufficient to capture interface topology. Results are also presented below for a more refined grid with $4000\times 2400$ cells. For the turbulent cases C4 and C5 the configuration and the grid resolution are similar to those used for the turbulent mixing layer reported in \cite{hawkes2007scalar}. The grid has $288\times336\times192$ cells in the streamwise, transverse and cross-stream directions, respectively, corresponding to 18.6 million cells in total. Since the real interface is infinitely sharp the smallest interfacial scales cannot not, in general, ever be fully resolved \cite{shinjo2010simulation,ling2017spray}. Following the widely adopted approach found in past publications \cite{agbaglah2017numerical,hasslberger2019flow,duret2012dns} the grid is chosen to ensure that the Kolmogorov scale and the interfacial boundary layers are adequately resolved. The cell size here is about 0.83 characteristic Kolmogorov length scale and the light fluid boundary layer (see Section\,\ref{sec:hypoLv}) is resolved by about 11 cells. It is also important to note that the focus of this work is on closure of sub-volume models and the results below demonstrate that they are numerically converged with the current grid resolution. To trigger the development of turbulence in C4 and C5 the initial velocity field is initialised with homogeneous turbulent perturbations with an amplitude of $0.05U$ and an integral length scale of $H_0/3$ in the region $-0.5H_0 < y < 0.5H_0$.
\begin{figure}
 \centering
 \subfigure[C1, $\tau=1$]{\includegraphics[clip=true, trim=390 300 390 300, height=0.195\textwidth]{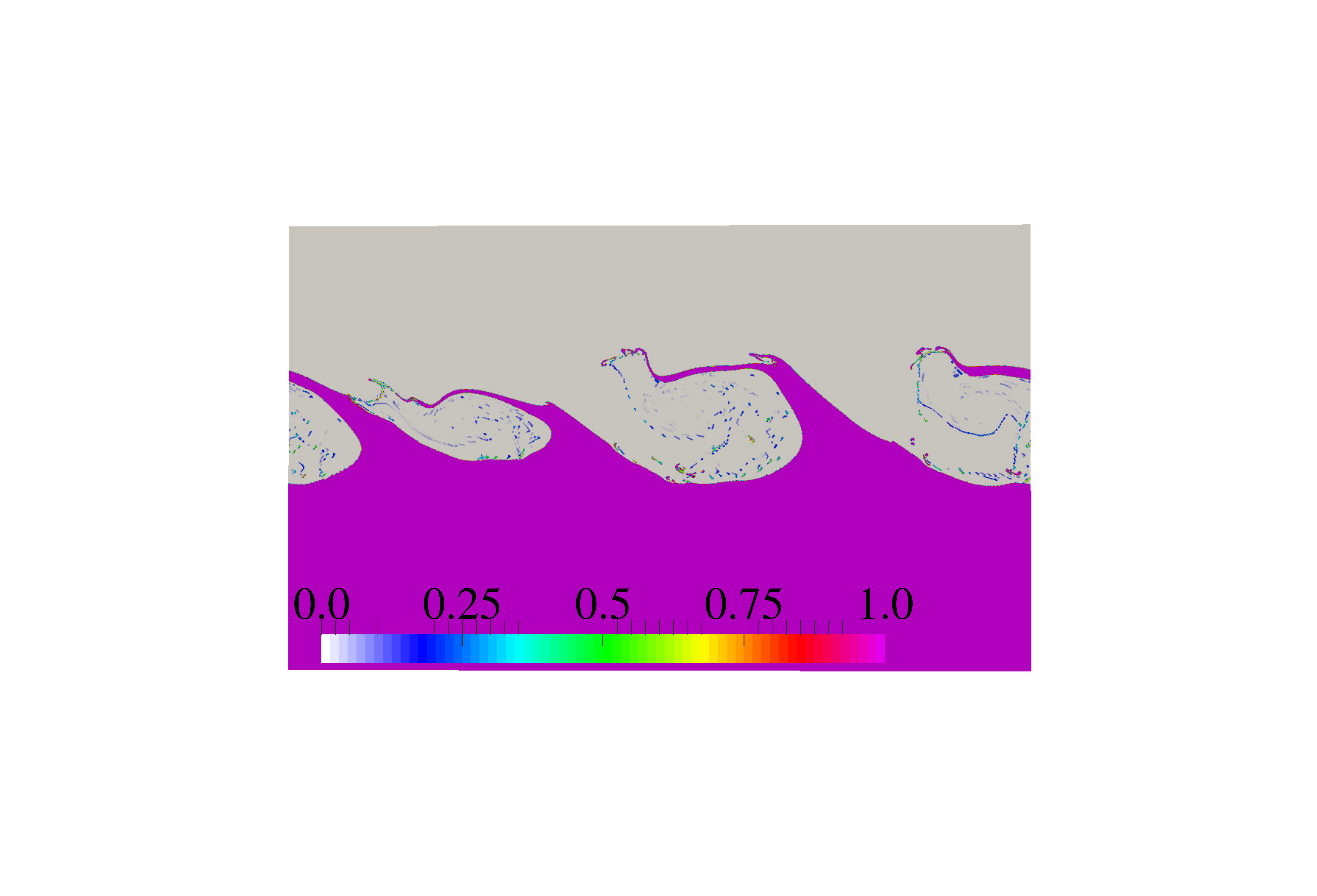}\label{fig:picRFSa}}
 \subfigure[C1, $\tau=1.6$]{\includegraphics[clip=true, trim=390 300 390 300, height=0.195\textwidth]{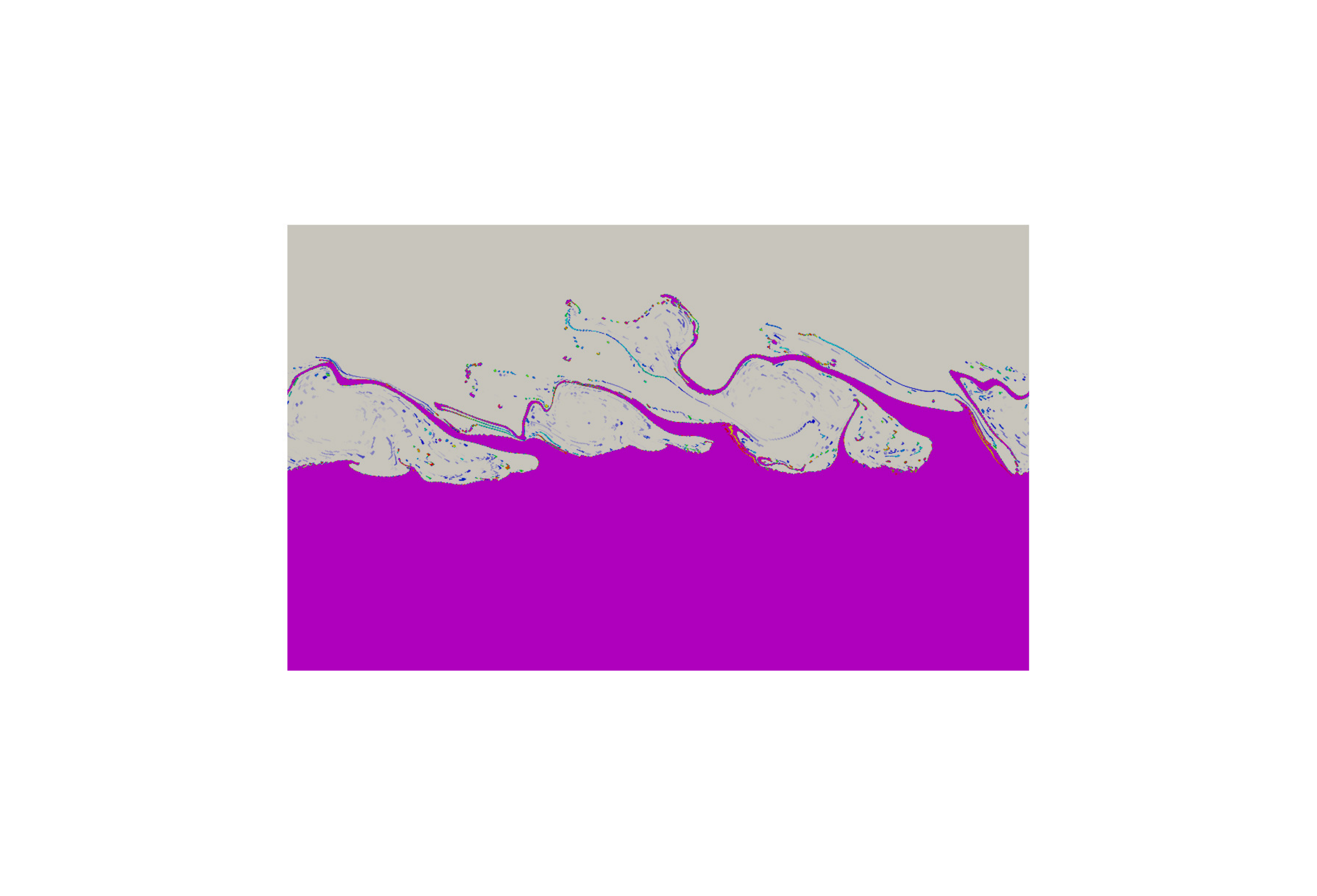}}
 \subfigure[C1, $\tau=2.2$]{\includegraphics[clip=true, trim=390 300 390 300, height=0.195\textwidth]{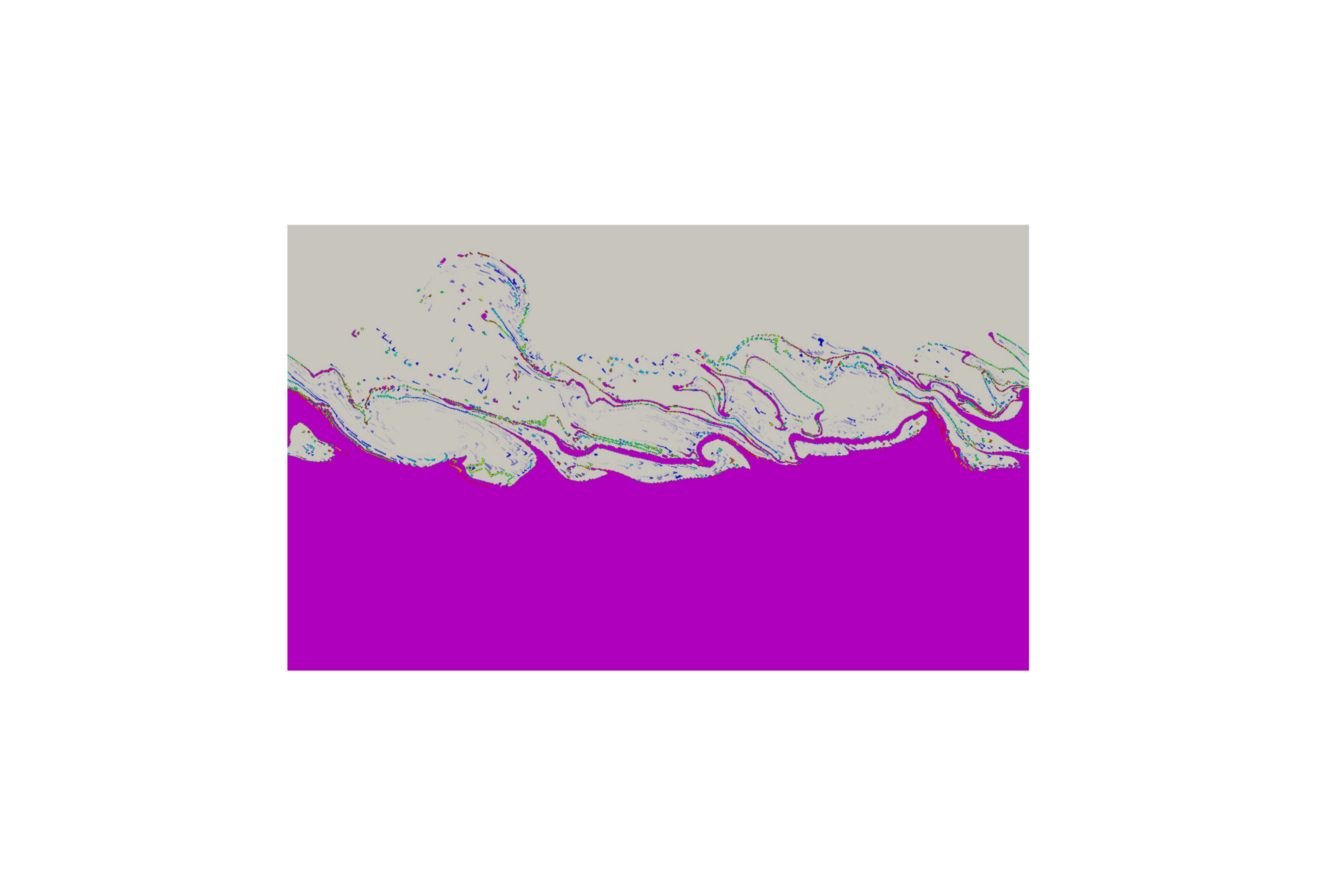}\label{fig:picRFSc}}

 \vspace{-0.2cm} 
 \subfigure[C2, $\tau=1.6$]{\includegraphics[clip=true, trim=390 300 390 300, height=0.195\textwidth]{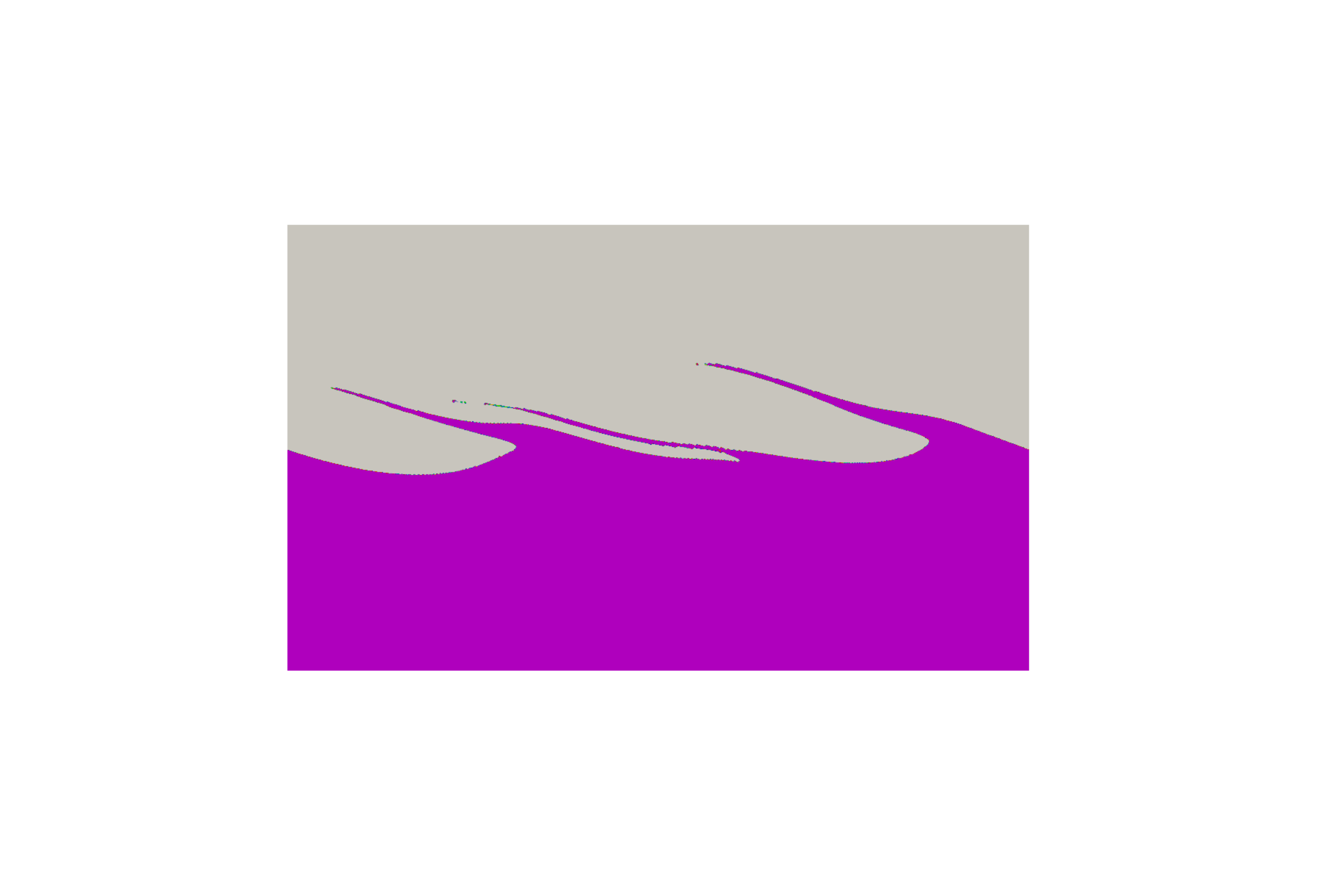}\label{fig:picRFSd}}
 \subfigure[C3, $\tau=1.6$]{\includegraphics[clip=true, trim=390 300 390 300, height=0.195\textwidth]{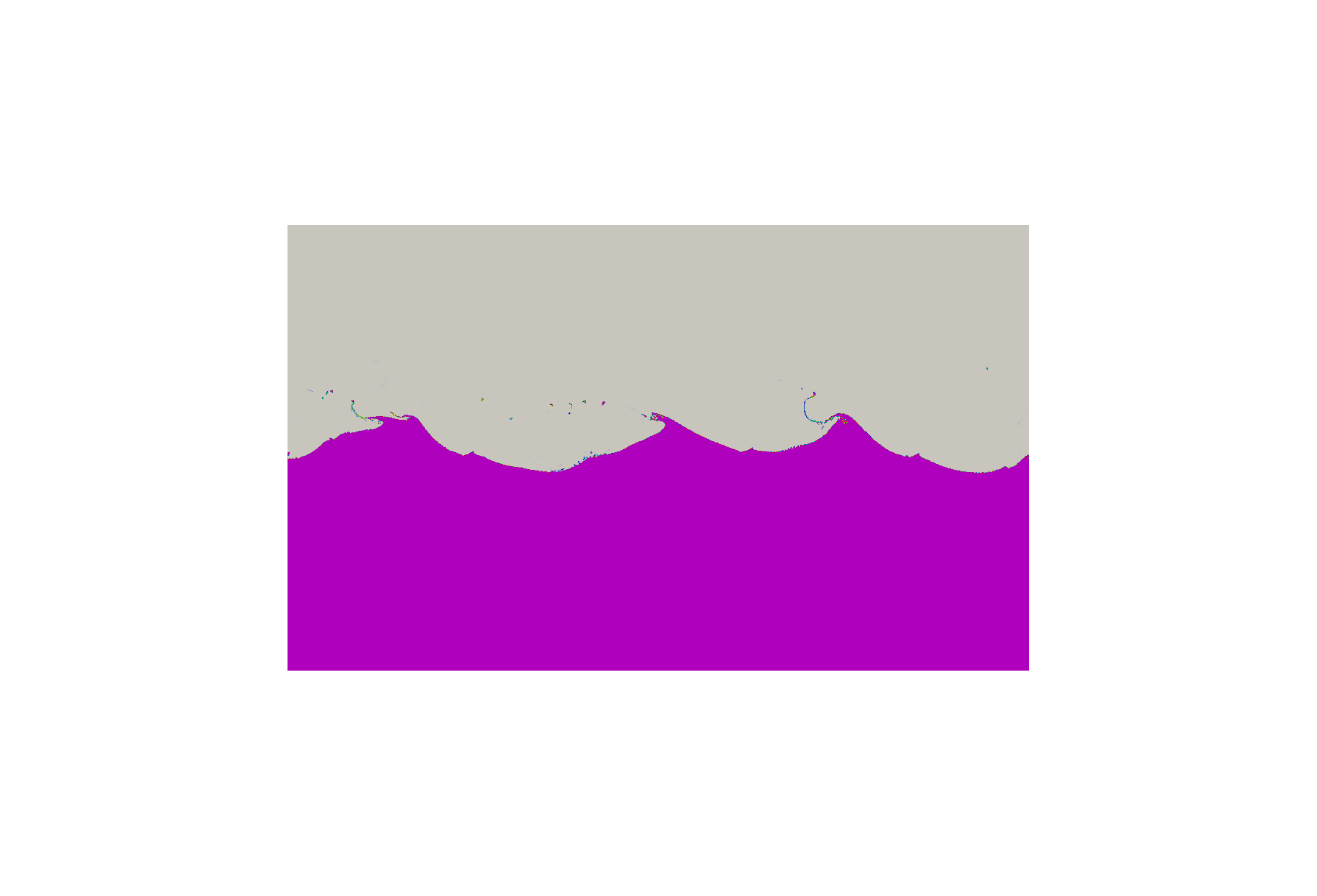}}
 
 
  \vspace{-0.4cm} 
 \subfigure[C4, $t_j=40$]{\includegraphics[clip=true, trim=400 0 400 100, height=0.4\textwidth]{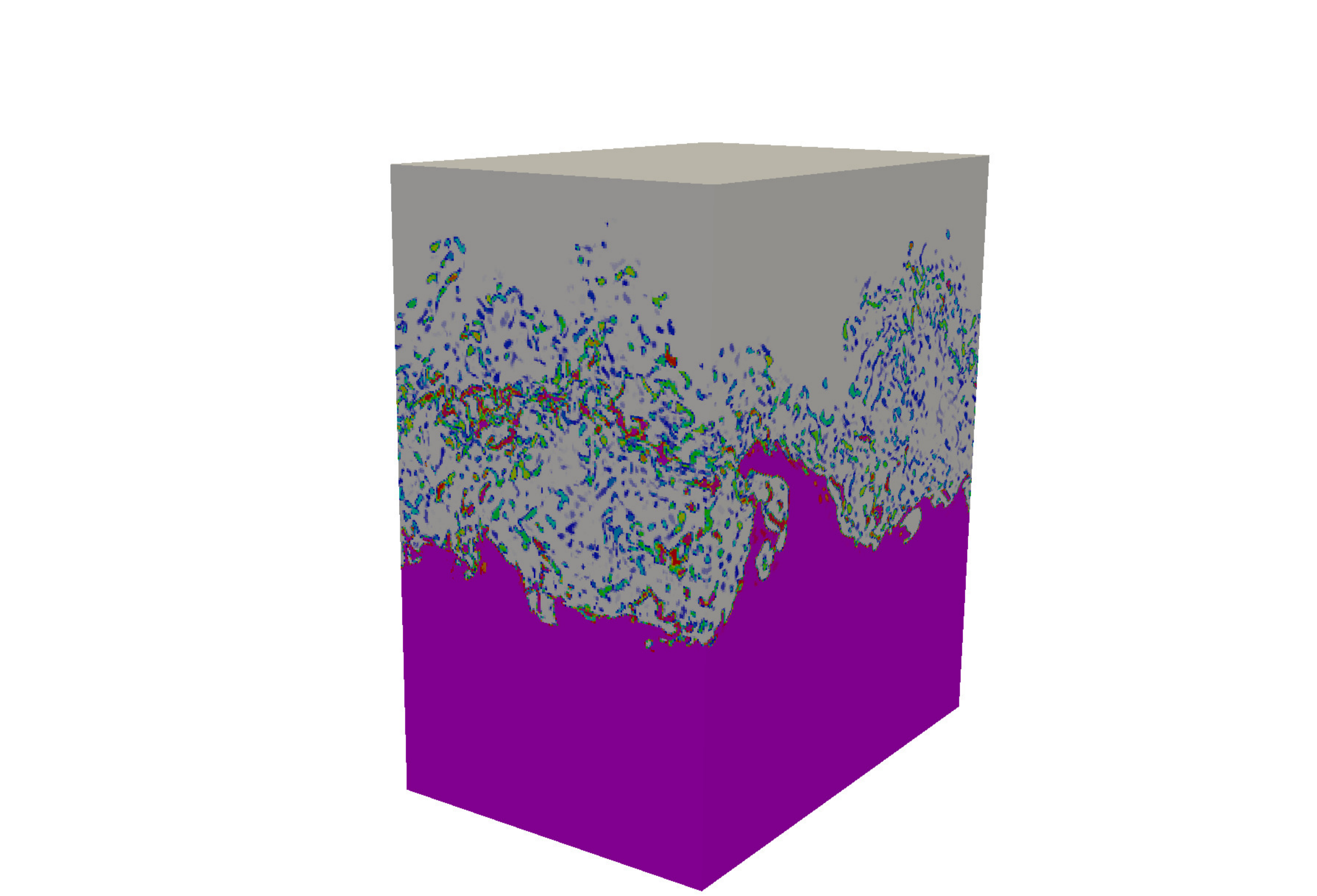}\label{fig:picRFSf}}
 \subfigure[C5, $t_j=40$]{\includegraphics[clip=true, trim=400 0 400 100, height=0.4\textwidth]{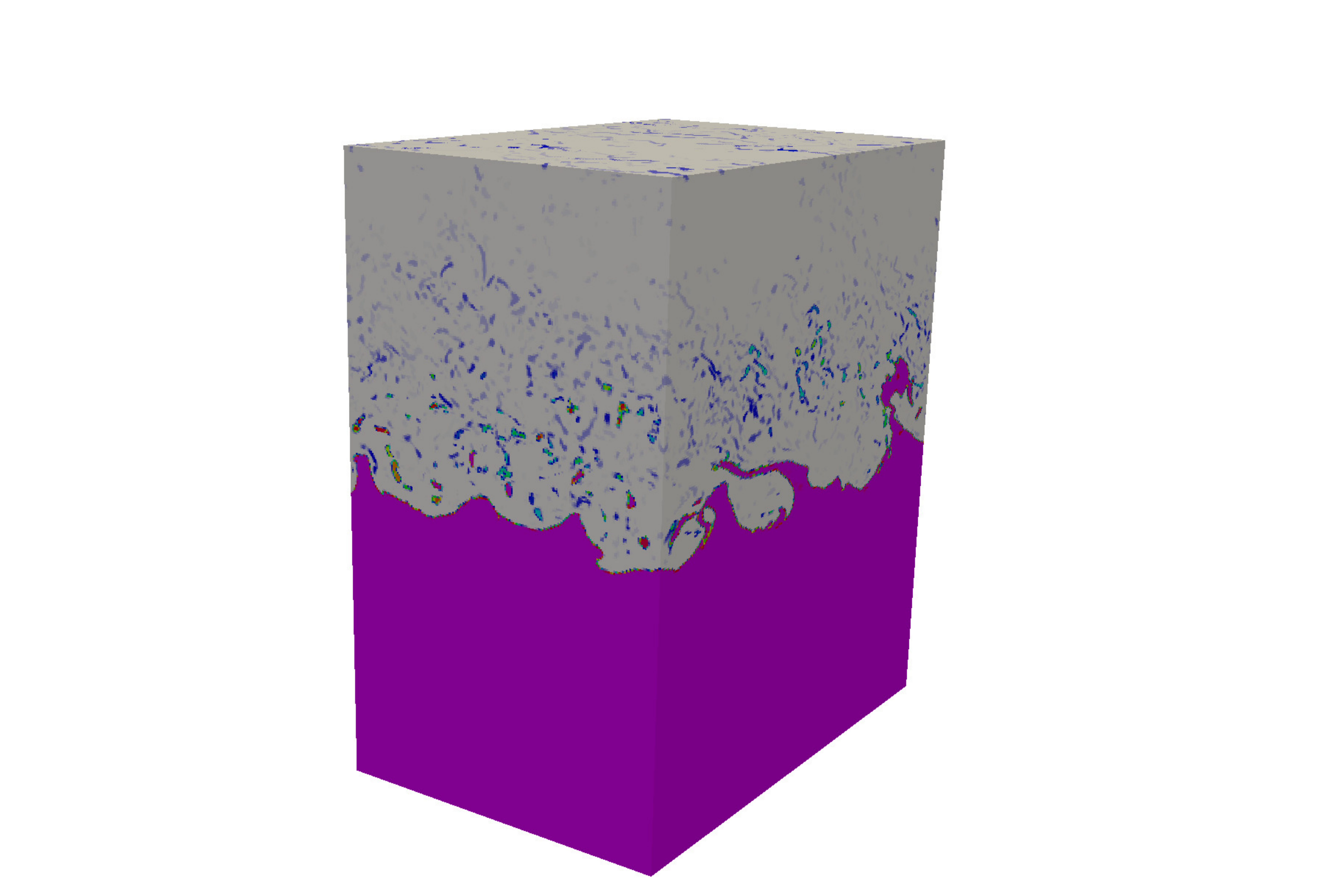}\label{fig:picRFSg}}
 
  \caption{Volume fraction fields for RFS of C1~-~C5.}\label{fig:picRFS}
\end{figure}
\\\\
\noindent Figure\,\ref{fig:picRFS} shows the volume fraction fields at three times instants, $\tau$=1, 1.6 and 2.2, for case C1 (Figs.\,\ref{fig:picRFSa}~-~\ref{fig:picRFSc}) and at a single point in time for C2~-~C5 (Figs.\,\ref{fig:picRFSd}~-~\ref{fig:picRFSg}). For C4 and C5 the time is normalised differently as $t_j = t/(H_0/U)$ \cite{hawkes2007scalar}. C1 falls into the laminar flow regime and the evolution of two-dimensional flow patterns can be observed. Similar to the visualisation in Hoepffner et al. \cite{hoepffner2011self}, asymmetric waves develop for this case with a moderate density ratio of 10. The vortices are generally unclosed but are covered by the elongated ligaments which stretch into the light fluid above the interface. During the temporal evolution, the number of ligaments increase due to the breakup of larger ones and rolling up new ones in the heavy fluid. When reducing the Reynolds number by increasing kinematic viscosity by one order of magnitude, case C2 is obtained. Compared with C1, case C2 has much sharper extended ligaments that are well preserved and do not break up because the larger kinematic viscosity dissipates the kinetic energy and the surface tension is sufficient to stabilise the ligament structures. For C3 with a much larger density ratio, the light fluid on top of the interface imparts a relatively smaller inertial drag force which is not sufficient to pull the heavy fluid into elongated ligaments. Similar behaviour is seen in \cite{hoepffner2011self}). Cases C4 and C5, exhibit obvious three-dimensional turbulent structures and the waves develop in both the streamwise and cross-stream directions. C5 has a much larger denisty ratio and, once again, the light fluid imparts relatively smaller inertial drag and the ligaments do not develop to the same length as they do in C4.

\subsection{\textit{A priori} assessment of sub-volume closures}\label{sec:apriori}
\noindent In this section, the statistics from the resolved flow simulations are used for the evaluation of the sub-volume closures through an \textit{a priori} analysis in which the model input parameters are taken from the RFS data and the model output quantity is compared to the equivalent RFS quantity. The EVD are not being solved in this stage of the validation. To avoid redundancy, only C1, C2 and C4 are considered here but later, when the EVD model is validated as a whole, all five cases are used.

\subsubsection{Sub-volume velocity fluctuations}\label{sec:Uprime}
\begin{figure}[h!]
\footnotesize 
 \centering
 \includegraphics[clip=true, trim=0 0 0 0, width=0.95\textwidth]{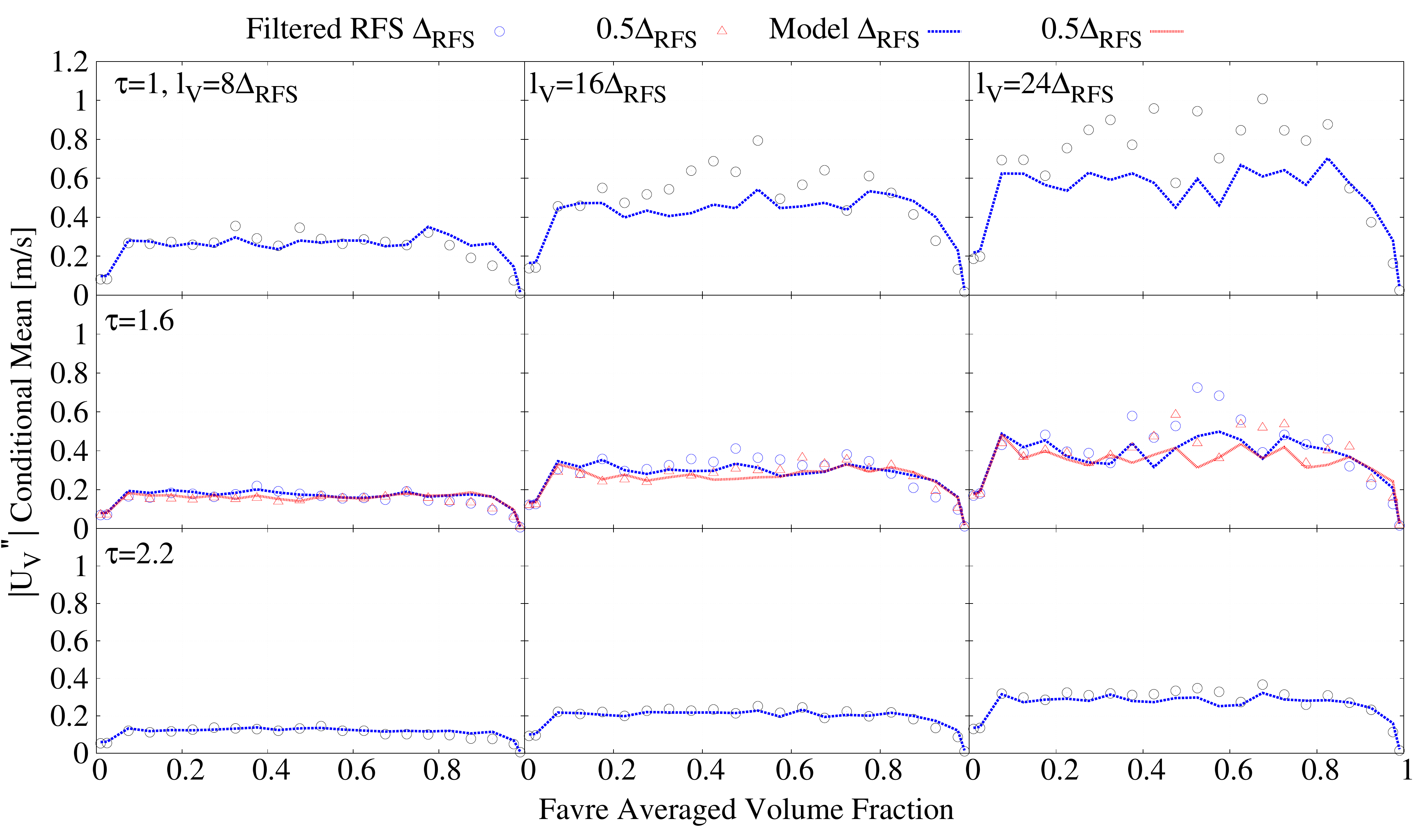}
\caption{RMS of sub-volume of velocity fluctuations conditionally averaged on volume fraction for C1 with three different values of $l_V$ and at three different times. Symbols represent filtered RFS data for two different grid resolutions, and lines represent model results using Eq.~(\ref{eq:magUprime}) with $C_{SL}=0.33$.}\label{fig:UrmsApriC1}
\end{figure}

\begin{figure}[h!]
\footnotesize 
 \centering
 \includegraphics[clip=true, trim=0 0 0 0, width=0.95\textwidth]{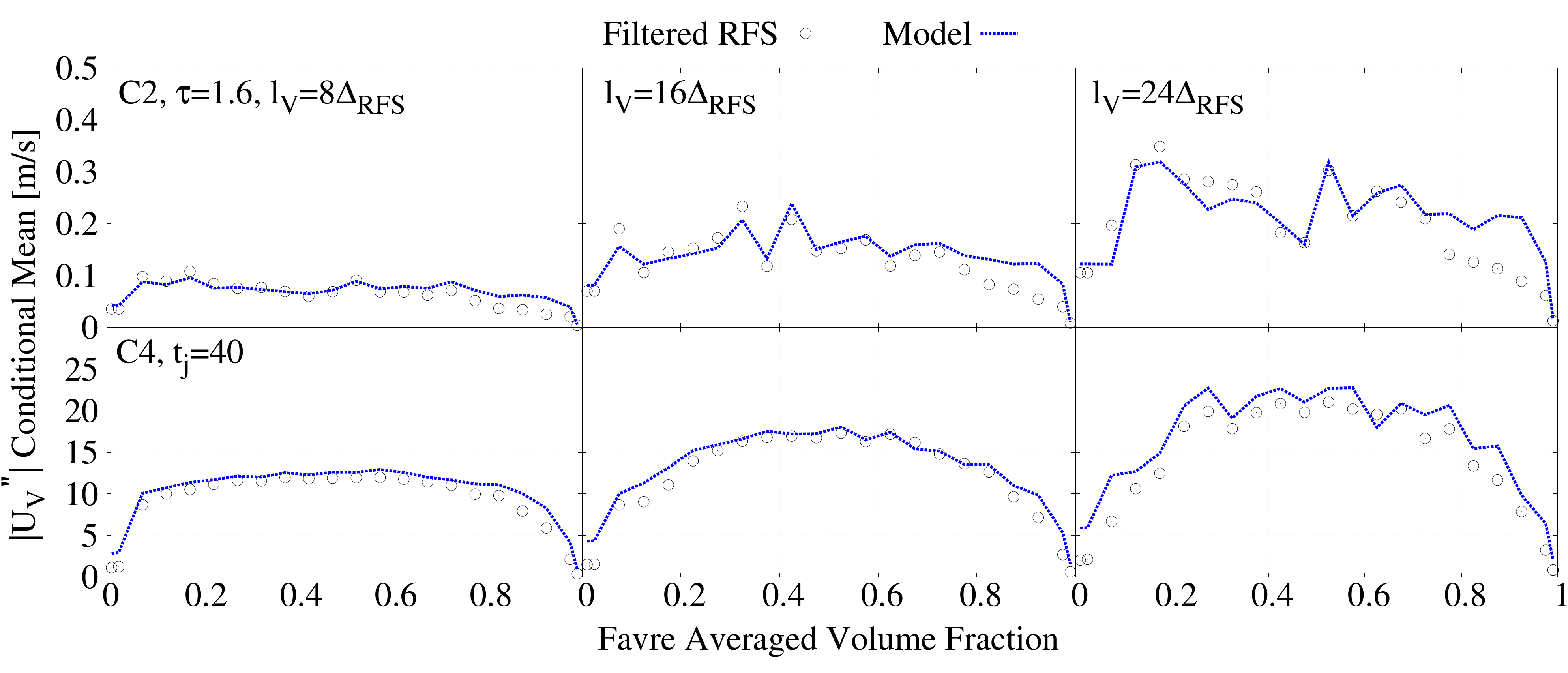}
\caption{RMS of sub-volume of velocity fluctuations conditionally averaged on volume fraction for C2 at $\tau=1.6$ (top row) and C4 at $t_j=40$ (bottom row) with three different values of $l_V$. Symbols represent filtered RFS data, and lines represent model results using Eq.~(\ref{eq:magUprime}) {with $C_{SL}=0.33$ for C2 (laminar) and $C_{SL}=0.66$ for C4 (turbulent)}.}\label{fig:UrmsApriC2C4}
\end{figure}

Figures\,\ref{fig:UrmsApriC1} and \ref{fig:UrmsApriC2C4} compare the model for sub-volume rms of velocity fluctuations, $\left|u_V^{\backprime\backprime}\right|$, given by Eq.~(\ref{eq:magUprime}) with the data extracted from resolved flow simulations for C1, C2 and C4. Since it appears in the closures for the explicit volume diffusion diffusion coefficient and viscosity, $D_V$ and $\nu_V$, and the volume averaged surface tension force, $\left|u_V^{\backprime\backprime}\right|$ is a critical part of the EVD model. The validity of the closure and the most suitable value for the model constant $C_{SL}$ are assessed through examination of mean values conditioned on $\wideparen{\alpha}$ which are presented in Fig.~\ref{fig:UrmsApriC1} for C1 with three different values of the explicit length scale, $l_v$, and at three time instants. For convenience $l_V$ is expressed as a multiple of $\Delta_{RFS}$, but in general it is selected independently of the grid. The modelled $\left|u_V^{\backprime\backprime}\right|$ 
shown by lines are for {$C_{SL}=0.33$} and are in good agreement with the RFS data shown by symbols for all $l_V$ and all times. The RFS data is filtered over {an} explicit volume of size $V \sim l_V^3$. To demonstrate that the RFS data is converged, statistical results at $\tau=1.6$ are also shown for the more refined grid with half the base grid scale, i.e. $0.5\Delta_{RFS}$. Although not shown, the grid convergence at the other time instants is of similar quality.

The validity of the model for $\left|u_V^{\backprime\backprime}\right|$ is shown for C2 and C4 with lower and higher Reynolds numbers, respectively, in Fig.\,\ref{fig:UrmsApriC2C4}. Again, the agreement with the RFS data is good and the results correctly scale as $l_V$ is varied. {It should be noted $C_{SL}=0.66$ is suggested for matching the results of C4 in turbulent flows. Higher $C_{SL}$ can be attributed to the turbulent fluctuations in addition to the fluctuations due to interface dynamics. Importantly, the function shape given by Eq.~(\ref{eq:magUprime}) is good and provides predictions independent of the volume size and laminar/turbulent flow regimes.} All of the results indicate that the highest values of $\left|u_V^{\backprime\backprime}\right|$ occur near the middle of the $\alpha$ range; that is, in the middle of the interfacial region where volume fraction inhomogeneity is greatest. $\left|u_V^{\backprime\backprime}\right|$ decreases to a non-zero value near $\wideparen{\alpha} = 0$ due to the existence of a relatively thick boundary layer which extends from the interface out into the pure light fluid stream.
Conversely, the boundary layer on the heavy fluid side of the interface is very thin \cite{marmottant2004spray} such that the laminar heavy fluid stream is almost not disturbed. Therefore, $\left|u_V^{\backprime\backprime}\right|$ approaches zero near $\wideparen{\alpha}=1$. With an increase in the volume length scale, $\left|u_V^{\backprime\backprime}\right|$ become larger.

\subsubsection{Sub-volume flux}
\begin{figure}[h!]
\footnotesize 
 \centering
 \includegraphics[clip=true, trim=0 0 0 0, width=0.95\textwidth]{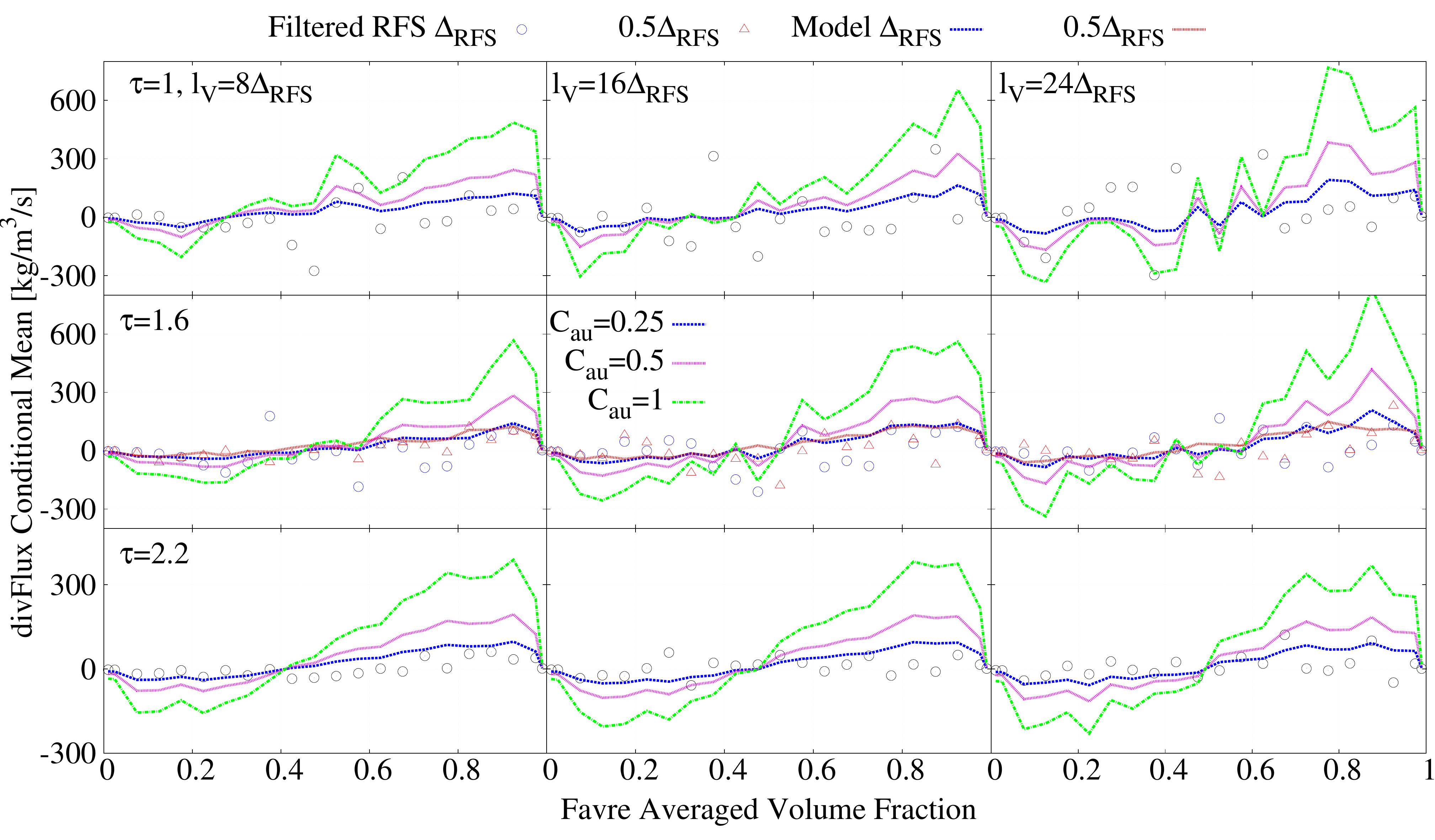}
\caption{Divergence of sub-volume flux conditionally averaged on volume fraction for C1 with three different values of $l_V$ and at three different times. Symbols represent filtered RFS data for two different grid resolutions, and lines represent model results using Eqs~(\ref{eq:svFlux}) and (\ref{eq:DV}) with $C_{\alpha u}=0.25$, $0.5$ and $1$.}\label{fig:FluxApriC1}
\end{figure}

\begin{figure}
\footnotesize 
 \centering
 \includegraphics[clip=true, trim=0 0 0 0, width=0.95\textwidth]{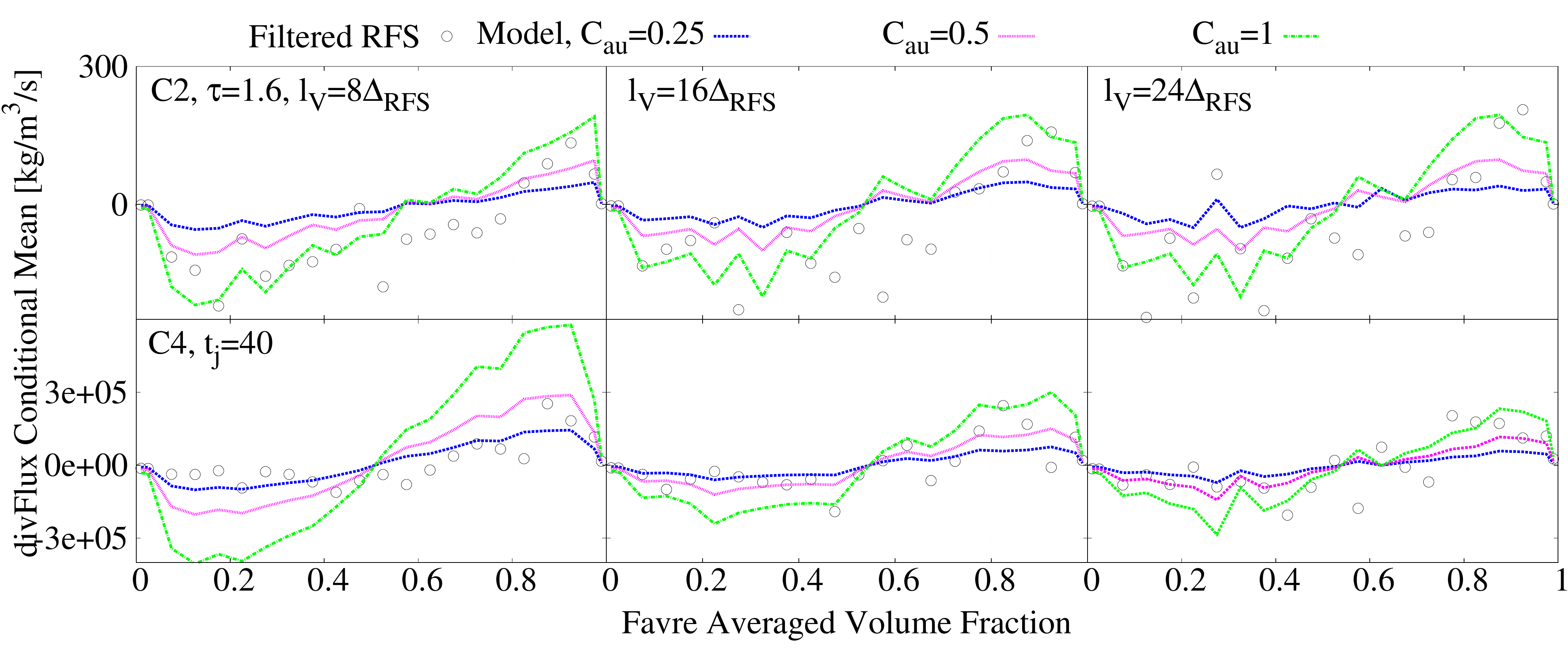}
\caption{Divergence of sub-volume flux conditionally averaged on volume fraction for C2 at $\tau=1.6$ (top row) and C4 at $t_j=40$ (bottom row) with three different values of $l_V$. Symbols represent filtered RFS data, and lines represent model results using Eqs~(\ref{eq:svFlux}) and (\ref{eq:DV}) with $C_{\alpha u}=0.25$, $0.5$ and $1$.}\label{fig:FluxApriC2C4}
\end{figure}

The sub-volume flux is modelled by the gradient diffusion model according to Eq.~(\ref{eq:svFlux}) which contains the explicit volume diffusion coefficient, $D_V$, modelled by Eq.~(\ref{eq:DV}) for which the constant $C_{\alpha u}$ requires calibration. The flux is a vector field and therefore to simplify the comparison between the model and the RFS data we take its divergence,
\begin{align}
\frac{\partial}{\partial x_i}J_{\alpha,i}^{va} = \frac{\partial}{\partial x_i}\widehat{\rho}\wideparen{\alpha^{\backprime\backprime} u_i^{\backprime\backprime}} = \frac{\partial}{\partial x_i}\left(-\widehat{\rho}D_V\frac{\partial{\wideparen{\alpha}}}{\partial x_i}\right).
\end{align}
This divergence conditionally averaged on $\wideparen{\alpha}$ is shown for both the model predictions and the filtered RFS data in Fig.\,\ref{fig:FluxApriC1} for C1 at three different times and in Fig.\,\ref{fig:FluxApriC2C4} for C2 and C4
at single time instants. The results are again shown for three different values of $l_V$. An additional set of results is shown for C1 at $\tau=1.6$ with the refined RFS grid scale, indicating grid convergence for this statistical quantity. {The sensitivity of the model to the constant $C_{\alpha u}$ in the range 0.25 to 1 is also presented and helps to demonstrate the congruence of the trends predicted by the model and given by the RFS data. The selection of $C_{\alpha u}$ is somewhat subjective and no single value produces the best match everywhere. In the remainder of this work we choose $C_{\alpha u}=0.25$ which appears to give the overall best fit of the model with the data. More importantly,} the model reproduces the correct trends with increasing $l_V$ and with volume fraction variations with the divergence of the sub-volume flux increasing from negative values at low $\wideparen{\alpha}$ to positive values at high $\wideparen{\alpha}$. The model also produces the correct trends over the wide range of Reynolds numbers between C2 and C4 {independent of laminar or turbulent regimes.} This confirms the general validity of the explicit volume diffusion concept although it is important to note that the model does not capture all of the detailed and sharp variations in the divergence, especially for C1 at $\tau=1$. This is caused by the presence of a relatively small number of intact long heavy-fluid ligaments along the interface (see Fig.\,\ref{fig:picRFSa}) that lead to insufficient statistical samples to produce a smooth profile. However, this is alleviated at later times as the shear layer develops and ligaments breakup and the model results at $\tau=2.2$ are very good. A similar finding can be made for C2 in Fig.\,\ref{fig:FluxApriC2C4} (1st row) corresponding to a lack of statistical variety of the ligaments (see Fig.\,\ref{fig:picRFSd}). 

\subsubsection{Sub-volume stress}
\begin{figure}[h!]
\footnotesize 
 \centering
 \includegraphics[clip=true, trim=0 0 0 0, width=0.95\textwidth]{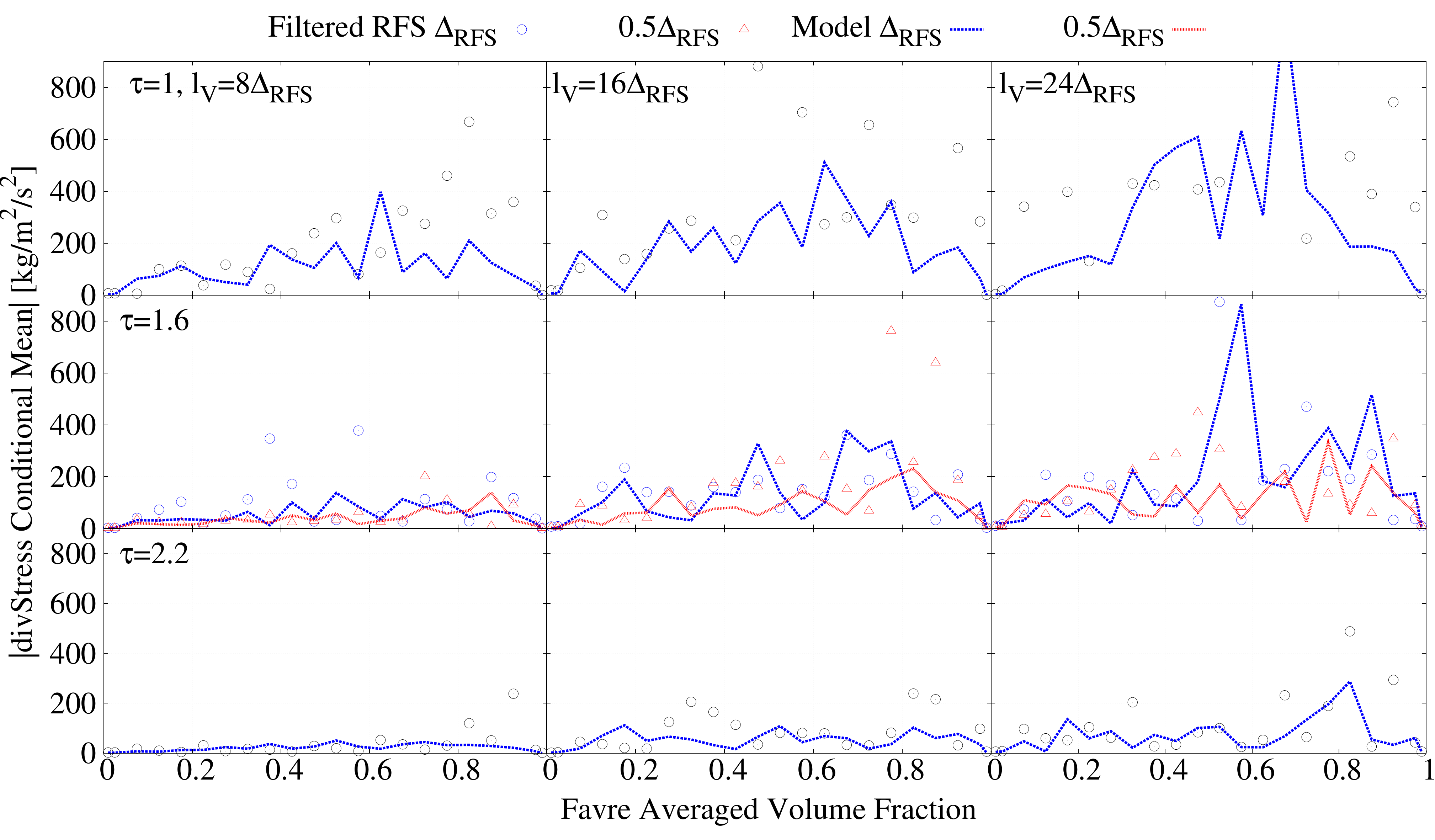}
\caption{Divergence of sub-volume stress conditionally averaged on volume fraction for C1 with three different values of $l_V$ and at three different times. Symbols represent filtered RFS data for two different grid resolutions, and lines represent model results using Eqs~(\ref{eq:subStress}) and (\ref{eq:nueff}) with $Sc_V=3$.}\label{fig:StressApriC1}
\end{figure}

\begin{figure}[h!]
\footnotesize 
 \centering
 \includegraphics[clip=true, trim=0 0 0 0, width=0.95\textwidth]{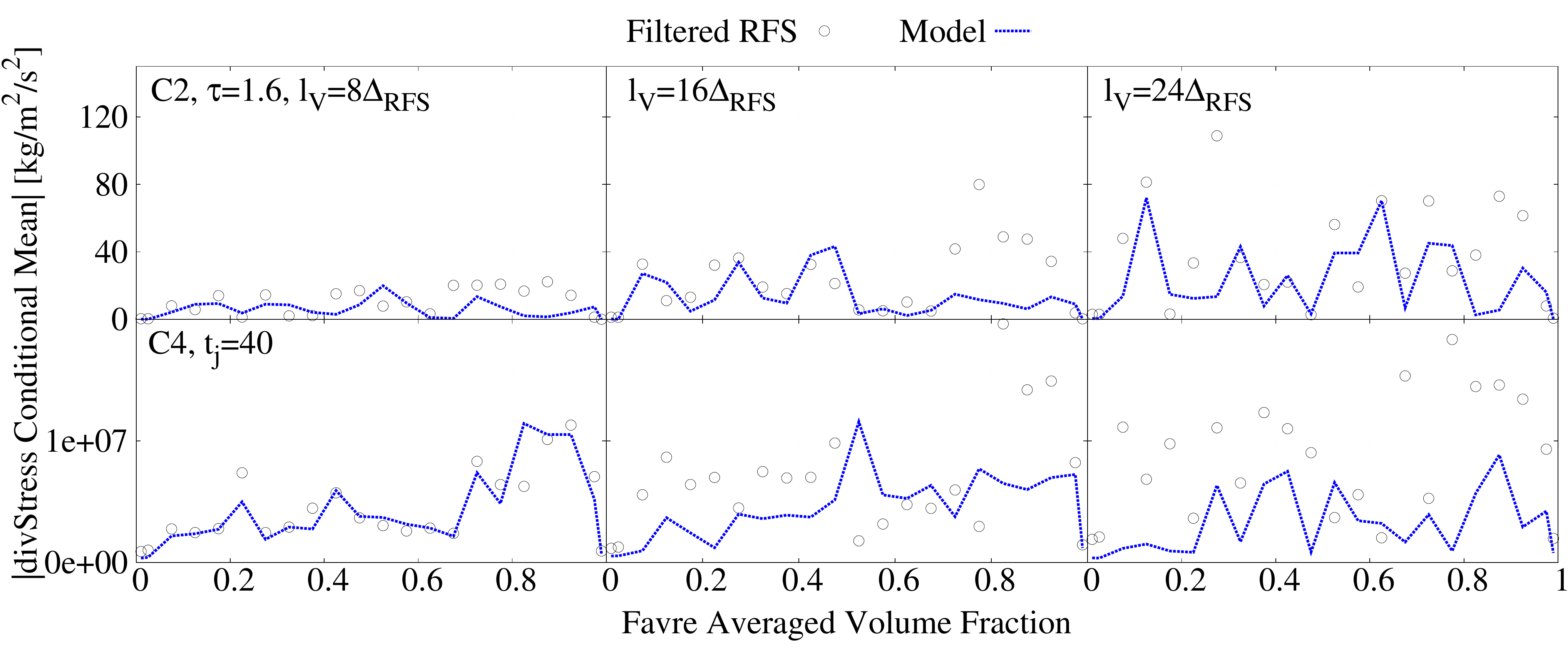}
\caption{Divergence of sub-volume stress conditionally averaged on volume fraction for C2 at $\tau=1.6$ (top row) and C4 at $t_j=40$ (bottom row) with three different values of $l_V$. Symbols represent filtered RFS data for two different grid resolutions, and lines represent model results using Eqs~(\ref{eq:subStress}) and (\ref{eq:nueff}) with $Sc_V=3$.}\label{fig:StressApriC2C4}
\end{figure}

\noindent The sub-volume stress is modelled according to Eq.~(\ref{eq:subStress}) introducing the explicit volume Schmidt number, $Sc_V$, and an effective turbulent viscosity, $\nu_t^{eff}$, that is given by Eq.~(\ref{eq:nueff}). Since the stress is a tensor, the divergence is obtained for comparison to the RFS data,
\begin{align}
\frac{\partial}{\partial x_j}\tau_{ij}^{va} = \frac{\partial}{\partial x_j} \widehat{\rho} \wideparen{u_i^{\backprime\backprime} u_j^{\backprime\backprime}} = \frac{\partial}{\partial x_j}\left[-2\widehat{\rho} \left(D_VSc_V+\nu_t^{eff}\right)\wideparen{S_{ij}}\right].
\end{align}
Figures\,\ref{fig:StressApriC1} and \ref{fig:StressApriC2C4}
show the divergence of the sub-volume stress conditionally averaged on $\wideparen{\alpha}$ for C1, C2 and C4 for three values of $l_V$. All model results are for $Sc_V=3$. The results C1 at $\tau=1.6$ are also shown for the RFS with double the resolution in all dimensions. Although there is more scatter in these results than for sub-volume velocity rms and sub-volume flux, the variation with refinement of the RFS grid remains relatively small indicating that the sub-volume stress statistics are generally converged. With a few statistical exceptions, the model results for C1 correctly predict the overall trends for the variation of the divergence of sub-volume flux with $\wideparen{\alpha}$ and with variations in $l_V$. for different volume length scales except for few protrusive points. The generally good performance of the model with a decrease in $Re$ (case C2) and an increase in $Re$ into the turbulent regime (case C4) is retained as shown in Fig.\,\ref{fig:StressApriC2C4}. For case C4, the sub-volume stress near the pure heavy and light fluid streams (i.e. $\wideparen{\alpha} \to 0$ and 1) approaches a non-zero value. There, $D_V \to 0$ but the sub-volume closure still provides satisfactory predictions since the effective turbulent viscosity is included for modelling as described by $\nu_t^{eff} $.

\subsubsection{Volume averaged surface tension force}
\begin{figure}[h!]
\footnotesize 
 \centering
 \includegraphics[clip=true, trim=0 0 0 0, width=0.95\textwidth]{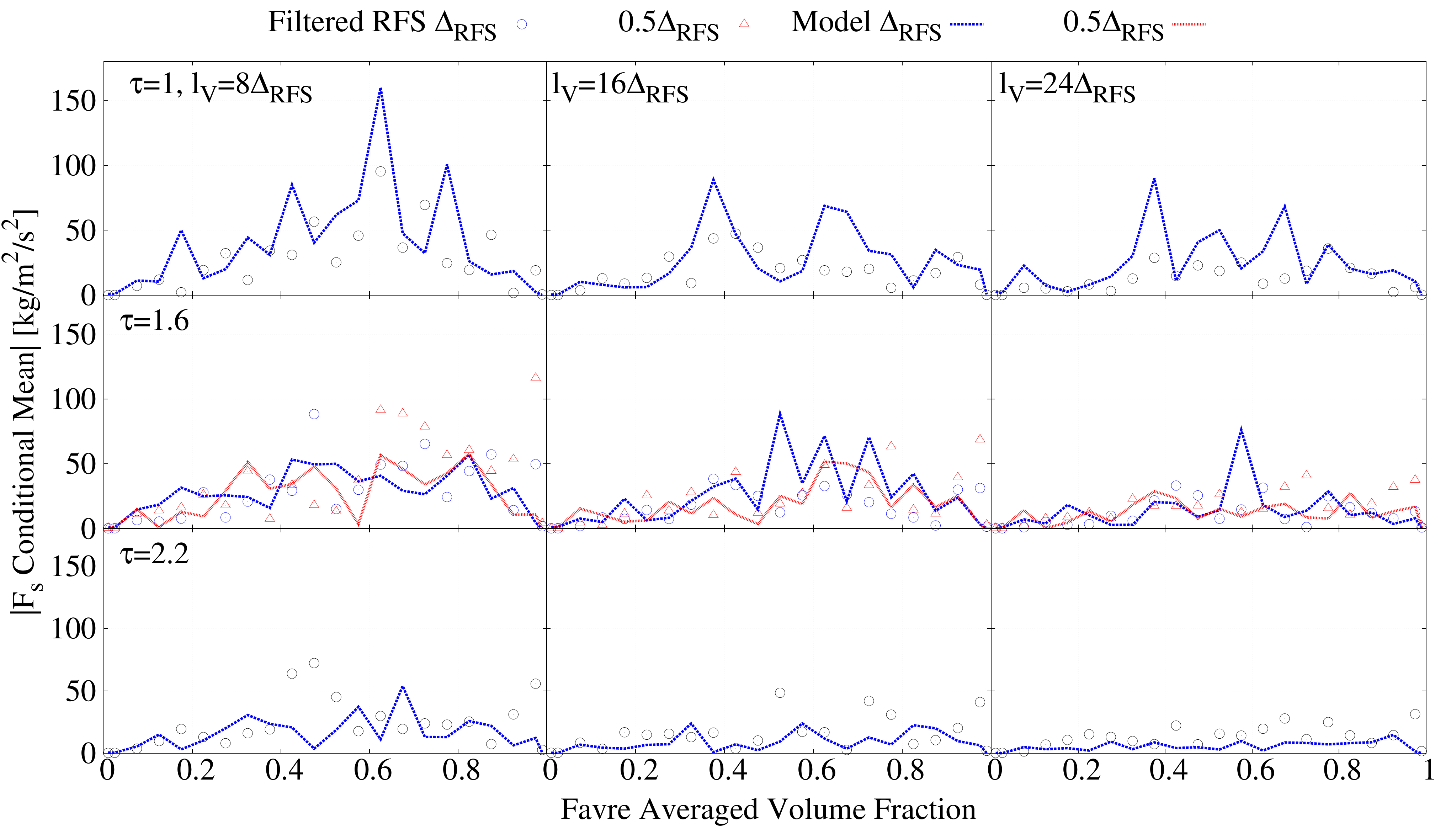}
\caption{Magnitude of the volume averaged surface tension force for C1 with three different values of $l_V$ and at three different times. Symbols represent filtered RFS data for two different grid resolutions, and lines represent model results using Eqs~(\ref{eq:Fs}) with $C_{sf}=1.8$.}\label{fig:FsApriC1}
\end{figure}

\begin{figure}[h!]
\footnotesize 
 \centering
 \includegraphics[clip=true, trim=0 0 0 0, width=0.95\textwidth]{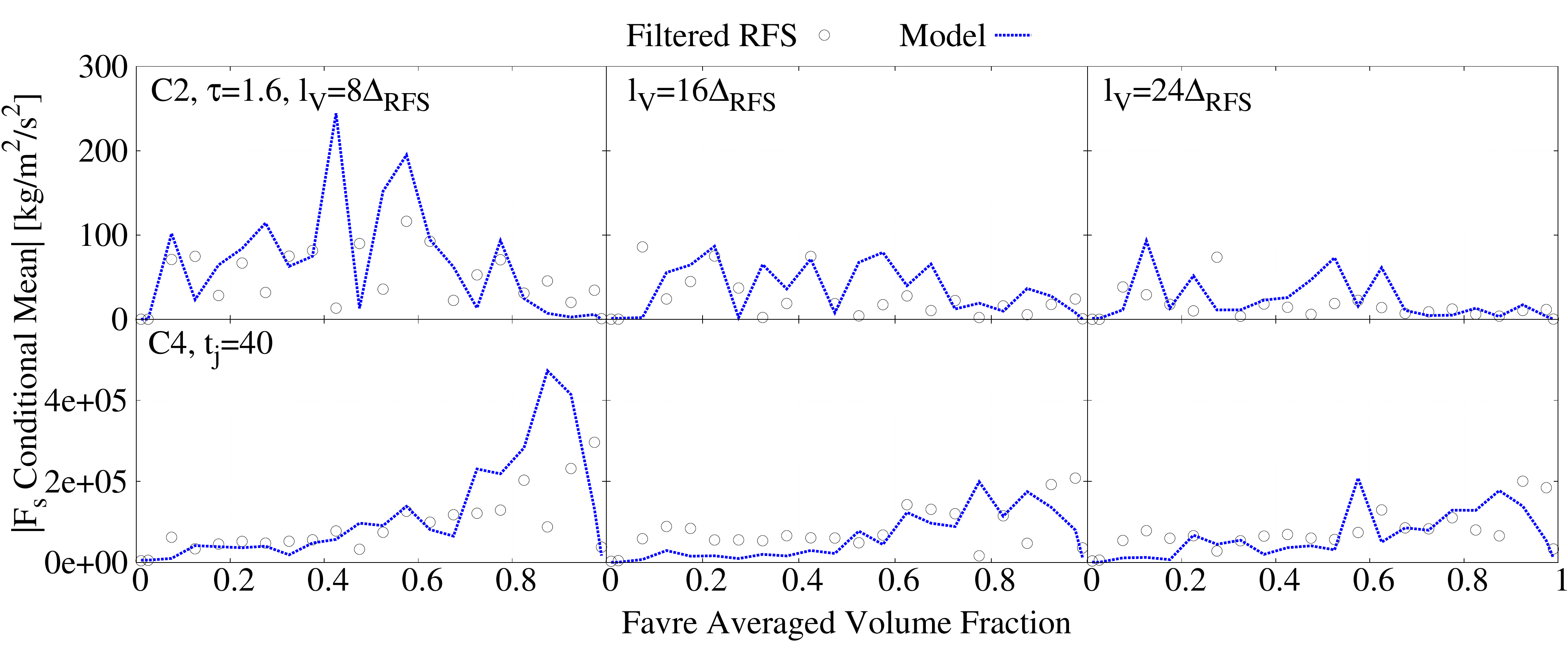}
\caption{Magnitude of the volume averaged surface tension force for C2 at $\tau=1.6$ (top row) and C4 at $t_j=40$ (bottom row) with three different values of $l_V$. Symbols represent filtered RFS data for two different grid resolutions, and lines represent model results using Eqs~(\ref{eq:Fs}) with $C_{sf}=1.8$.}\label{fig:FsApriC2C4}
\end{figure}

\noindent The final model to be validated by \textit{a priori} analysis is the closure for volume averaged surface tension force that is given by Eq.~(\ref{eq:Fs}). The assessment is conducted based on the magnitude of the force. Figures\,\ref{fig:FsApriC1} and \ref{fig:FsApriC2C4} show results for the force magnitude conditionally averaged on $\wideparen{\alpha}$ for C1, and C2 and C4, respectively. All results are with the constant $C_{sf}=1.8$. For C1, once again refinement of the RFS grid is checked and the results confirm
low sensitivity and convergence of the conditional statistics of this volume averaged surface tension force. The model has satisfactory agreement with the RFS data and the trend with increasing $l_V$. Similar quality agreement is observed for C1, C2 and C4 with different Reynolds numbers. For the turbulent C4 flow, the magnitude of volume averaged surface tension is relatively small for $\wideparen{\alpha}<0.5$. As seen in Fig.\,\ref{fig:picRFSf}, heavy fluid fragments are dispersed throughout the light fluid stream above the interface and upon integrating these elements over the explicit volumes both the average volume fraction and average surface tension force are relatively small, as expected. In the mixed region but closer to the heavy fluid side (i.e where $\wideparen{\alpha} \to 1$), the force magnitude increases due to the intact interface. These values decrease with increasing $l_V$ and return to zero inside the interior of the heavy fluid where $\wideparen{\alpha} = 1$. For the laminar flow cases, C1 and C2, the elongate ligaments are the dominant flow patterns and consequently several irregular peaks arise in the volume averaged surface tension force statistics near the interface region.

\subsection{Results of EVD simulations}\label{sec:aposteriori}

\noindent Having tested the sub-volume closures by \textit{a priori} analysis, the next step is to conduct full EVD simulations for cases C1~-~C5 using the CFD implementation described in Section\,\ref{sec:numericalImp}. Comparisons between the EVD results and filtered RFS data are made, but unlike the \textit{a priori} analysis above, in the present section no model input parameters are taken form the RFS data. Numerical convergence of the EVD model is demonstrated by reducing the numerical cell size while keeping the explicit volume scale, $l_V$, constant. The accuracy of EVD relative to the RFS is discussed and sensitivity to variations in $l_v$ is also investigated.

\subsubsection{Selecting the explicit volume length scale}\label{sec:hypoLv}
\noindent 
In the \textit{a priori} analysis in Section~\ref{sec:apriori}, the validity of the sub-volume closures was demonstrated for different volume scales varying as a multiple of the grid scale of RFS. However, in practice $l_V$ needs to be selected based on the physical scales of the flow. The characteristic physical length scales related to the evolution of interfacial flows include wave lengths defined longitudinally along the interface, wave amplitudes and interface boundary layer thicknesses. Additionally, there are scales associated with turbulence (if it exists). During the growth of interfacial instabilities, explicit volume diffusion smears the volume fraction in the direction normal to the interface and the wave amplitude is therefore an obvious physical length scale to which the explicit volume length scale, $l_V$, should be associated. Both the wave length and amplitude are functions of the interface boundary layer thickness that develops in both fluids normal to the interface. Marmottant and Villermaux \cite{marmottant2004spray} have found that the interfacial shear instabilities are controlled mainly by the boundary layer thickness in the light fluid. This boundary thickness is defined by \cite{marmottant2004spray}
\begin{equation}
\zeta\cong 5.6H{Re_l}^{-0.5}
\end{equation}
in laminar flows and is given by
\begin{equation}\label{eq:deltahTurb}
\zeta \cong 74H{Re_l}^{-0.75}
\end{equation}
in turbulent flows. Here, $Re_l=U_l H/\nu_l$ with $H$ being the outer scale of the shear layer as explained in Section\,\ref{sec:aprioriSetup}, and $U_l$ and $\nu_l$ denote the the velocity and kinematic viscosity of the light fluid, respectively. The boundary thickness for the cases investigated here is computed and listed in Table\,\ref{tab:delta} along with its ratio to the initial wavelength, $\lambda_1$ or $\lambda_2$, and relative to RFS grid scale, $\Delta_{RFS}$. Previous studies \cite{desjardins2013direct,agbaglah2017numerical} have indicated that at least two numerical cells are required to resolve the light fluid boundary layer and to accurately predict interfacial wave dynamics and the table shows that the RFS database has very well resolved interface boundary layers.
For EVD simulations, we suggest that the explicit volume length scale be selected such that $l_V  \leq \zeta$ since the alternative (i.e. $l_V > \zeta$) would result in excessive smearing and artificial expansion of the boundary layer and potentially changed interface dynamics.
\begin{table}[h!]
  \begin{center}
\def~{\hphantom{0}}
  \begin{tabular}{lccccc}
  \toprule
    Cases  & C1  & C2 & C3 & C4 & C5 \\[3pt]
       \midrule  
    $\zeta$ & 0.12 & 0.31 & 0.057 & 0.00035 & 0.00033\\
    $\zeta/\lambda_1$ & 0.12 & 0.31 & 0.057 &         &        \\
    $\zeta/\lambda_2$ &      &      &       & 0.103   & 0.096\\
    $\zeta/\Delta_{RFS}$ & 96 & 248 & 45.6 & 11.7 & 11\\
    $l_V/\Delta_{RFS}$ & 20, 40, 80 & 40, 200 & 40 & 8 & 8\\
       \bottomrule
     \end{tabular}
  \caption{Theoretical boundary thickness, $\zeta$, expressed as ratios of the the initial wave lengths $\lambda_1$ and $\lambda_2$ for the laminar and turbulent cases, respectively, and $\Delta_{RFS}$ as well as $l_V/\Delta_{RFS}$.}
  \label{tab:delta}
  \end{center}
\end{table} 
{Finer grids are needed to obtain numerical convergence when reducing $l_V$ that leads to high computational cost. The initial wave lengths, $\lambda_1$ and $\lambda_2$, are artificial scales (see Section~\ref{sec:aprioriSetup} and Fig.~\ref{fig:picSetup}) which are different from the characteristic wave length scales. The ratio of the characteristic wave length to the boundary layer thickness is constant and is proportional to $\left(\frac{\rho_h}{\rho_l}\right)^{0.5}$ \cite{marmottant2004spray,singh2020instability}.} Table\,\ref{tab:delta} also lists $l_V$ normalised by $\Delta_{RFS}$ for the five cases. In the first set of results we use $l_V/\Delta_{RFS}= 40$ for the laminar cases C1, C2 and C3, and $l_V/\Delta_{RFS}= 8$ for the two turbulent cases. Results for of $l_V/\Delta_{RFS} =20$ and 80 for C1 and $l_V/\Delta_{RFS}=200$ for C2 are also presented to test sensitivity.

\subsubsection{EVD simulations of turbulent shear layer cases}\label{sec:shearEVD}
\begin{figure}[h!]
 \centering
 \textbf{\footnotesize C1, $\tau=1.6$}\par\medskip
 \vspace{-0.2cm}
 \subfigure[RFS, $\alpha^{\Delta}$]{\includegraphics[clip=true, trim=390 300 390 300, height=0.195\textwidth]{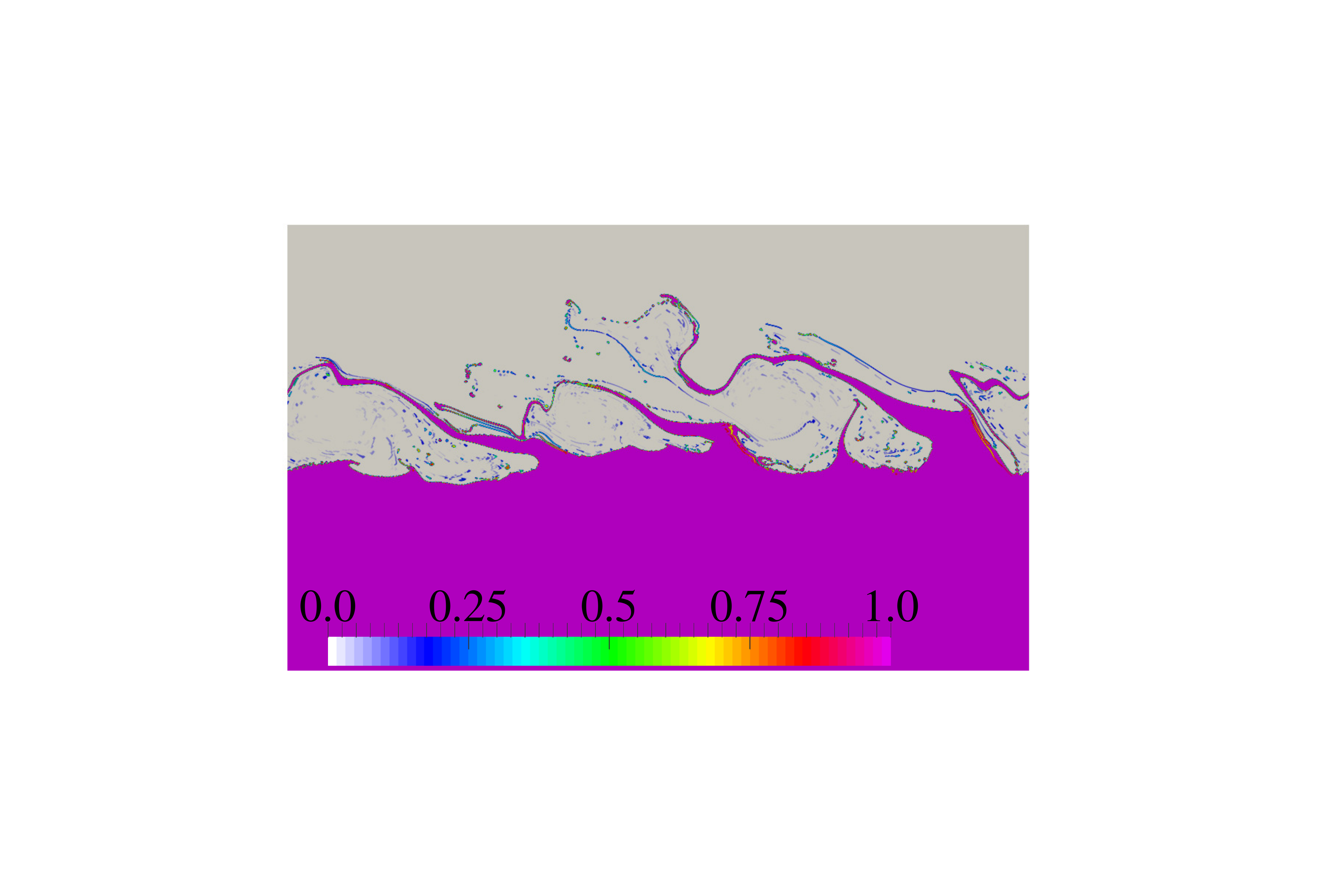}\label{picRLS}}
 \subfigure[EVD, $\Delta=2\Delta_{RFS}$, $\widehat{\alpha}^{\Delta}$]{\includegraphics[clip=true, trim=390 300 390 300, height=0.195\textwidth]{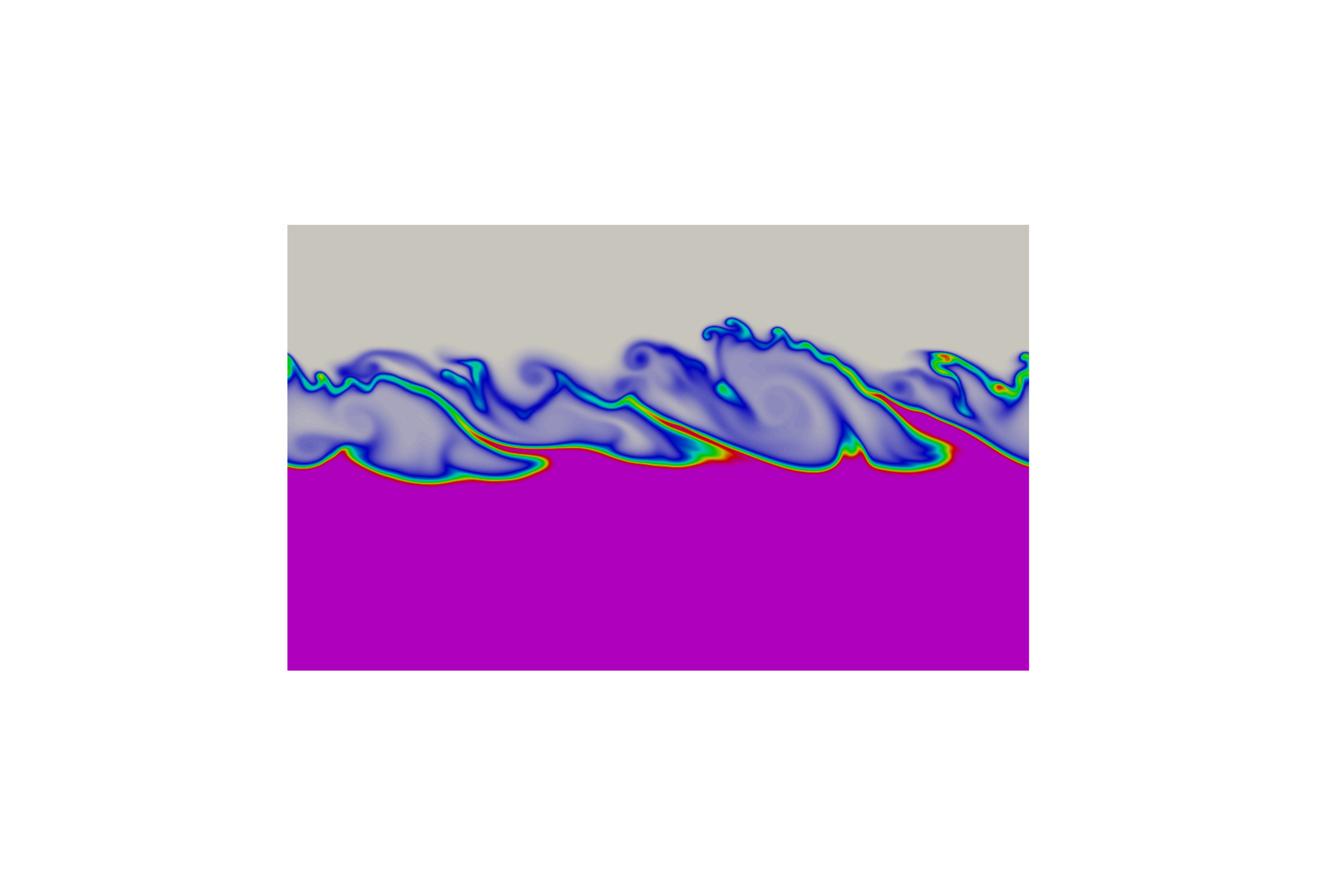}\label{picEVD2}}
 \subfigure[EVD, $\Delta=4\Delta_{RFS}$, $\widehat{\alpha}^{\Delta}$]{\includegraphics[clip=true, trim=390 300 390 300, height=0.195\textwidth]{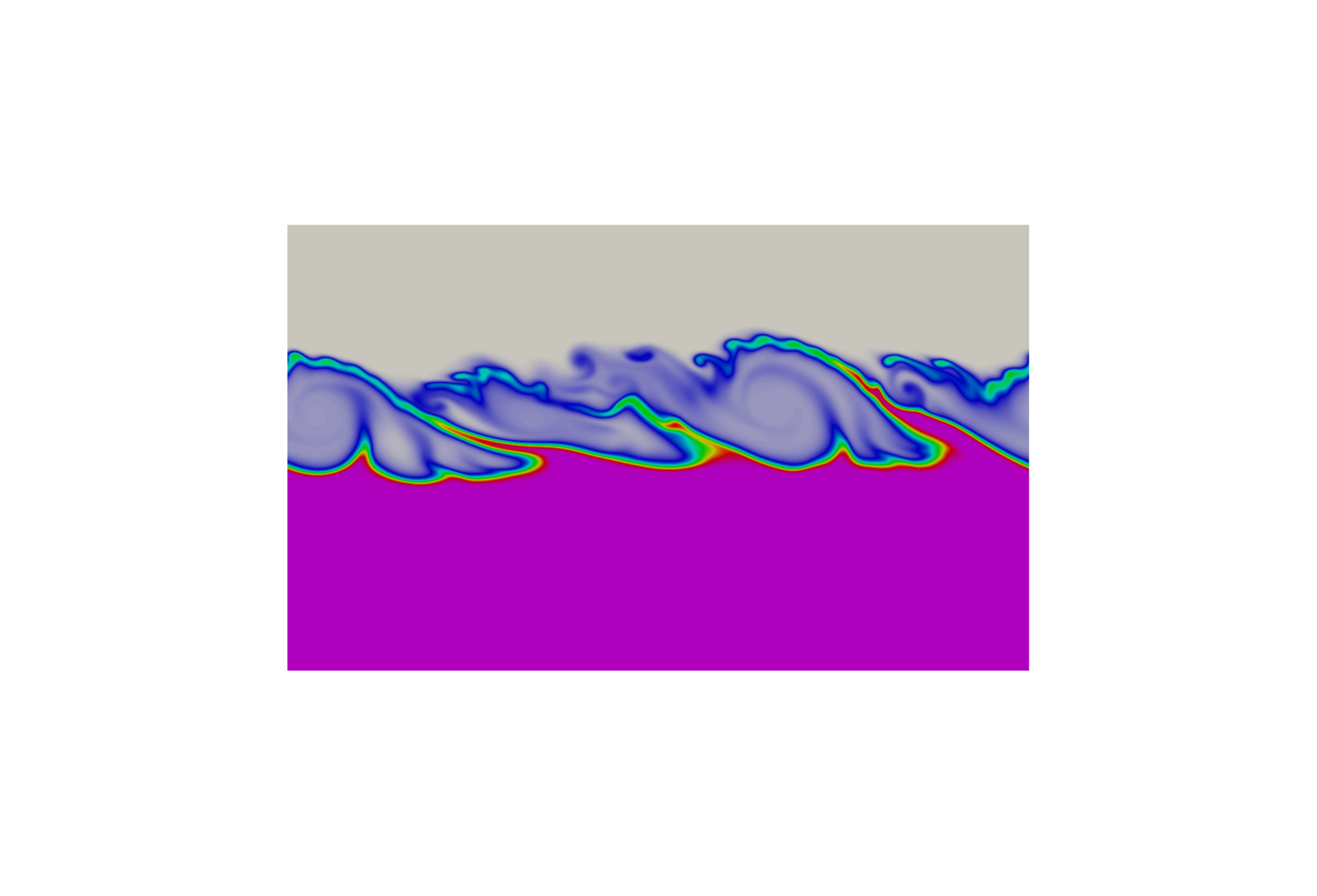}\label{picEVD4}}
 
 \subfigure[RFS, $\widehat{\alpha}^V$]{\includegraphics[clip=true, trim=390 300 390 300, height=0.195\textwidth]{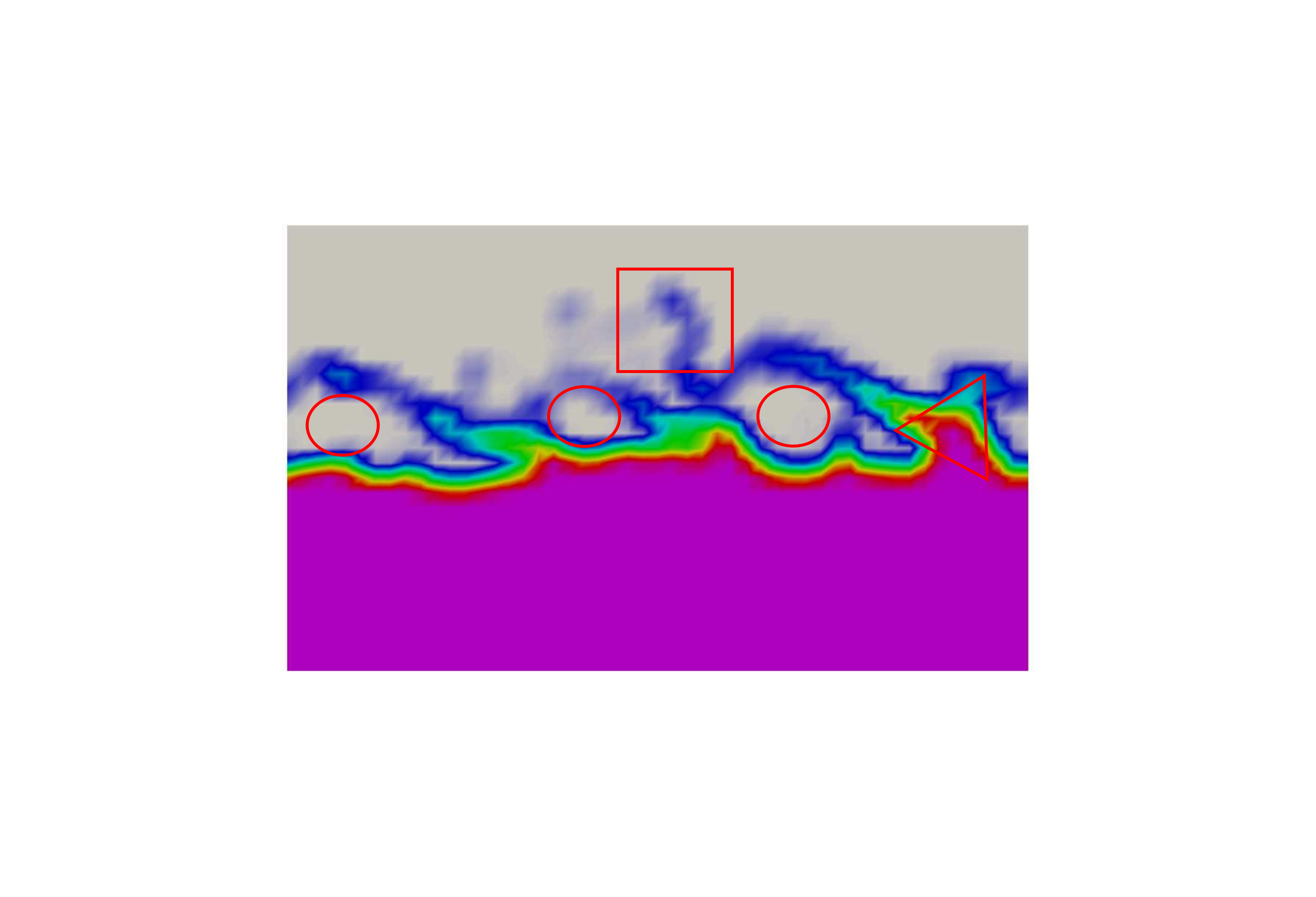}\label{picRLSV}} 
 \subfigure[EVD, $\Delta=2\Delta_{RFS}$, $\widehat{\alpha}^V$]{\includegraphics[clip=true, trim=390 300 390 300, height=0.195\textwidth]{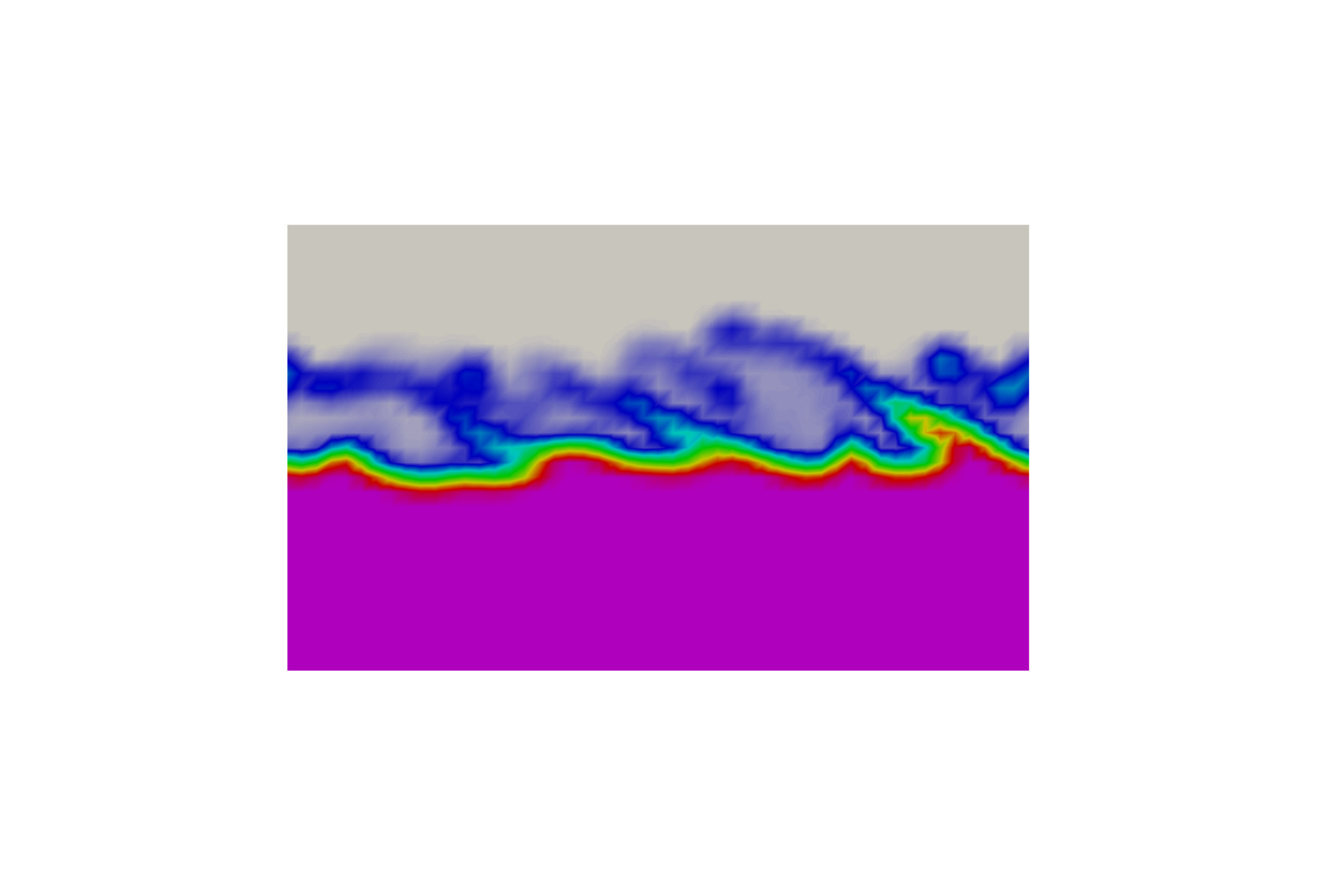}\label{picEVDV2}} 
 \subfigure[EVD, $\Delta=4\Delta_{RFS}$, $\widehat{\alpha}^V$]{\includegraphics[clip=true, trim=390 300 390 300, height=0.195\textwidth]{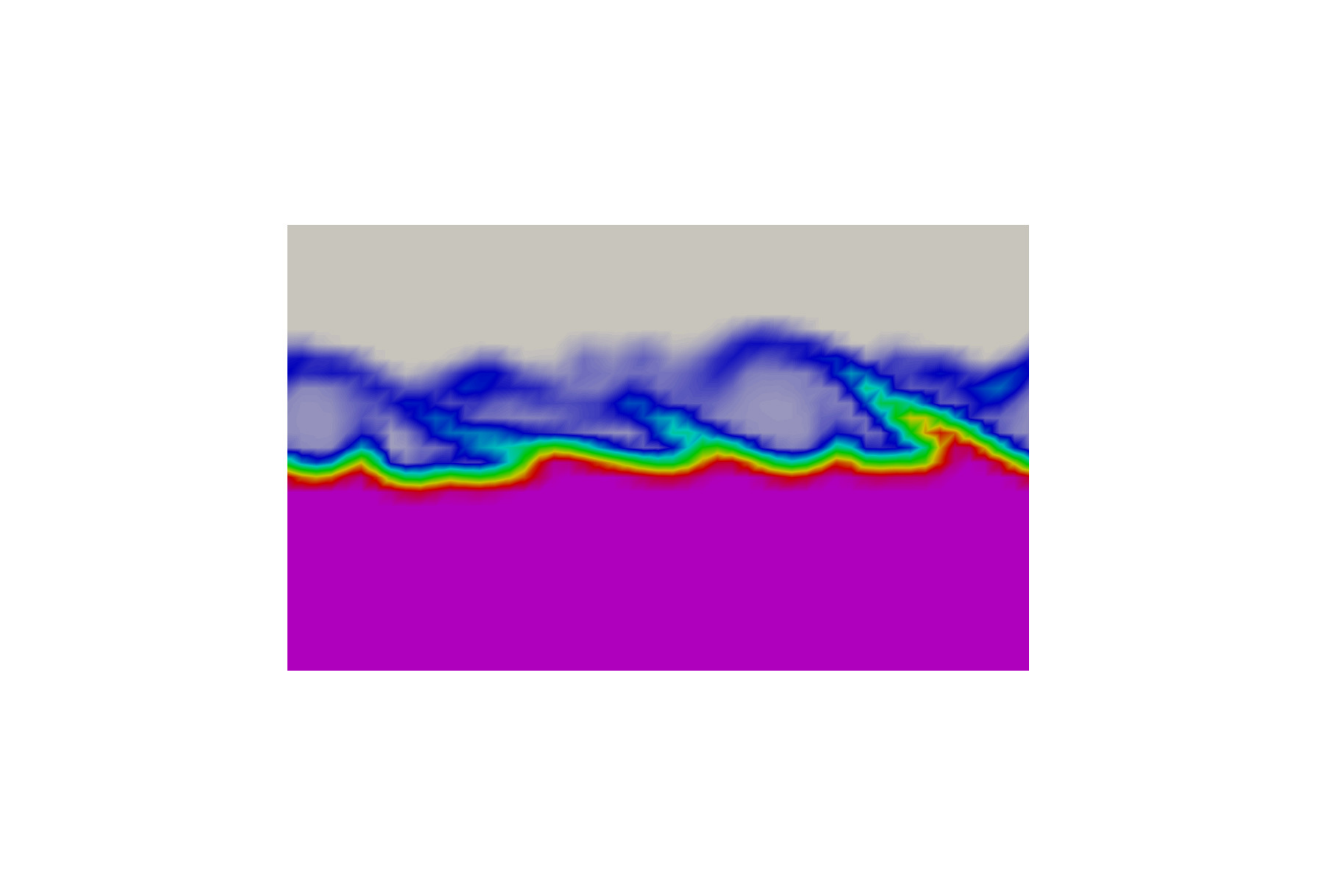}\label{picEVDV4}}
  \caption{Grid based and volume integrated volume fraction fields for C1 using RFS and EVD with $l_V=40\Delta_{RFS}$ and $\Delta=2\Delta_{RFS}$ and $\Delta=4\Delta_{RFS}$.}\label{fig:picRLSV}
\end{figure}

\begin{figure}[h!]
 \centering
 \textbf{\footnotesize C1, $\tau=1.6$ }\par\medskip
 \vspace{-0.2cm}
 \subfigure[EVD, $\Delta=2\Delta_{RFS}$, $\widehat{\alpha}^{\Delta}$]{\includegraphics[clip=true, trim=360 300 365 300, height=0.2\textwidth]{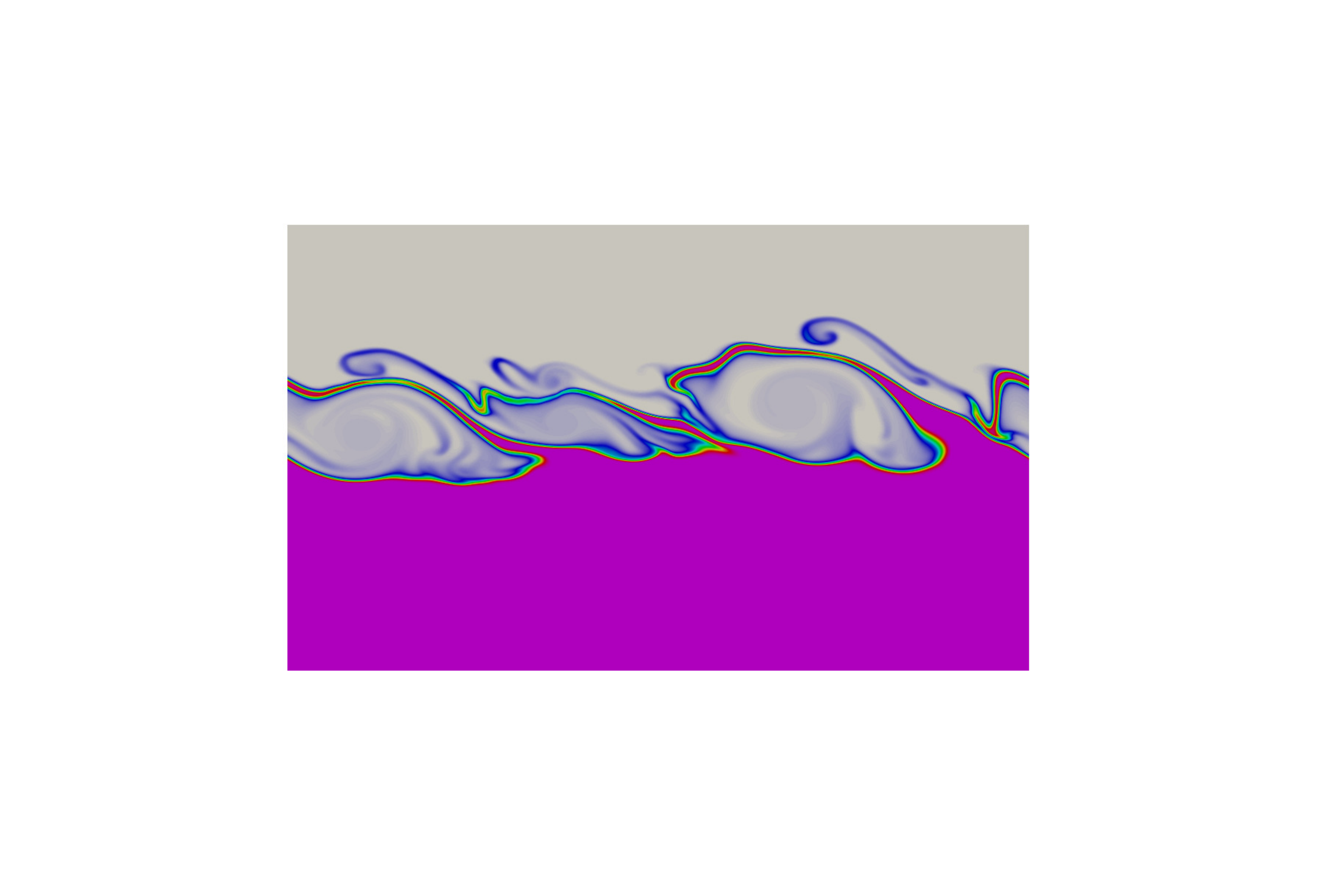}\label{picEVDV20a}}
 \subfigure[EVD, $\Delta=4\Delta_{RFS}$, $\widehat{\alpha}^{\Delta}$]{\includegraphics[clip=true, trim=340 300 385 300, height=0.2\textwidth]{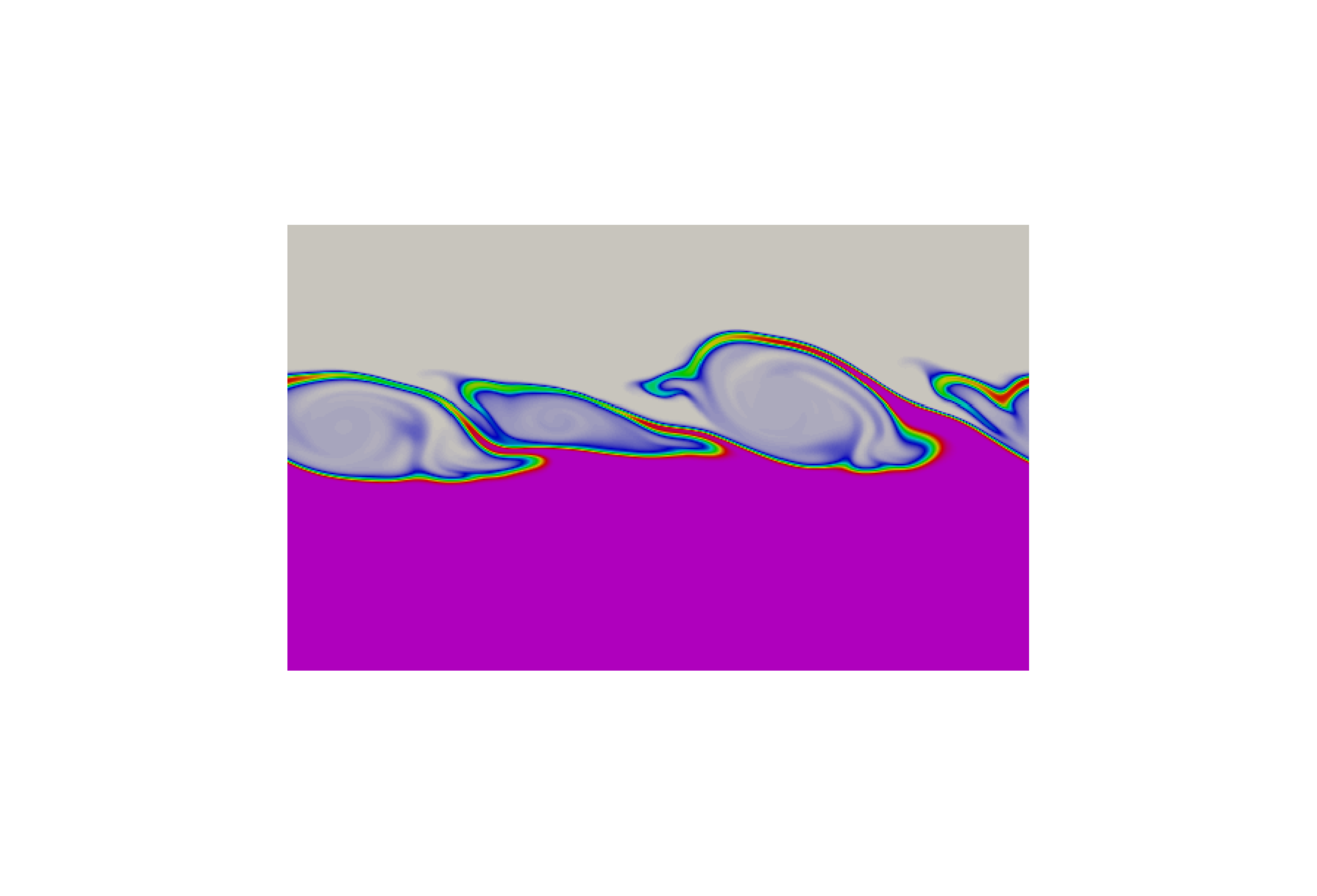}\label{picEVDV20b}}  
 
  \caption{Grid based volume fraction fields for C1 using EVD with $l_V=20\Delta_{RFS}$ and $\Delta=2\Delta_{RFS}$ and $\Delta=4\Delta_{RFS}$. 
  }\label{fig:picVOFVD}
\end{figure}

\begin{figure}
\footnotesize 
 \centering
 \includegraphics[clip=true, trim=0 0 0 0, width=0.95\textwidth]{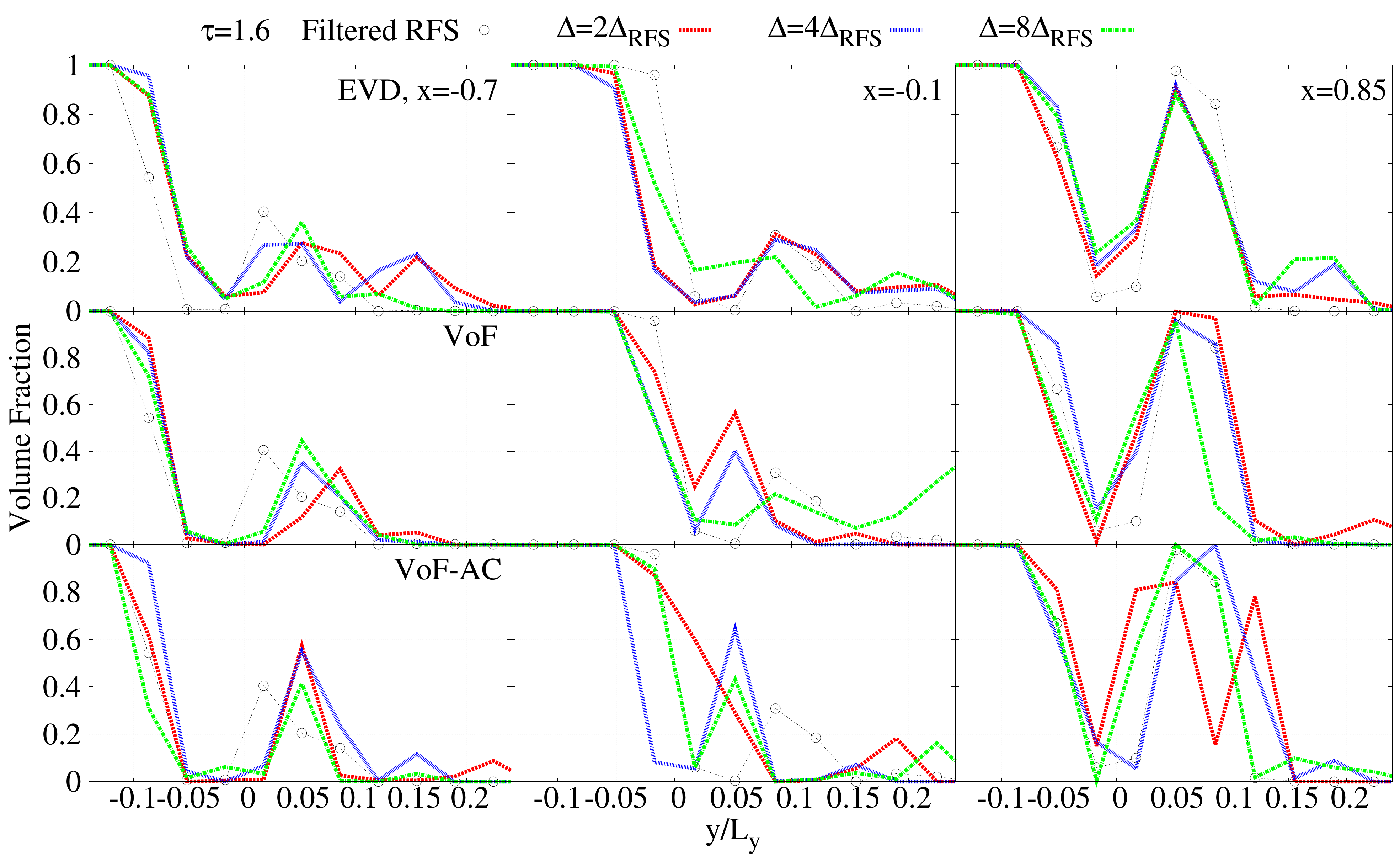}
\caption{Volume integrated volume fraction along three transverse lines by EVD with $l_V=40\Delta_{RFS}$, VoF, VoF-AC and RFS for C1 at $\tau=1.6$. The y axis is normalised by the transverse width of the domain, $L_y$.}\label{fig:AlphaXC1t2}
\end{figure}

\begin{figure}
\footnotesize 
 \centering
 \includegraphics[clip=true, trim=0 0 0 0, width=0.95\textwidth]{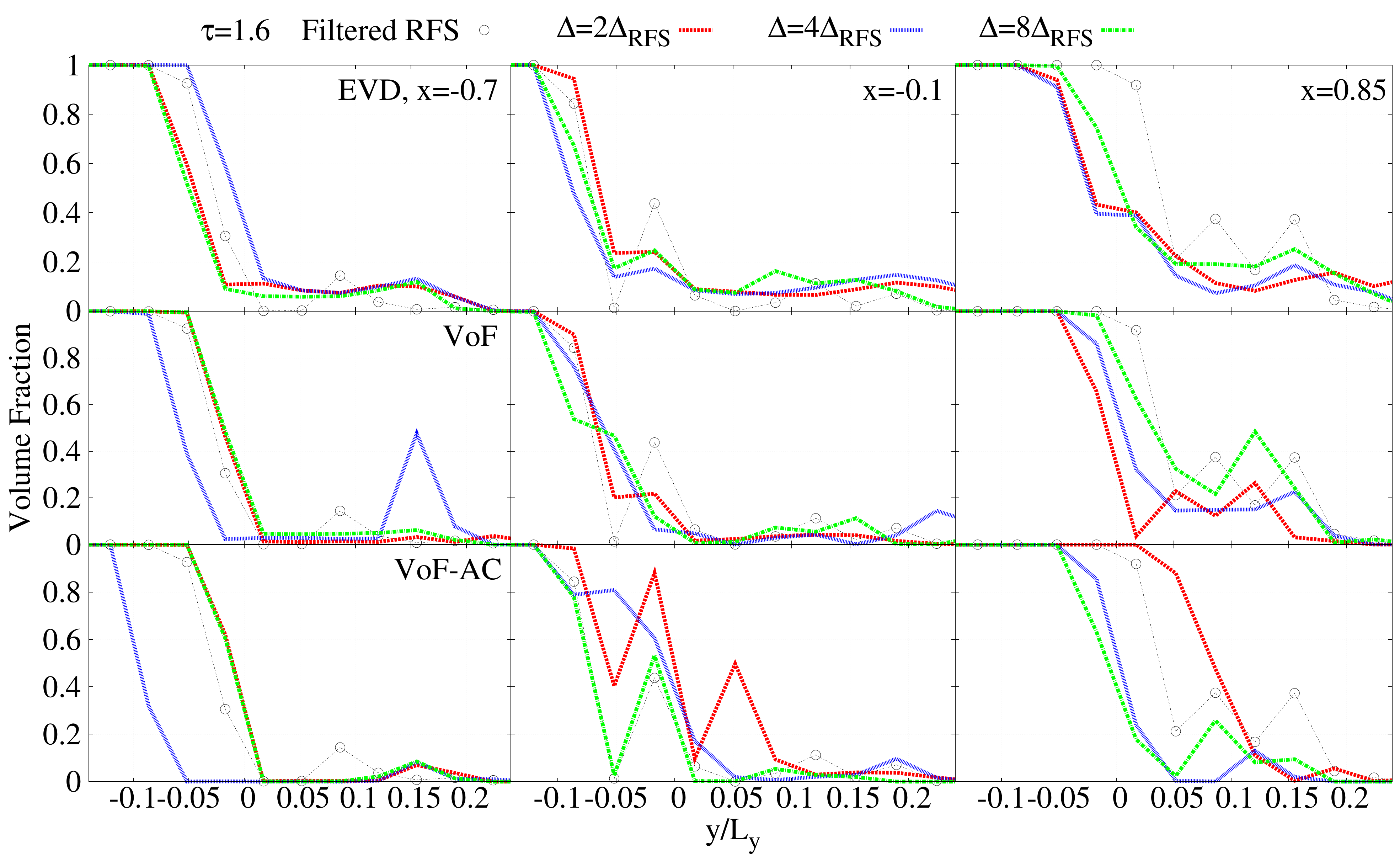}
\caption{Volume integrated volume fraction along three transverse lines by EVD with $l_V=40\Delta_{RFS}$, VoF, VoF-AC and RFS for C1 at $\tau=2.2$. The y axis is normalised by the transverse width of the domain, $L_y$.}\label{fig:AlphaXC1t3}
\end{figure}

\noindent Figure\,\ref{fig:picRLSV} shows qualitative comparisons of the volume fraction fields from both RFS and EVD simulations for C1 at $\tau=1.6$. These EVD simulations are for $l_V=40\Delta_{RFS}$ and different grid sizes are used; $\Delta = 2\Delta_{RFS}$ and $\Delta = 4\Delta_{RFS}$. Figures\,\ref{picRLS}~-~\ref{picEVD4} present the volume fraction fields on the numerical grids, $\widehat{\alpha}^\Delta$. Inspection of Figs.\,\ref{picEVD2} and \ref{picEVD4} which are the EVD results with different grids, shows that the ligament and vortex structures are{, overall,} very similar as is expected of numerically converged model results. A quantitative analysis of convergence is conducted below. In the bottom row, Figs\,\ref{picRLSV}~-~\ref{picEVDV4}, the RFS and EVD results have been integrated over the volume scale of $l_V=40\Delta_{RFS}$ to obtain the volume integrated volume fraction, $\widehat{\alpha}^V$. The ligaments are smeared by explicit volume diffusion but keep their general form. The overall similarity between the RFS and EVD integrated fields is rather good, but three minor {differences} are briefly discussed. Firstly, the ligament marked by the red square box is more elongated into the upper stream in the RFS. For EVD with $\Delta=2\Delta_{RFS}$ (Fig.\,\ref{picEVDV2}), one ligament appears in the corresponding location nearby but does not extend as far vertically, while for $\Delta=4\Delta_{RFS}$ (Fig.\,\ref{picEVDV4}), a much flatter ligament appear at that location. Secondly, the red triangular box indicates a thick and curved liquid core whereas both EVD results have a relatively thin and flat protrusion. Thirdly, the three vortices marked by red circles are filled with the light fluid ($\wideparen{\alpha} \sim 0$) in the RFS but in the EVD explicit volume diffusion has smeared the volume fraction fields and $\wideparen{\alpha} > 0$.
\\\\
\noindent As a contrast to Fig.~\ref{fig:picRLSV}, the volume fraction fields for EVD with the explicit length scale halved to $l_V=20\Delta_{RFS}$ are shown in Fig.\,\ref{fig:picVOFVD} for two different grids, $\Delta=2\Delta_{RFS}$ and $\Delta=4\Delta_{RFS}$. Significant grid sensitivity is evident and convergence is not obtained. Specifically, for $\Delta=4\Delta_{RFS}$ the three vortices are closed but for $\Delta=2\Delta_{RFS}$ three ligaments stretched out from the lower positions and extend well into the upper stream. The reason for the lack of numerical convergence is well understood. As $l_V$ is reduced so to is the explicit volume diffusion and the numerical diffusion, which is always present, is relatively large and dominant. 
These results indicate that $l_V$ needs to be carefully chosen as discussed above.
\\\\
\noindent Quantitative assessment of numerical convergence and predictive accuracy against the RFS data are now conducted for the EVD model with $l_V=40\Delta_{RFS}$ alongside the standard VoF model without artificial compression (VoF) and VoF with artificial compression (VoF-AC) which are commonly used in simulations that may be under-resolved. Volume integrated volume fraction along three transverse lines normal to the interface at $x=$-0.7, -0.1, 0.85, are shown for C1 at $\tau=1.6$ in Fig.\,\ref{fig:AlphaXC1t2}.
At $x=$ -0.7, the grid sensitivity is generally good but with some scatter for $0<y/L_y<0.15$. In contrast the VoF convergence at that location is poor with a significant shift in the profile for the most refined VoF grid. At $x=$ -0.1, the convergence of the EVD with refinement from the coarsest grid, $\Delta=8\Delta_{RFS}$, to the finest grid, $\Delta=2\Delta_{RFS}$, is clearly demonstrated. The coarse grid result is an outlier but the results for $\Delta=2\Delta_{RFS}$ and $4\Delta_{RFS}$ are very similar to each other and capture the volume averaged RFS data quite well especially in the multi-layered ligament region of the flow $y/L_y>0.05$, although there is some underprediction for -$0.05<y/L_y<0$. For VoF at $x=$ -0.1 no numerical convergence is evident as the grid is refined. At $x=0.85$ the EVD results are once again rather insensitive to the grid except for some statistical error for $y/L_y>0.12$ where the RFS data is overpredicted. The VoF convergence is also reasonable at $x=0.85$ although not as good as for the EVD for which even the $\Delta=8\Delta_{RFS}$ is converged. VoF-AC presents a slightly better convergence at $x=$ -0.7 than VoF since the peak values at $y/L_y=0.05$ for $\Delta=2\Delta_{RFS}$ and $4\Delta_{RFS}$ coincide. However, the profiles of VoF-AC at the other two positions diverge evidently. Compared with the results of VoF, the grid resolutions even impose a stronger effect on the results.
\\\\
\noindent The comparisons of transverse volume fraction profiles for C1 at a later time, $\tau=2.2$, are shown in Fig.\,\ref{fig:AlphaXC1t3}. At $x=$ -0.7, EVD converges for the three grids from $y/L_y>0$ onwards but is not as strong at $-0.1<y/L_y<0$, although the trends are consistent and in good agreement with the RFS. An 
apparent divergence occurs over the range $0.1<y/L_y<0.2$ for VoF. At $x=$ -0.1, the EVD exhibits good overall numerical convergence between the $\Delta=2\Delta_{RFS}$ and $4\Delta_{RFS}$ grids, however, VoF results diverge over the spatial range -$0.05<y/L_y<0$ and $y/L_y>0.15$. The EVD profiles are smoother than the volume averaged RFS data and this is artifact of the explicit volume diffusion. However, the EVD generally captures the spatial variations and the locations of crests and troughs in the volume fraction profile that result from the presence of ligaments. At $x=0.85$, the convergence vanishes for VoF but is preserved well for EVD. Similarly, the profiles given by EVD are smooth and are qualitatively consistent with the RFS data although trough and peak amplitudes are underpredicted due to explicit volume diffusion. An apparent underestimate can be seen in the spatial range -$0.05<y/L_y<0.05$ where the volume diffusion leads to a smoother interface region than the RFS. Again VoF-AC gives evident {diverging} results.
\\\\

\begin{figure}[h!]
 \centering
 \textbf{\footnotesize C2,C3\,\,\,$\tau=1.6$}\par\medskip
 \vspace{-0.cm}
 \subfigure[C2, RFS, $\alpha^{\Delta}$]{\includegraphics[clip=true, trim=390 300 390 300, height=0.195\textwidth]{C2alphaInterRhoDif10R1t2-eps-converted-to.pdf}}
 \subfigure[EVD, $\Delta=2\Delta_{RFS}$, $\widehat{\alpha}^{\Delta}$]{\includegraphics[clip=true, trim=390 300 390 300, height=0.195\textwidth]{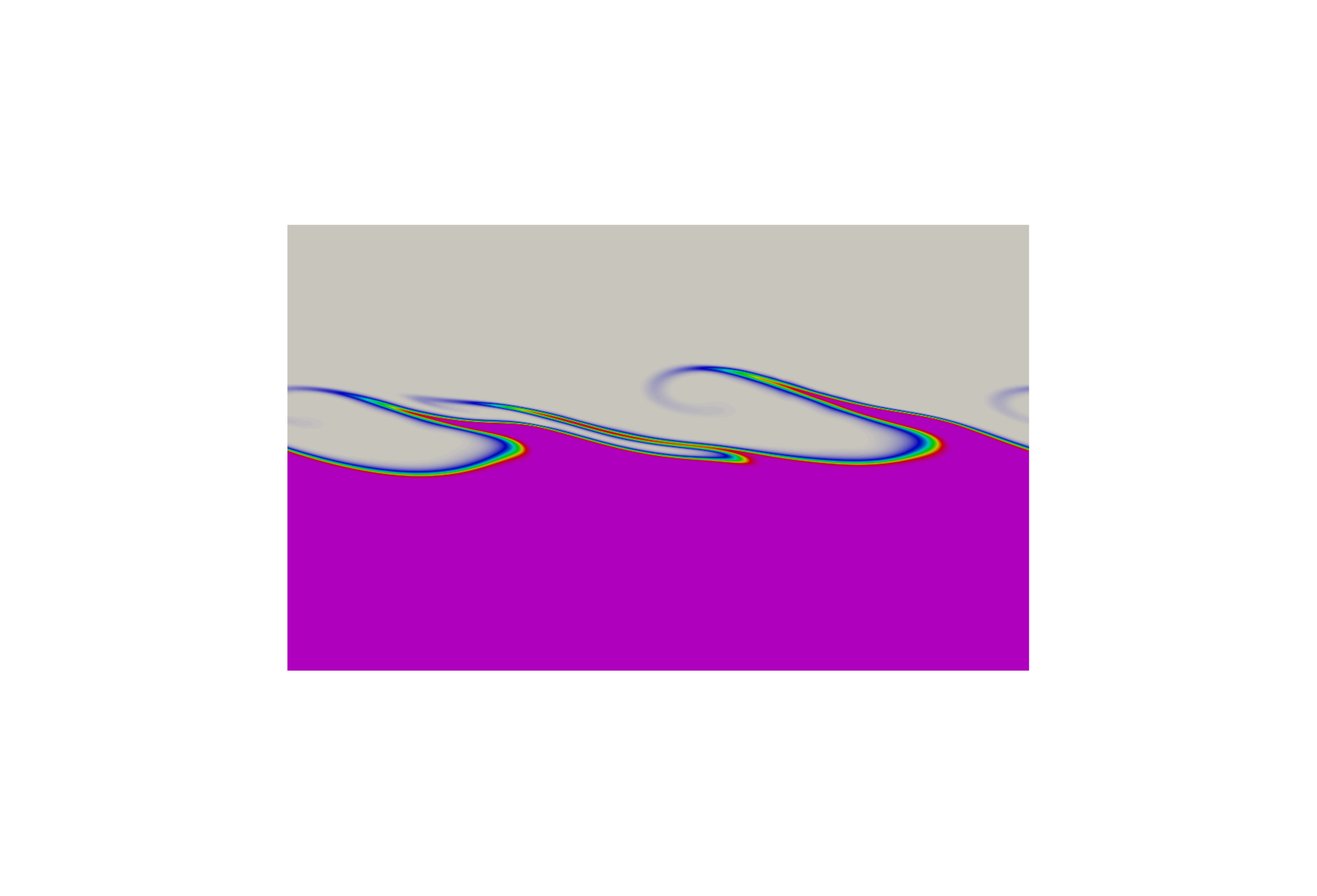}}
 \subfigure[EVD, $\Delta=4\Delta_{RFS}$, $\widehat{\alpha}^{\Delta}$]{\includegraphics[clip=true, trim=390 300 390 300, height=0.195\textwidth]{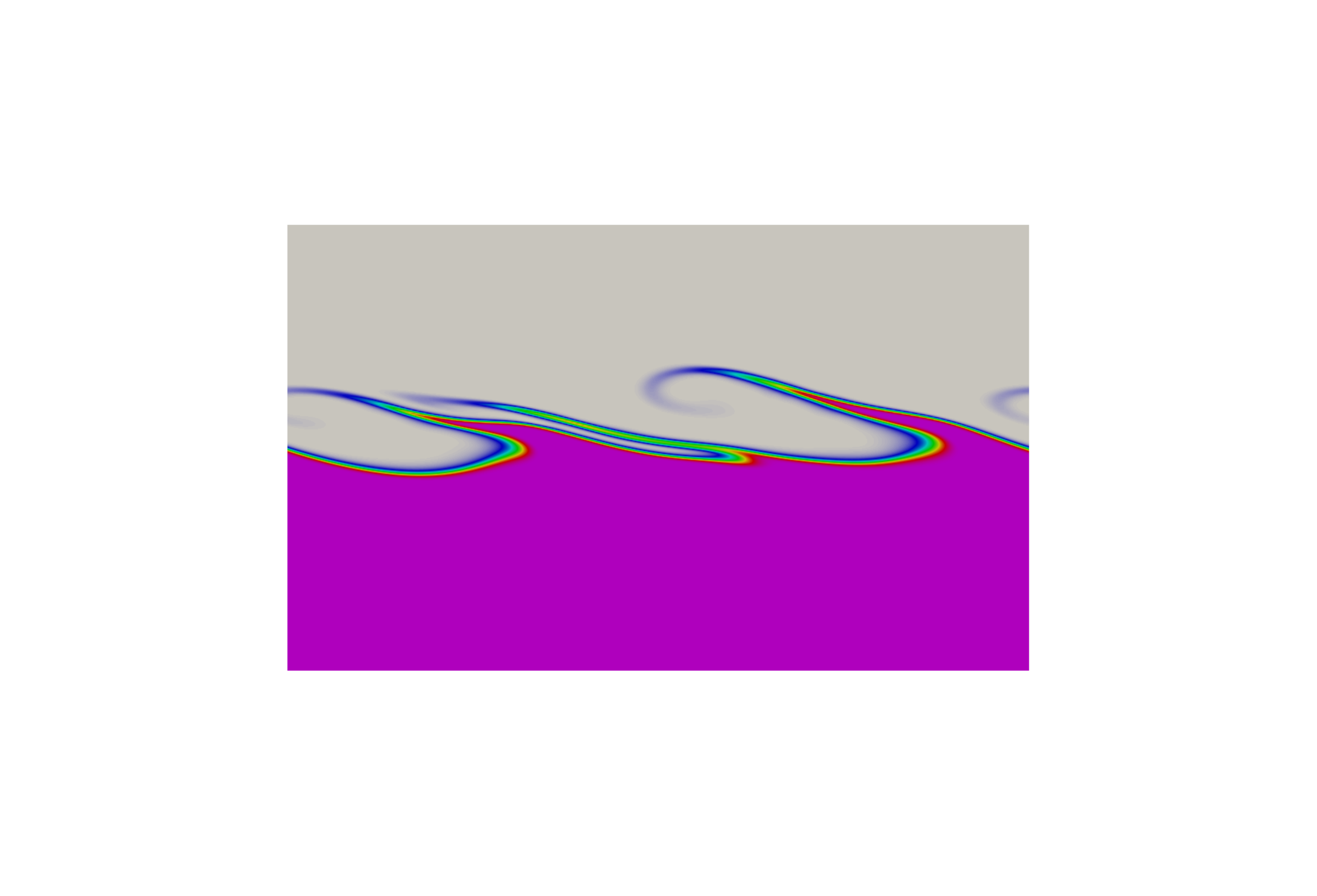}}  
 
 \subfigure[C3, RFS, $\alpha^{\Delta}$]{\includegraphics[clip=true, trim=390 300 390 300, height=0.195\textwidth]{alphaInterRhoDif100R1t2-eps-converted-to.pdf}}   
 \subfigure[EVD, $\Delta=2\Delta_{RFS}$, $\widehat{\alpha}^{\Delta}$]{\includegraphics[clip=true, trim=390 300 390 300, height=0.195\textwidth]{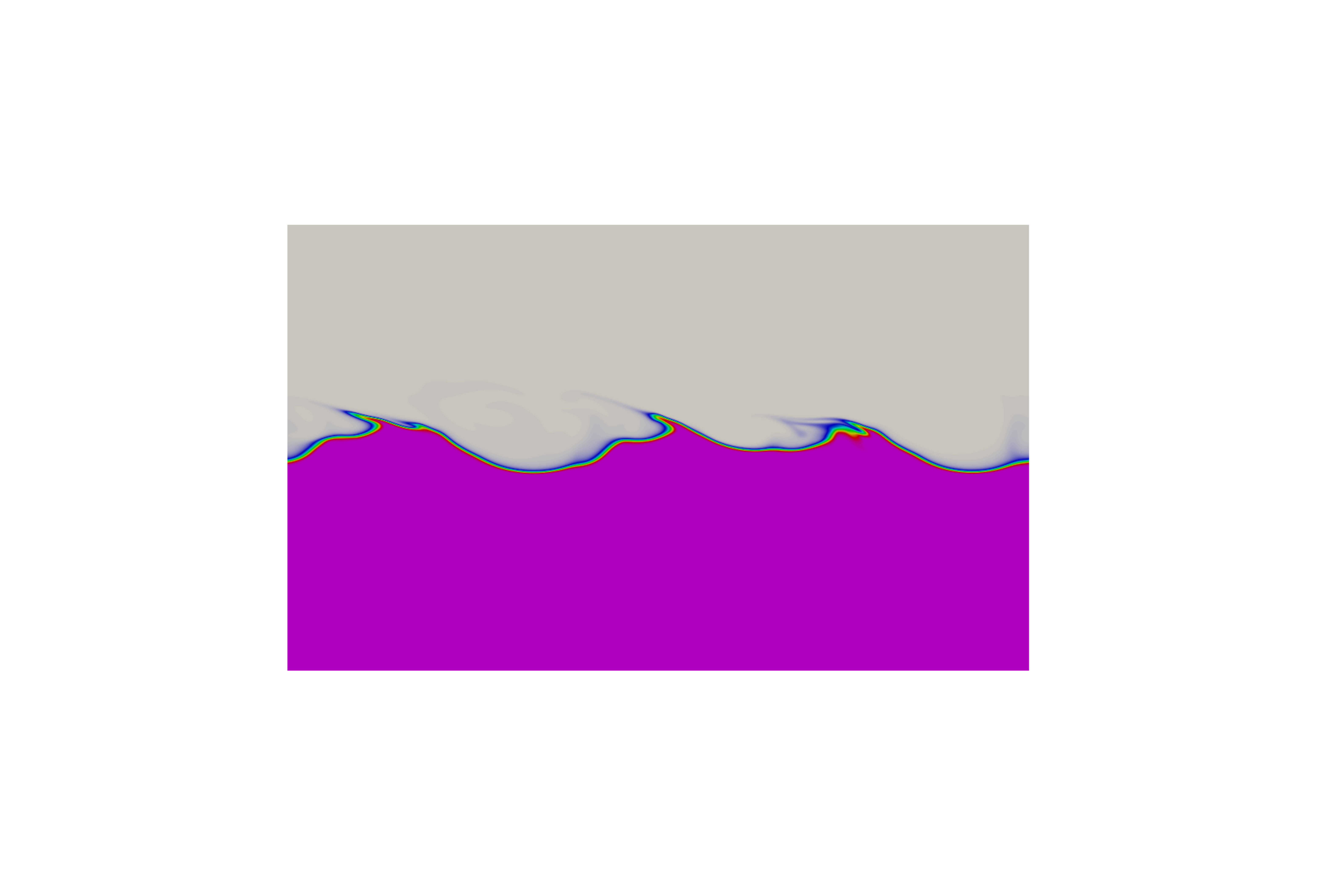}}  
 \subfigure[EVD, $\Delta=4\Delta_{RFS}$, $\widehat{\alpha}^{\Delta}$]{\includegraphics[clip=true, trim=390 300 390 300, height=0.195\textwidth]{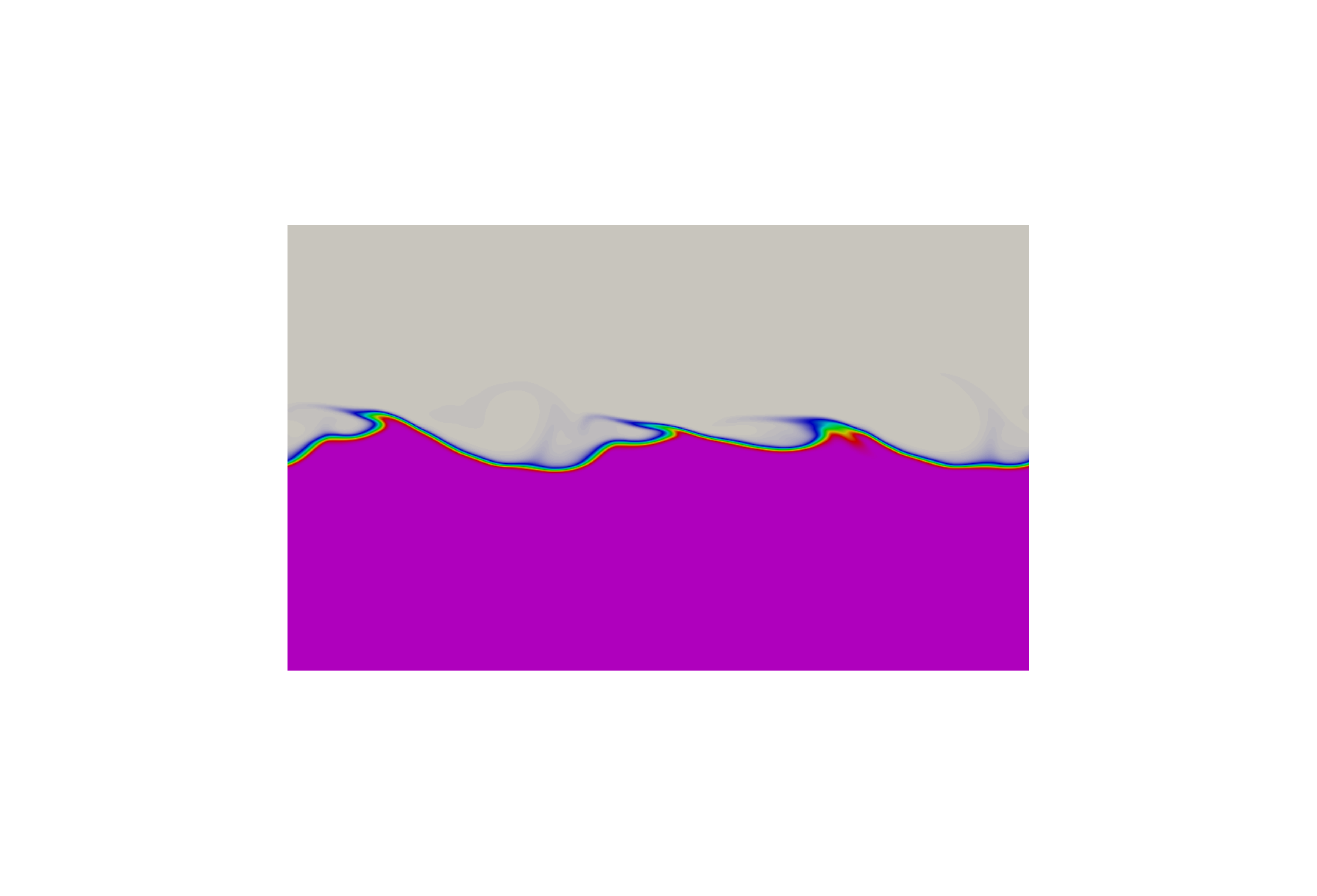}}  
  \caption{
  Grid based and volume integrated volume fraction fields for C2 (1st row) and C3 (2nd row) using RFS and EVD with $l_V=40\Delta_{RFS}$ and $\Delta=2\Delta_{RFS}$ and $\Delta=4\Delta_{RFS}$.}\label{fig:picC2C3Post}
\end{figure}

\begin{figure}
\footnotesize 
 \centering
 \includegraphics[clip=true, trim=0 0 0 0, width=0.95\textwidth]{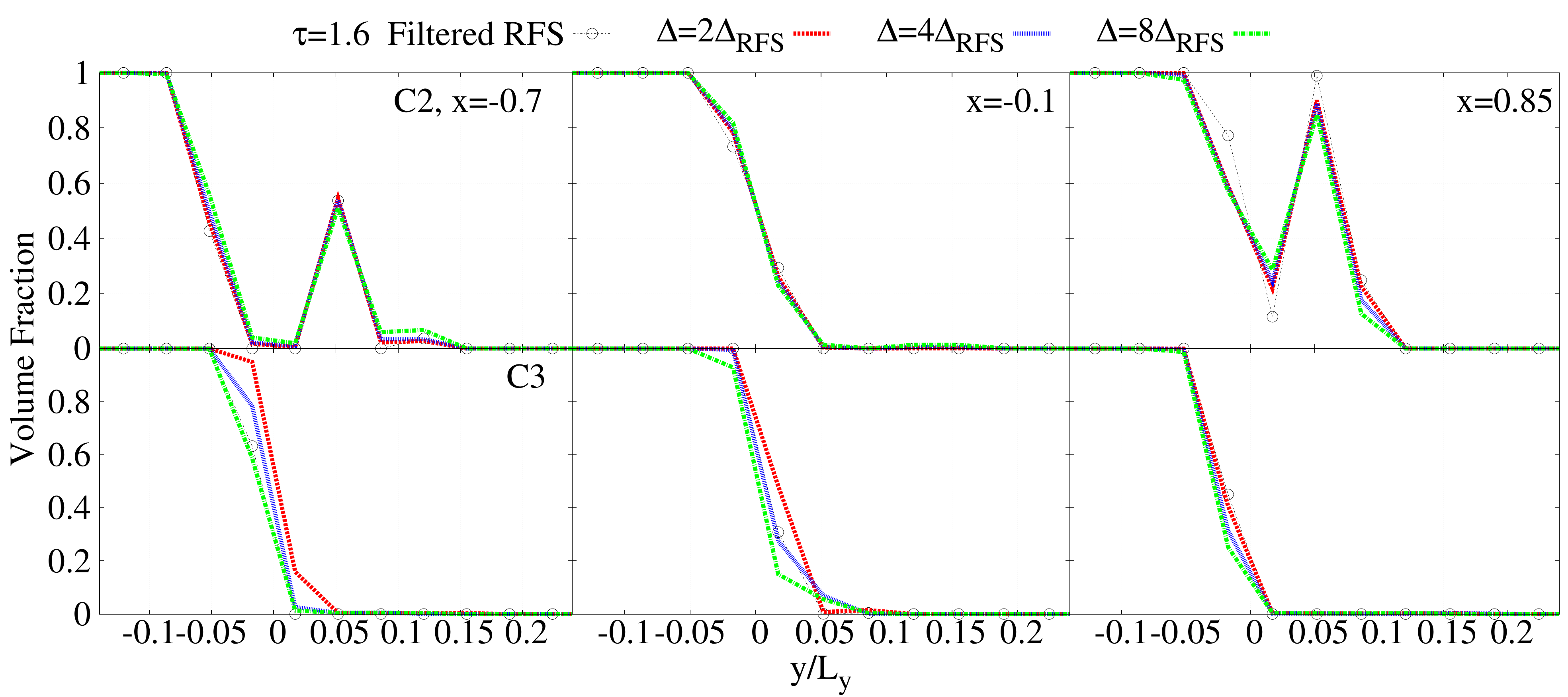}
\caption{Volume integrated volume fraction along three transverse lines by EVD with $l_V=40\Delta_{RFS}$ and RFS for C2 and C3 at $\tau=1.6$. The y axis is normalised by the transverse width of the domain, $L_y$.}\label{fig:AlphaXC2C3t2}
\end{figure}

\noindent Figure\,\ref{fig:picC2C3Post} shows images of the instantaneous volume fraction fields for C2 with a lower Reynolds number than C1 and C3 with a higher density ratio
than C1. Comparisons are made between EVD with $l_V=40\Delta_{RFS}$ on two different grids, $\Delta=2\Delta_{RFS}$ and $\Delta=4\Delta_{RFS}$, and RFS data. A high degree of similarity can be seen between both EVD results and the RFS fields for the wave patterns and ligament positions. One noticeable difference occurs for C2 in which the tips of ligaments in the EVD simulations are long and have curved profiles. Quantitative comparisons are shown in Fig.\,\ref{fig:AlphaXC2C3t2} and the profiles at $x=$ -0.7, -0.1 and 0.85 are examined as was the case for C1. Overall the results indicate very good numerical convergence with refinement of the grid from $\Delta =8\Delta_{RFS}$ to $\Delta =2\Delta_{RFS}$. The grid independence at $x=$ -0.7 for C3 is not as strong. The rate of convergence in these cases is faster than for case C1 as seen by the fact that even the coarsest grid, $\Delta=8\Delta_{RFS}$, produces consistent results. Furthermore, it can be seen that the EVD solutions are accurate relative to the RFS data. 
\\\\

\begin{figure}[h!]
\footnotesize 
 \centering
 \includegraphics[clip=true, trim=0 0 0 0, width=0.7\textwidth]{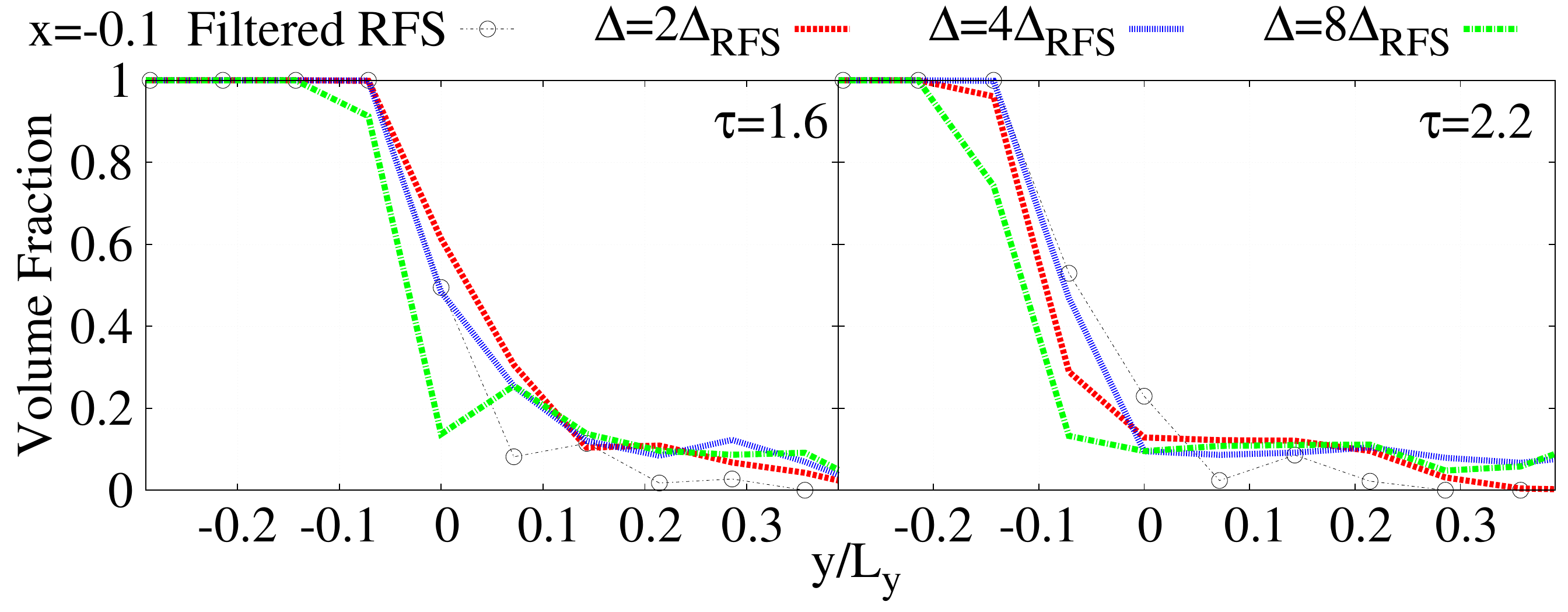}
\caption{Volume integrated volume fraction at $x=$ -0.1 by EVD with $l_V=80\Delta_{RFS}$ and RFS for C1 at $\tau=1.6$ and 2.2. The y axis is normalised by the transverse width of the domain, $L_y$.}\label{fig:alphaXV80-0.1}
\end{figure}

\begin{figure}[h!]
\footnotesize 
 \centering
 \includegraphics[clip=true, trim=0 0 0 0, width=0.7\textwidth]{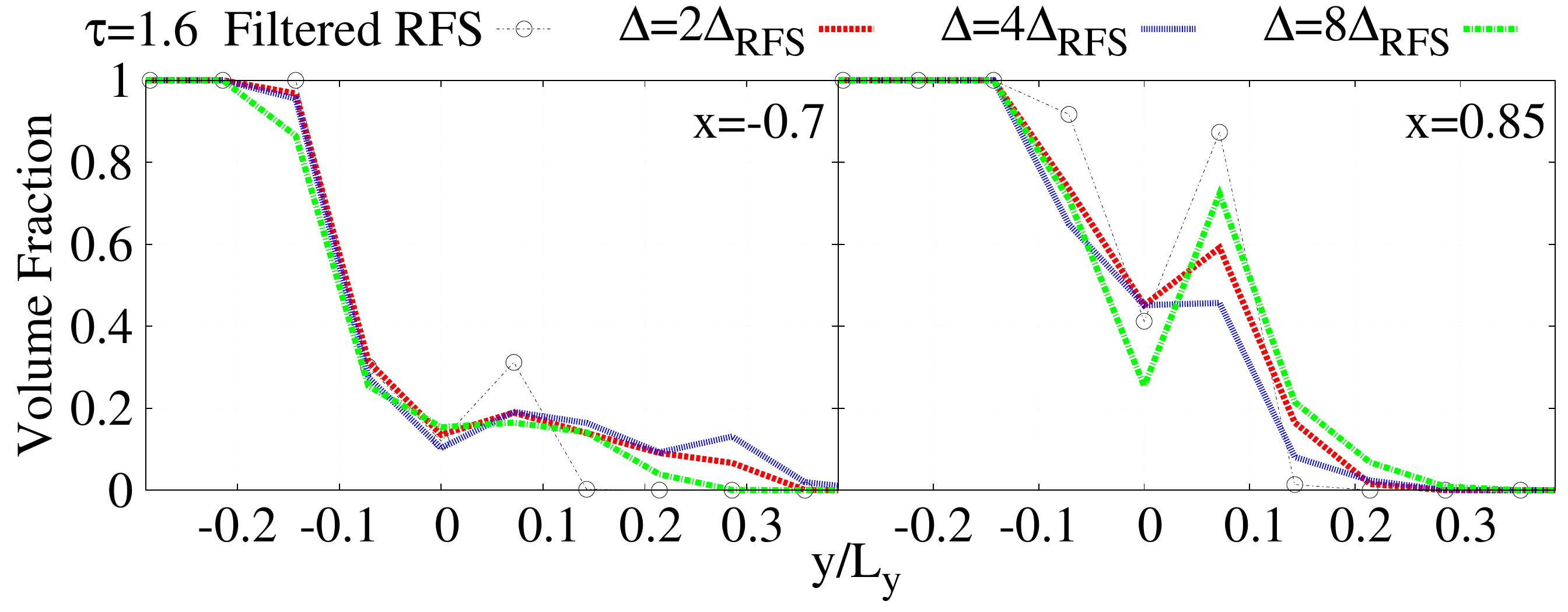}
\caption{Volume integrated volume fraction at $x=0.85$ and -0.7 by EVD with $l_V=80\Delta_{RFS}$ and RFS for C1 at $\tau=1.6$. The y axis is normalised by the transverse width of the domain, $L_y$.}\label{fig:alphaXV80tau1.6}
\end{figure}

\begin{figure}[h!]
\footnotesize 
 \centering
 \includegraphics[clip=true, trim=0 0 0 0, width=0.7\textwidth]{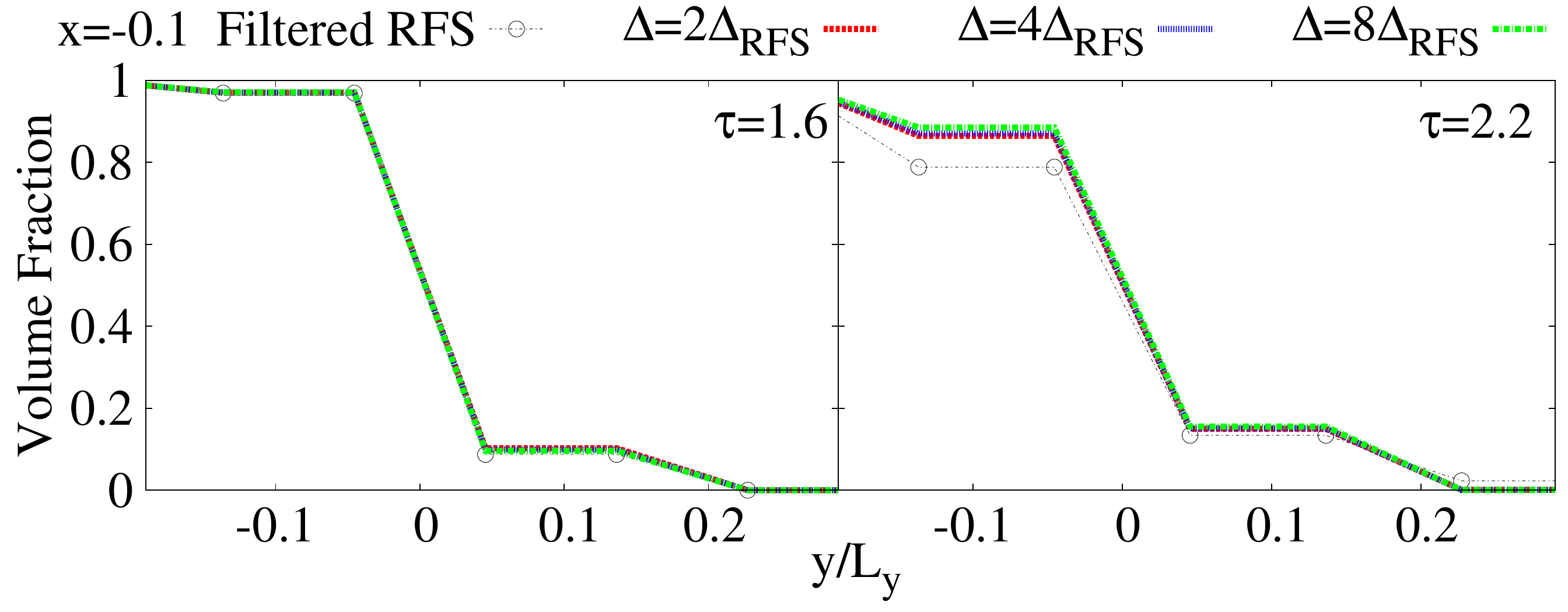}
\caption{Volume integrated volume fraction at $x=$ -0.1 by EVD with $l_V=200\Delta_{RFS}$ and RFS for C2 at $\tau=1.6$ and 2.2. The y axis is normalised by the transverse width of the domain, $L_y$.}\label{fig:alphaC2V200}
\end{figure}

\noindent 
A further brief check on the sensitivity to the value of $l_V$ is conducted for C1 and C2, this time by increasing $l_V$ towards the boundary layer thickness, $\zeta$ (see Table~\ref{tab:delta}), whereas above a reduction in $l_V$ was considered. Profiles of volume fraction for C1 with $l_V=80\Delta_{RFS}$ are presented for different times and locations in Figs\,\ref{fig:alphaXV80-0.1} and \ref{fig:alphaXV80tau1.6}.
The results for $\Delta$=2$\Delta_{RFS}$ and 4$\Delta_{RFS}$ show good similarity and convergence with grid refinement is evident. Agreement with the RFS is also good. For C2 with a lower Reynolds number, grid convergence for $l_V=200\Delta_{RFS}$ is very convincing (see Fig.~\ref{fig:alphaC2V200}). Overall, increasing $l_V$ does not essentially affect the numerical convergence{, but rather smooths} 
the volume fraction fields and thickens the resolved interface. Further analysis of sensitivity to $l_V$ is conducted against for the airblast ethanol spray jet {as shown} in Section\,\ref{sec:ESvalid}.

\begin{figure}[h!]
\footnotesize 
 \centering
 \subfigure[C4, $l_V=8\Delta_{RFS}$]{\includegraphics[clip=true, trim=0 0 0 0, height=0.28\textwidth]{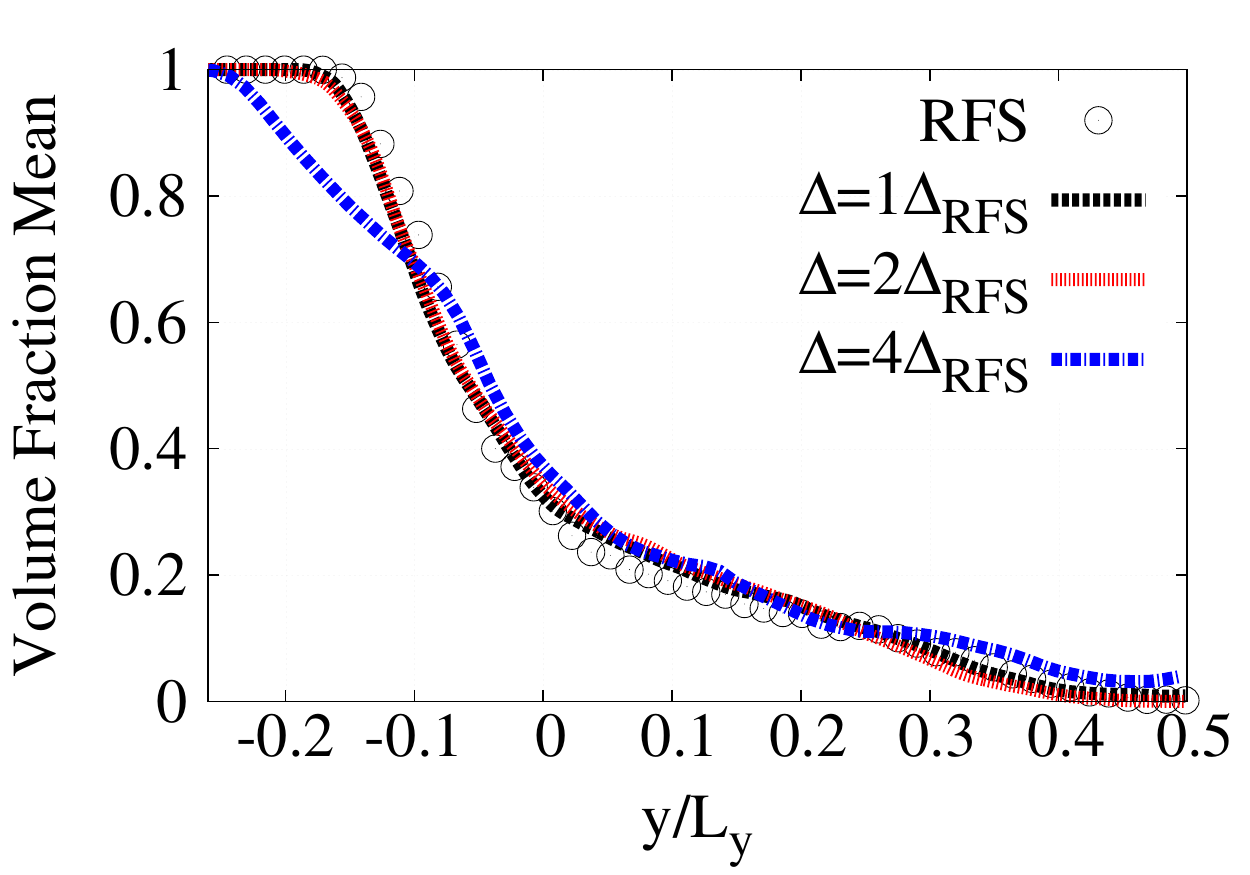}}
 \subfigure[C5, $l_V=8\Delta_{RFS}$]{\includegraphics[clip=true, trim=0 0 0 0, height=0.28\textwidth]{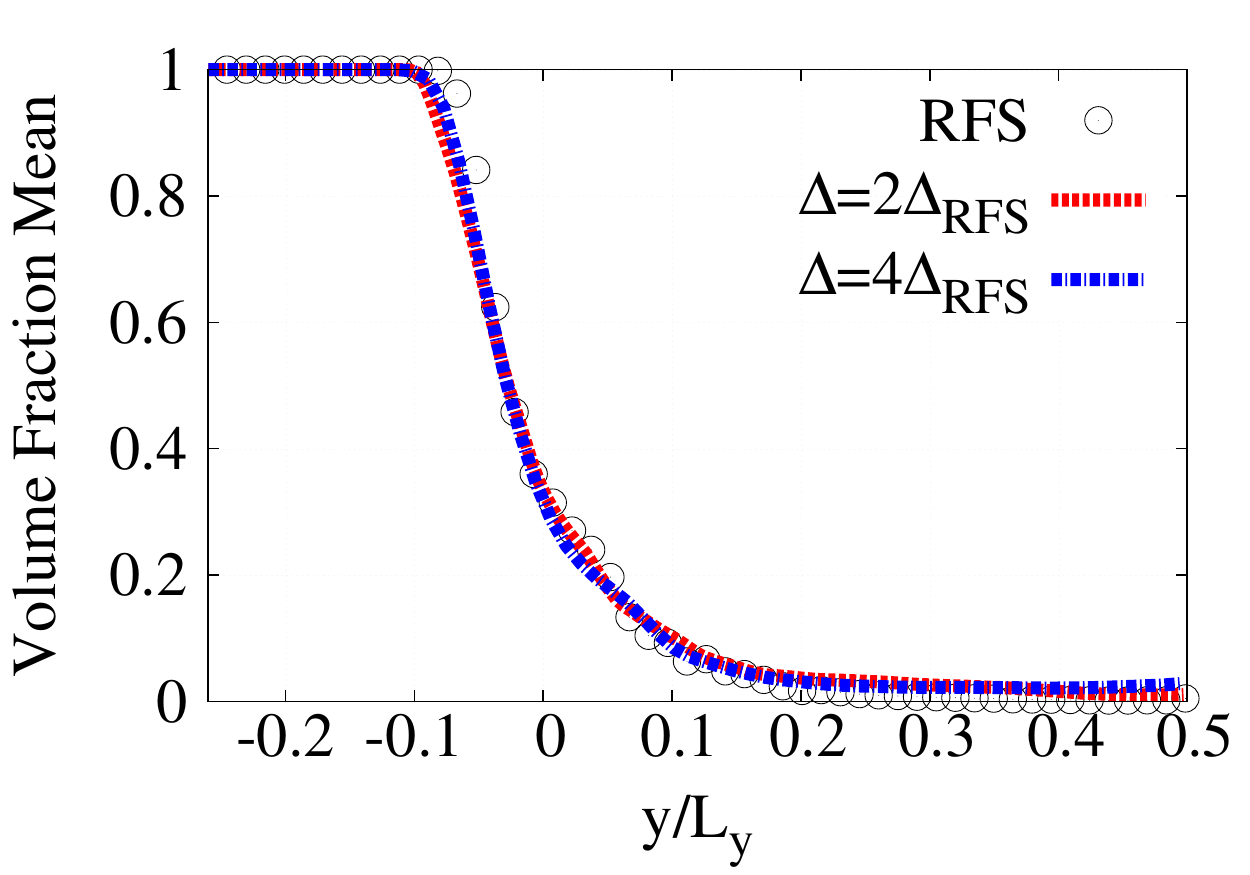}} 
\caption{Mean volume fraction in the transverse direction for C4 and C5 at $t_j=60$ from RFS and EVD.}\label{fig:Alpha3DMeanYC4C5}
\end{figure}

\noindent For the EVD simulations of the two cases in the turbulent regime, C4 and C5, the volume length scale is $l_V=8\Delta_{RFS}$. Numerical convergence is tested for the average volume fraction integrated over the statistically homogeneous $x-z$-plane as presented in Fig.\,\ref{fig:Alpha3DMeanYC4C5}. For C4 with a moderate density ratio of 10, the EVD results converge for numerical cell sizes no larger than $2\Delta_{RFS}$ and give comprehensive predictions of the RFS data. For $\Delta=4\Delta_{RFS}$ a large underestimate is observed for $y/L_y<$ -0.1 and a slight overprediction starts from $y/L_y>$ -0.1 onwards. This indicates that the lower resolution EVD results cause the heavy fluid to roll up producing more ligaments and fragments in the upper stream. For C5 with a larger density ratio of 100, EVD converges rapidly with grid refinement and gives very good agreement with the RFS.

\section{Validation of EVD based on an airblast ethanol spray on Sydney needle spray burner}\label{sec:ESvalid}

\subsection{Experimental and computational setup}\label{sec:expES2}
\noindent The EVD approach is now validated for a turbulent airblast ethanol spray jet case that has been investigated experimentally with the two-angle backlit imaging technique on the Sydney needle spray burner \cite{singh2020instability,singh2020volume}. The ethanol is injected from a needle with diameter $d_j=686~\mu$m and the concentric airblast tube has diameter $D=10$~mm. In the case investigated here, called N-ES2, the needle and airblast exit planes are flush. Key flow parameters and fluid physical properties are shown in Table\,\ref{tab:expSet} where $U_a$ and $U_e$ represent the mean injection velocities of the air and ethanol, respectively. {Density and kinematic viscosity of liquid ethanol and air are also presented. The surface tension coefficient of the ethanol and air interface is set to be $\sigma=0.02239$kg/s$^2$. The jet Weber number based on the interfacial drift velocity is 35 that is calculated by $We=\rho_a\left(U_a-U_e\right)^2 d_j/\sigma$.}
\begin{table}[h!]
  \begin{center}
\def~{\hphantom{0}}
  \begin{tabular}{lcccccccc}
  \toprule
    Case  & $U_a$(m/s)  & U$_e$(m/s) & $\rho_a$(kg/m$^3$) & $\rho_e$(kg/m$^3$) & $\nu_a$(m$^2$/s) & $\nu_e$(m$^2$/s) & $\sigma$(kg/s$^2$) & $We$ \\[3pt]
    N-ES2 & 36 & 4.3 & 1.178 & 789 & 1.567e-05 & 1.521e-06 & 0.02239 & 35 \\ 
    \bottomrule  
  \end{tabular}
  \caption{Parameter setup for the experimental airblast ethanol spray, N-ES2}
  \label{tab:expSet}
  \end{center}
\end{table}

A three-dimensional cylindrical domain is created for computations with axial length of $L_x=50mm$ and diameter of $D_{in}=5mm$. To reduce the computational cost, the diameter of the computational domain is half the experimental airblast diameter, $D$, and this is sufficient to cover the spray primary breakup region without undue effects from the numerical boundary conditions. The turbulence intensity and integral length scale of airblast are taken from the experimental data as $0.11U_a$ and $0.125D$, respectively. A weaker turbulence intensity of $0.05U_e$ is imposed on the liquid jet which is a typical turbulent pipe flow value (experimental data is not available) and the integral length scale is again set to $0.125d_j$. The inflow turbulent perturbations for both the air and ethanol streams are generated by a digital filter \cite{klein2003digital}.

From Eq.~(\ref{eq:deltahTurb}) where the characteristic length scale is the width of the annular gap of the jet, the characteristic boundary thickness on the gas side of the interface is approximated as $\zeta=198~\mu$m. For the base case simulations the explicit volume length scale in the interface region is $l_V=190~\mu$m which is just smaller than $\zeta$. This current definition of $l_V$ is linked to the boundary thickness on the gas side of the interface at the vicinity of the jet exit. In the primary breakup region, the boundary thickness tends to become larger in the downstream such that the volume length scale can resolve it well. Three different grid resolutions containing 1.1, 2.3 and 4.5 million cells are used to discretise the EVD transport equations. {The liquid jet diameter and the annular gap are discretised by 14, 18 and 24 cells, and 40, 50 and 62 cells, respectively. The expansion ratio of the largest and smallest cell size in the radial direction is 1.5 which leads to characteristic cell widths near the interface being $49~\mu$m, $38~\mu$m and $30~\mu$m. The axial length of the domain is resolved by 1000, 1250 and 1575 cells, respectively, that ensures the axial lengths of the cells to be almost identical to the characteristic cell widths near the interface.} The corresponding grid to explicit length ratios are $\Delta/l_V = $0.26, 0.2 and 0.16, respectively. Another larger value $l_V=320~\mu$m is tested below which is larger than $\zeta$ and for this case $\Delta/l_V$ values for the three numerical grids are 0.154, 0.12, 0.096. The larger $l_V$ will increase explicit volume diffusion and should accelerate numerical convergence on coarser grids. {However, it} could excessively smear the interface {and is} expected to give a less accurate prediction.

\subsection{Results}

\begin{figure}
\footnotesize 
 \centering
  \subfigure[1.1M, $\Delta=0.26l_V$, $\widehat{\alpha}^{\Delta}$]{\includegraphics[clip=true, trim=200 490 200 445, width=0.9\textwidth]{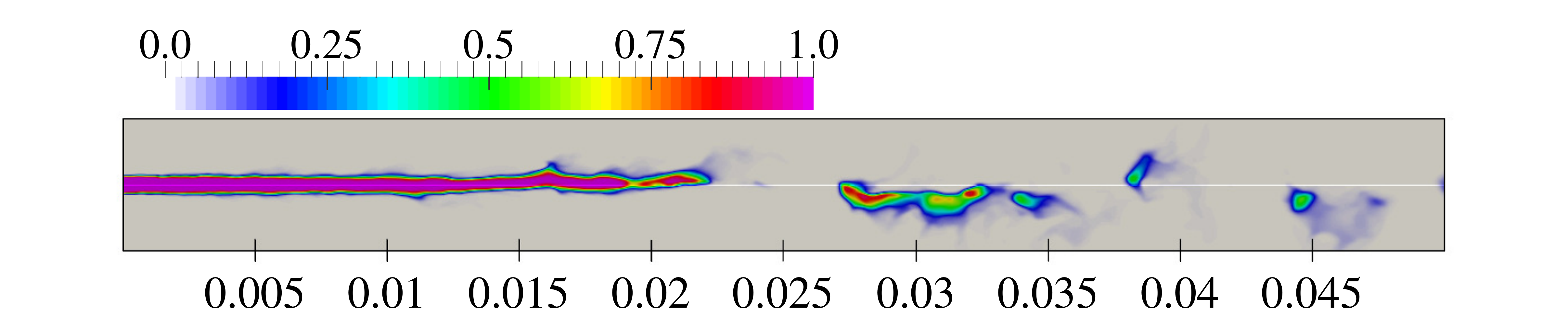}\label{fig:ES2Alpha1}}
  \subfigure[2.3M, $\Delta=0.2l_V$, $\widehat{\alpha}^{\Delta}$]{\includegraphics[clip=true, trim=200 540 200 542, width=0.9\textwidth]{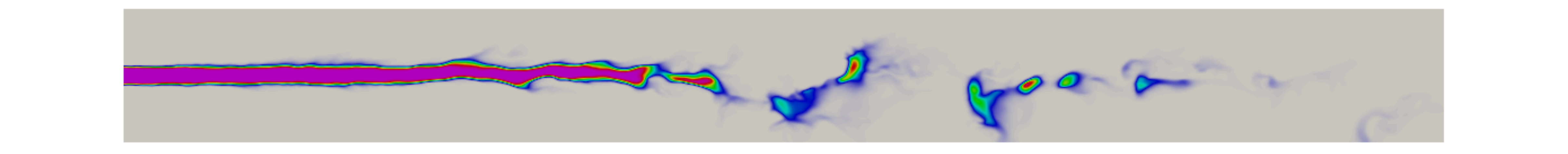}\label{fig:ES2Alpha2}}
  \subfigure[4.5M, $\Delta=0.16l_V$, $\widehat{\alpha}^{\Delta}$]{\includegraphics[clip=true, trim=200 540 200 545, width=0.867\textwidth]{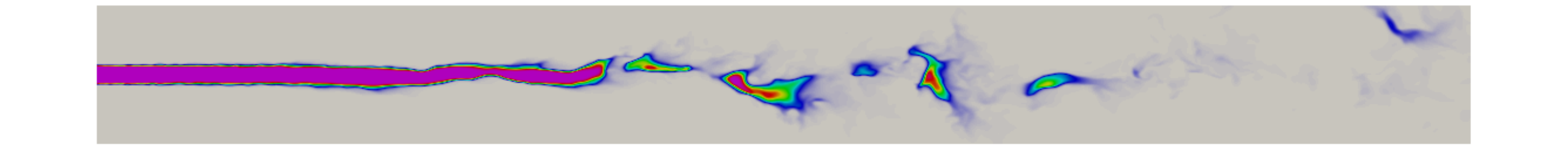}\label{fig:ES2Alpha3}}  
  \subfigure[2.3M, $\Delta=0.2l_V$, $\wideparen{\alpha}^{\Delta}$]{\includegraphics[clip=true, trim=200 540 200 542, width=0.9\textwidth]{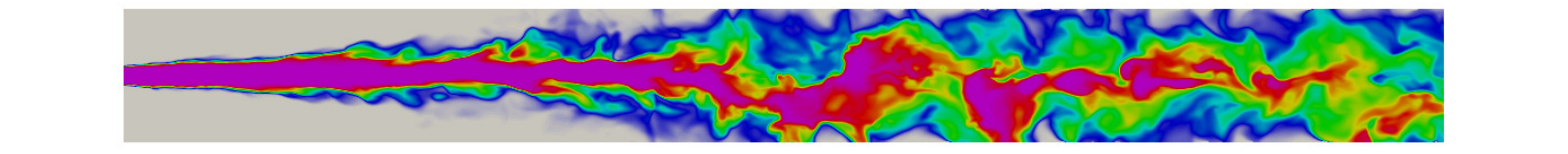}\label{fig:ES2rhoAlpha2}} 
  \subfigure[2.3M, $\Delta=0.2l_V$, $\widehat{\alpha}^V$]{\includegraphics[clip=true, trim=200 540 200 542, width=0.9\textwidth]{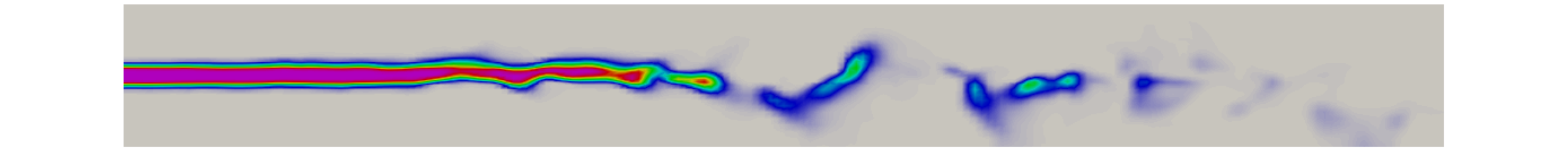}\label{fig:ES2AlphaV2}}
  
\caption{Comparisons of instantaneous volume fraction fields on different LES grids for N-ES2. The $x$-axis is added to the bottom of Fig.\,\ref{fig:ES2Alpha1} with the unit of meter. }\label{fig:ES2Alpha}
\end{figure}

\noindent Figure\,\ref{fig:ES2Alpha} shows instantaneous volume fraction fields along the centre plane of the computational domain. Figures\,\ref{fig:ES2Alpha1}~-~\ref{fig:ES2Alpha3} present the grid-based Reynolds volume averaged volume fractions, $\widehat{\alpha}^{\Delta}$, on the three LES grids. The Favre volume averaged volume fraction, $\wideparen{\alpha}^{\Delta}$, is the quantity that is solved directly and is shown in Fig.\,\ref{fig:ES2rhoAlpha2} for the 2.3M cell grid, while $\widehat{\alpha}^{\Delta}$ are the physically interesting fields that are obtained from $\wideparen{\alpha}^{\Delta}$ via Eq.~(\ref{eq:rhoAlpha}). There are stark differences between the $\widehat{\alpha}^{\Delta}$ and $\wideparen{\alpha}^{\Delta}$ fields in Fig.\,\ref{fig:ES2Alpha} is owing to the large density ratio of about 670. In these images the interface instability gradually grows and the liquid core tends to deviate from the centre line at about $x=0.0125$~m. At about $x=0.02$~m the liquid core breaks up into large irregular objects and subsequently into various ligaments and small liquid objects. The volume integrated volume fraction, $\widehat{\alpha}^V$, for 2.3M is presented in Fig.\,\ref{fig:ES2AlphaV2}. Compared with the grid-based volume fraction fields, the interfaces are considerably smoothed due to volume averaging but the other features remain visible.

\begin{figure}
\footnotesize 
 \centering
  \subfigure[2.3M, $\Delta=0.2l_V$, ${D_V}^{\Delta}$]{\includegraphics[clip=true, trim=200 490 200 445, width=0.9\textwidth]{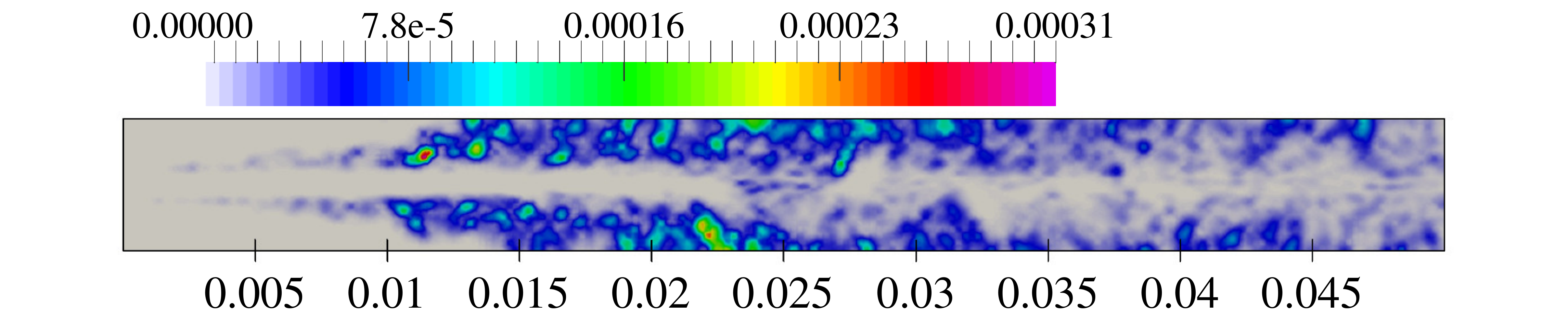}}
  \subfigure[2.3M, $\Delta=0.2l_V$, ${{\nu_t}^{eff}}^{\Delta}$]{\includegraphics[clip=true, trim=200 540 200 445, width=0.9\textwidth]{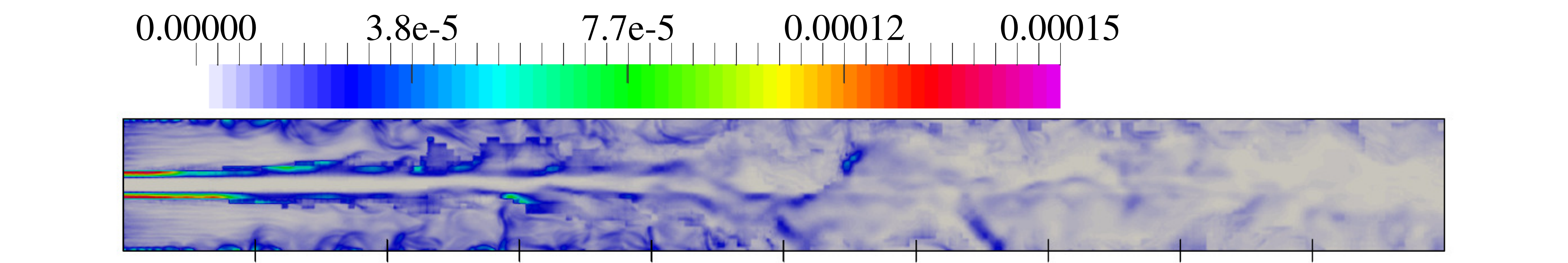}}
  
\caption{Instantaneous explicit volume diffusion coefficient and effective turbulent viscosity on the LES grid with 2.3 million cells for N-ES2.}\label{fig:ES2DvNut}
\end{figure}

Figure\,\ref{fig:ES2DvNut} shows instantaneous, grid-based fields for the explicit volume diffusion coefficient given by Eq.\,(\ref{eq:DV}) and effective turbulent viscosity given by Eq.\,(\ref{eq:nueff}).
The former {increases smoothly} between the inflow boundary on the left {(where it is zero)} and $x=0.01m$. More specifically, the evolution of $D_V$ in this near nozzle region of the jet is mainly controlled by the coherent structure function $\left|{Q}/{E}\right|$ (see Eq.\,(\ref{eq:DV})). As discussed in Section\,\ref{sec:modelFlux}, due to the non-existent or very low curvature along this continuous interface region leads to $\left|{Q}/{E}\right| \to 0$ and thus $D_V \to 0$ as well. However, there is high turbulent shear velocity due to the airblast that leads to high values of $\nu_t^{eff}$ which produces instability growth and the gradual onset of the curvature which in turn increases $D_V$. Downstream of $x=0.01m$ there is rapid increase of $D_V$ in the interfacial region and this is sustained until about $x=0.03m$ coinciding {with} the breakup of liquid jet into large ligaments. There are large inhomogeneities {in} the volume fraction field over this part of the flow leading to high values of $\sqrt{{\wideparen{\alpha}}\left(1-{\wideparen{\alpha}}\right)}$ and high strain rates that produce high values of $D_V$. For $x>0.03m$, the explicit volume diffusion coefficient decreases to low values due to small turbulence fluctuations and small volume fraction gradients.

\begin{figure}[h!]
\footnotesize 
 \centering
 \includegraphics[clip=true, trim=0 0 0 0, width=0.95\textwidth]{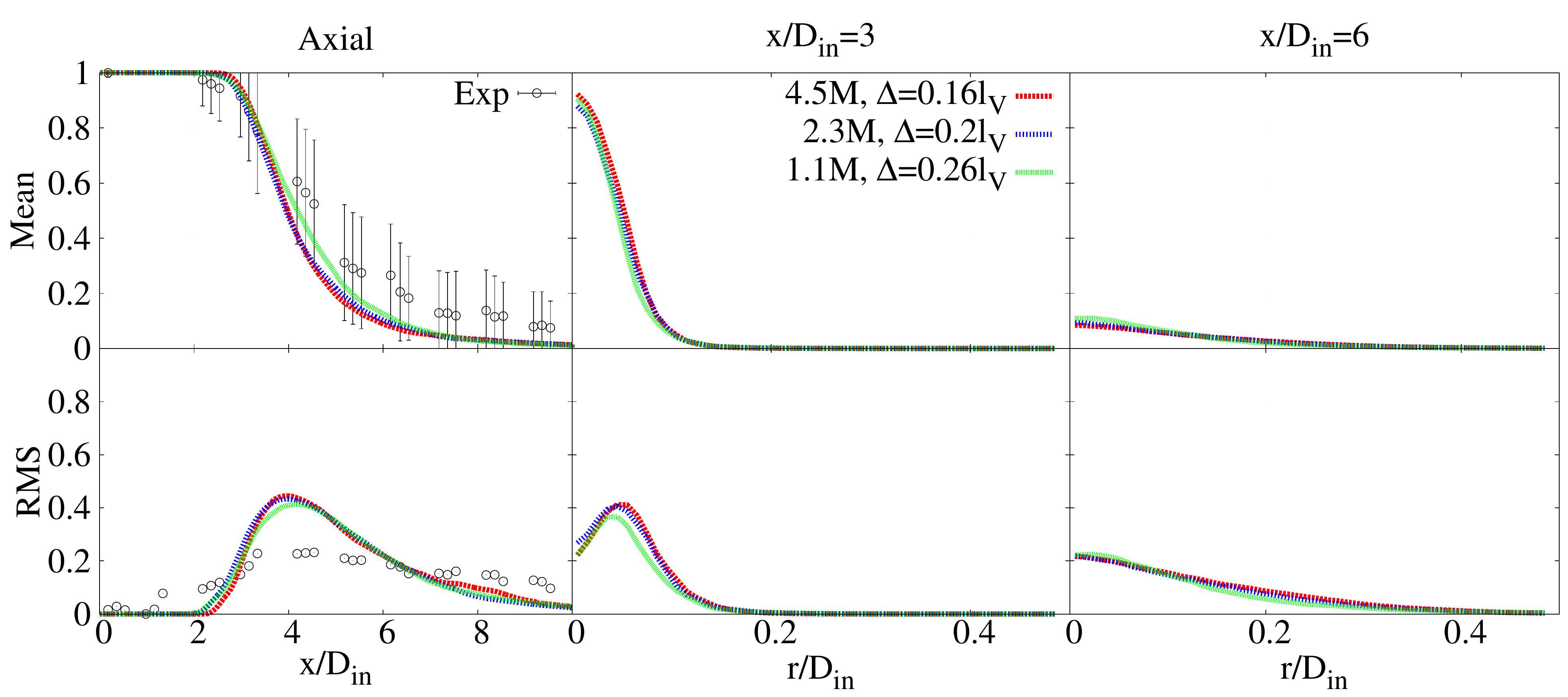}
\caption{Axial and radial profiles of mean (1st row) and rms (2nd row) of volume fraction for N-ES2. Radial profiles are shown at two axial stations normalised by the diameter of the computational domain, $D_{in}$. Experimental data \cite{singh2020volume} includes error bars. The explicit volume length scale is $l_V=190~\mu$m.}\label{fig:ethanolAlphaLv1}
\end{figure}

\noindent Figure\,\ref{fig:ethanolAlphaLv1} shows axial and radial profiles of mean and rms of volume fraction for $l_V = 190~\mu$m compared with experimental data. The rms magnitudes are consistent with the experimental error bars provided for the mean data. Overall there is excellent convergence of axial and radial fields with grid refinement with the 2.3M and 4.5M lines essentially on top of each other for both the mean and rms fields. These converged results provide satisfactory predictions of the spray breakup and the evolution of the mean volume fraction along the axial direction. There is underestimation from $x/D_{in}=4$ onwards but the predictions remain within the experimental uncertainty. The position of the highest rms is very well predicted at about $x/D_{in}=4$ but the magnitude of the peak value is overpredicted. Upstream and downstream of this peak value, the predictions of rms are generally good.

\begin{figure}[h!]
\footnotesize 
 \centering
 \includegraphics[clip=true, trim=0 0 0 0, width=0.95\textwidth]{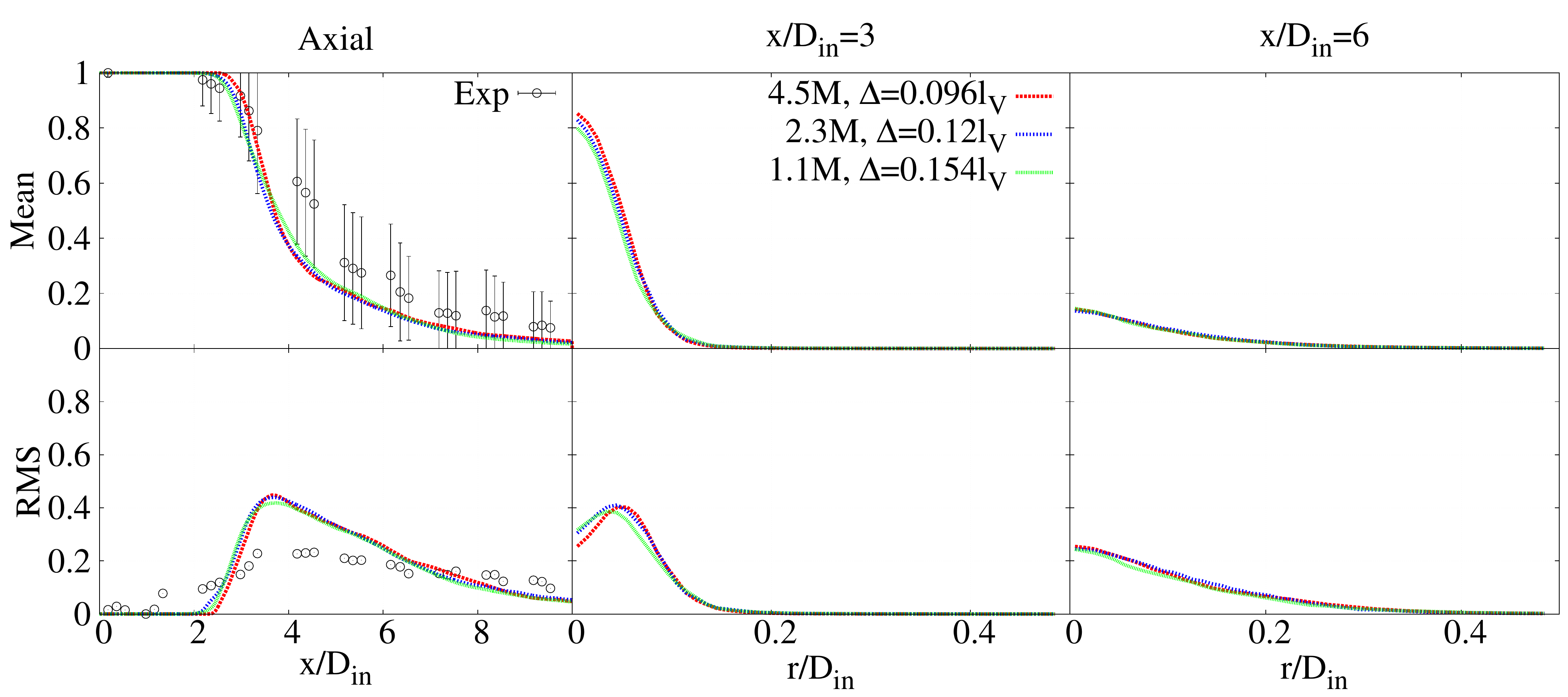}
\caption{Axial and radial profiles of mean (1st row) and rms (2nd row) of volume fraction for N-ES2. Radial profiles are shown at two axial stations normalised by the diameter of the computational domain, $D_{in}$. Experimental data \cite{singh2020volume} includes error bars. The explicit volume length scale is $l_V=320~\mu$m.}\label{fig:ethanolAlphaLv2}
\end{figure}

\noindent To test sensitivity to $l_V$, Fig.\,\ref{fig:ethanolAlphaLv2} presents results for simulations with $l_V=320~\mu$m which is larger than the boundary layer thickness,
$\zeta=198~\mu$m. Once again convergence with grid refinement is clearly demonstrated for both the mean and rms profiles. Due to $\Delta/l_V$ being smaller in this case than for the case with $l_V=190~\mu$m, the magnitude of numerical diffusion compared with explicit volume diffusion is expected to be smaller. {This is evident} in Fig.\,\ref{fig:ethanolAlphaLv2} {where} even the 1.1M cells case is in excellent agreement with the results for the two more refined grids, especially for the axial mean profile. However, {for $l_V=320~\mu$m,} the rate of decay of volume fraction is too fast and in the region $3<x/D_{in}<5$ there is more noticeable underprediction of the mean relative to the data and the predictions dip outside the margin of error. A visible deviation beyond the error limits of experimental measures is approximately located in $3<x/D_{in}<5$. It is likely that with this much larger value of $l_V$ the smearing of the interface and boundary layer is excessive causing the liquid volume fraction to decay too rapidly.

\section{Conclusions}
\noindent This paper has formulated and validated a new model for interfacial flows based on a novel explicit volume diffusion (EVD) concept. The governing equations have been derived by volume averaging the volume of fluids (VoF) transport equations over a physically-defined volume length scale, $l_V$, that is independent of the numerical grid scale. Unlike conventional turbulence filtering, the sub-volume fluctuations due to both interface dynamics and turbulence (if it exists) are attenuated in the EVD model through the process of volume averaging which introduces unclosed terms into the transport equations for the sub-volume flux, sub-volume stress and volume averaged surface tension force. A gradient diffusion closure for the sub-volume flux was proposed which makes use of the relations for a bimodal probability density function (PDF) for the sub-volume volume fraction fields which have a sharp interface between the two fluids. The derivation reveals that the sub-volume flux exists in both turbulent and laminar flows because it is induced by inhomogeneous volume fraction fields and the mean drift velocity. The gradient diffusion model includes an explicit volume diffusion coefficient that is linked to the physical volume scale, sub-volume fluctuations of volume fraction, volume filtered strain rate and an additional coherent structure function based on the normalised Q-criterion. The sub-volume stress was closed by introducing an explicit volume viscosity that is correlated with the explicit volume diffusion coefficient by a model Schmidt number. This acts in the interface region and, in turbulent flows, it is augmented by an effective turbulent viscosity away from the interface. The volume averaged surface tension force is modelled based on fractal properties of wrinkled sub-volume interfaces.

These newly derived closures were initially validated and their constants were calibrated by an \textit{a priori} analysis of resolved flow simulations (RFS) for a series two-dimensional laminar and three-dimensional turbulent interfacial shear flows. This analysis demonstrated the functionally correct behaviour of the model closures for variations in Reynolds number, density ratio and $l_V$. The EVD model and CFD implementation was subsequently used for simulating the interfacial shear flows without input from the RFS. The numerical grids were refined while keeping $l_V$ constant to demonstrate numerical convergence. {Numerical convergence can be achieved if $l_V$ is sufficiently large compared with the grid spacing. Higher viscosity and large density ratio are likely to lead to faster convergence.} The length scale $l_V$ is a physical model input parameter and it is suggested being no larger than the boundary layer thickness on the light fluid side of the interface. As long as $l_V$ and the ratio of $l_V$ to the grid scale is sufficiently large to ensure volume diffusion overwhelming numerical diffusion, varying $l_V$ does not essentially affect the numerical convergence.

Finally, the EVD method was validated against an experimental airblast ethanol spray jet. The explicit volume diffusion coefficient was shown to have the correct physical behaviour and is zero in regions of the flow with no interface curvature. Sub-volume viscosity is not zero, however, due to turbulent fluctuations of velocity at the sub-volume scale and this contributes to the growth of interfacial instabilities and curvature. Downstream of this point the explicit volume diffusion was shown to increase and be sustained over the parts of the flow with significant liquid breakup and large volume fraction fluctuations. Numerical convergence was demonstrated for the airblast case through refinement of the numerical grid with constant $l_V$. Agreement of the predictions with the experimental data for both the mean and rms of volume fraction along the jet axis was good for a case with $l_V$ a little smaller than the characteristic light fluid boundary layer thickness. For another case with larger $l_V$, the explicit volume diffusion coefficient increased and numerical convergence with grid refinement was stronger. However, larger errors relative to the experimental mean volume fraction measurements were present, especially in the liquid core breakup region of the flow.  
\\\\

\noindent{\bf Acknowledgements\bf{.}} 

\noindent The financial support by the Australian Research Council is acknowledged. We are grateful for use of the HPC facilities at the University of Sydney. The authors also thank Professor Andreas Kronenburg (University of Stuttgart) and Dr Konstantina Vogiatzaki (University of Brighton) for instrumental early discussions about the volume averaging concept.
\\\\

\noindent{\bf Declaration of Interests\bf{.}} 

\noindent The authors report no conflict of interest.
\\\\


\appendix

\section{}\label{appA}

\begin{figure}[h!]
 \centering
 \subfigure[AC, $\tau=1$]{\includegraphics[clip=true, trim=390 300 390 300, height=0.195\textwidth]{alphaInterRhoDif10R1t1-eps-converted-to.pdf}}
 \subfigure[$\tau=1.6$]{\includegraphics[clip=true, trim=390 300 390 300, height=0.195\textwidth]{alphaInterRhoDif10R1t2-eps-converted-to.pdf}}
 \subfigure[$\tau=2.2$]{\includegraphics[clip=true, trim=390 300 390 300, height=0.195\textwidth]{alphaInterRhoDif10R1t3-eps-converted-to.pdf}}

 \subfigure[isoAdv, $\tau=1$]{\includegraphics[clip=true, trim=390 300 390 300, height=0.195\textwidth]{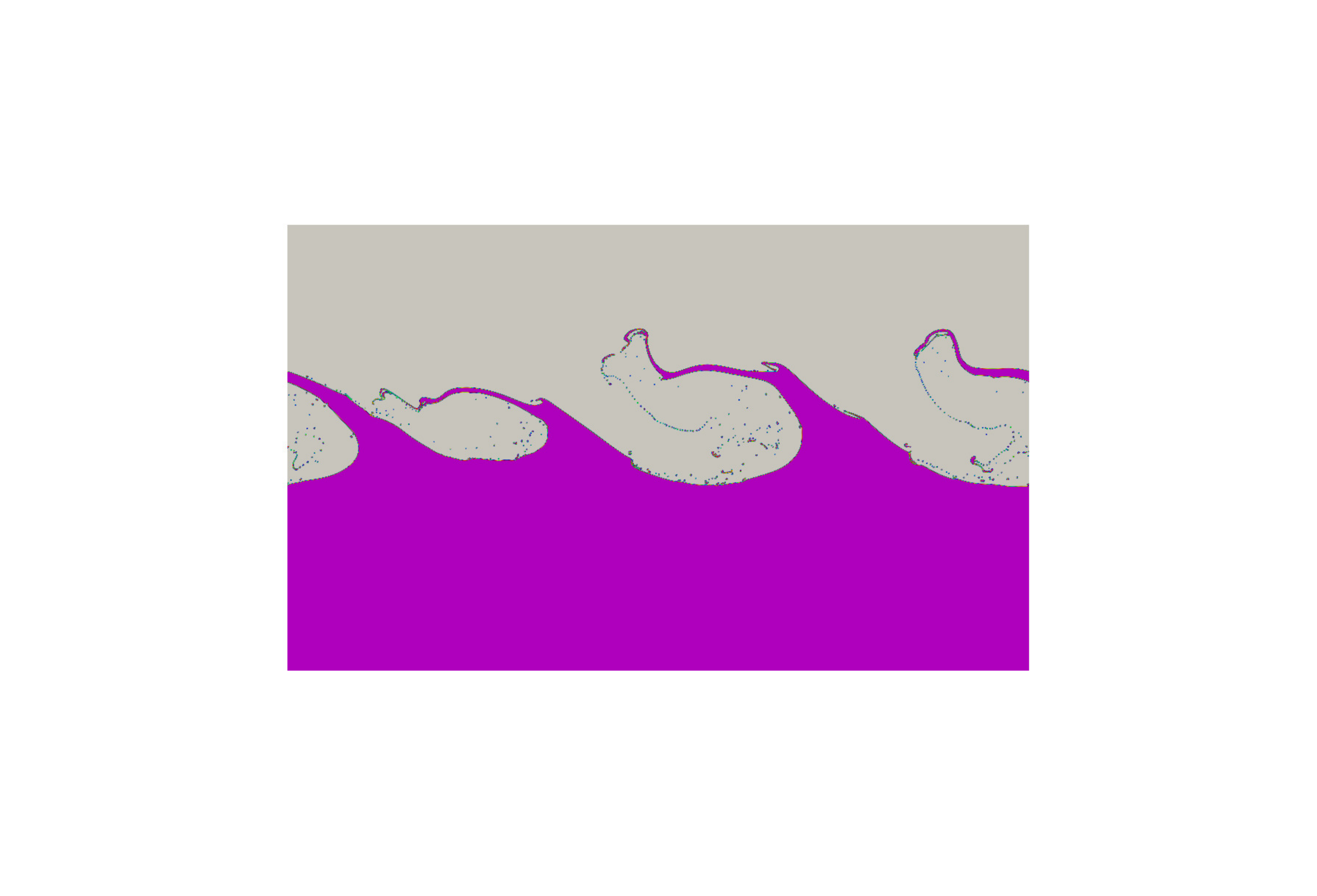}}
 \subfigure[$\tau=1.6$]{\includegraphics[clip=true, trim=390 300 390 300, height=0.195\textwidth]{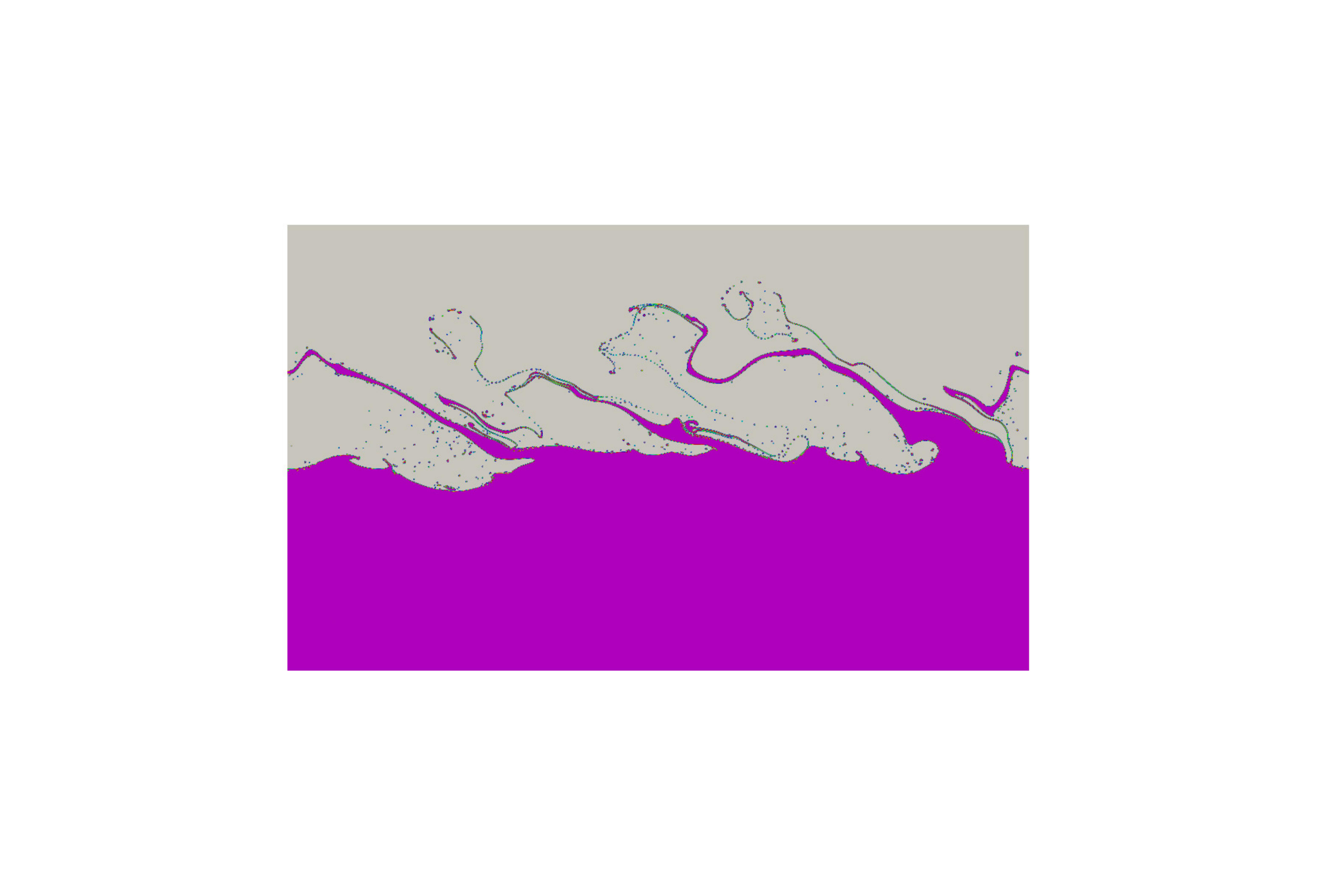}}
 \subfigure[$\tau=2.2$]{\includegraphics[clip=true, trim=390 300 390 300, height=0.195\textwidth]{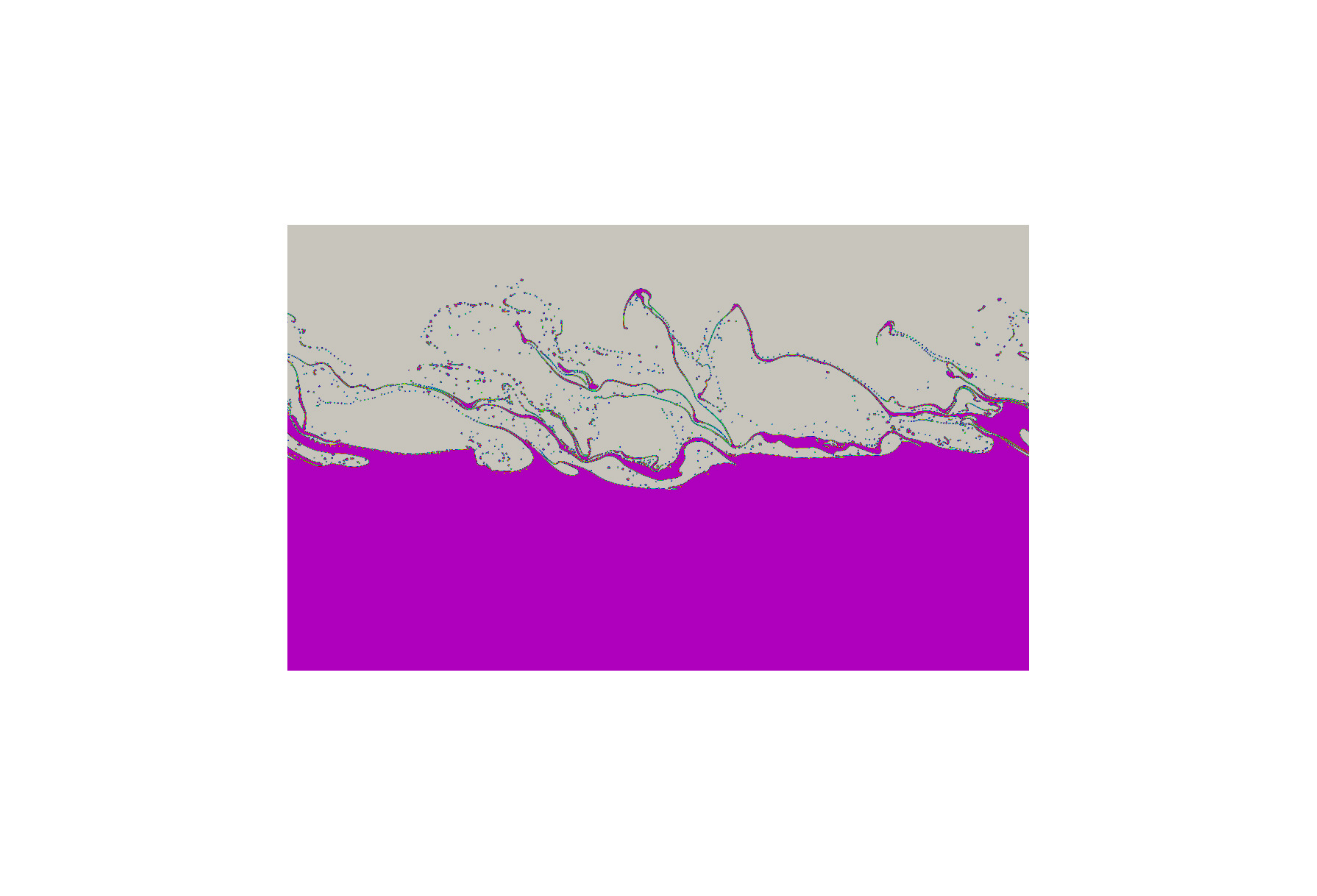}}

  \caption{Comparisons of the instantaneous volume fraction fields for C1 in resolved flow simulations using VoF-AC (1st row) and isoAdvector (2nd row).}\label{fig:picIsoAc}
\end{figure}

\begin{figure}[h!]
\footnotesize 
 \centering
 \subfigure[]{\includegraphics[clip=true, trim=0 0 0 0, height=0.28\textwidth]{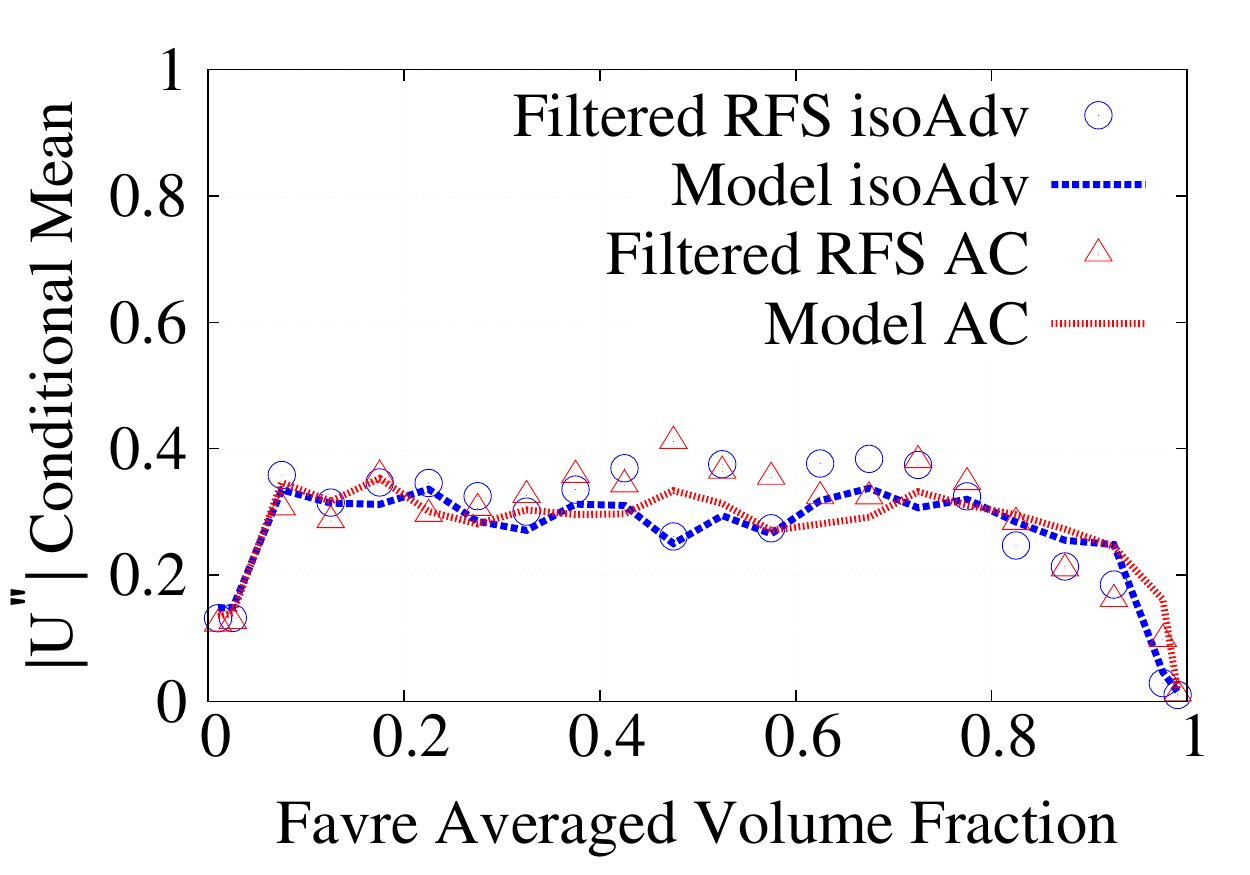}}
 \subfigure[]{\includegraphics[clip=true, trim=0 0 0 0, height=0.28\textwidth]{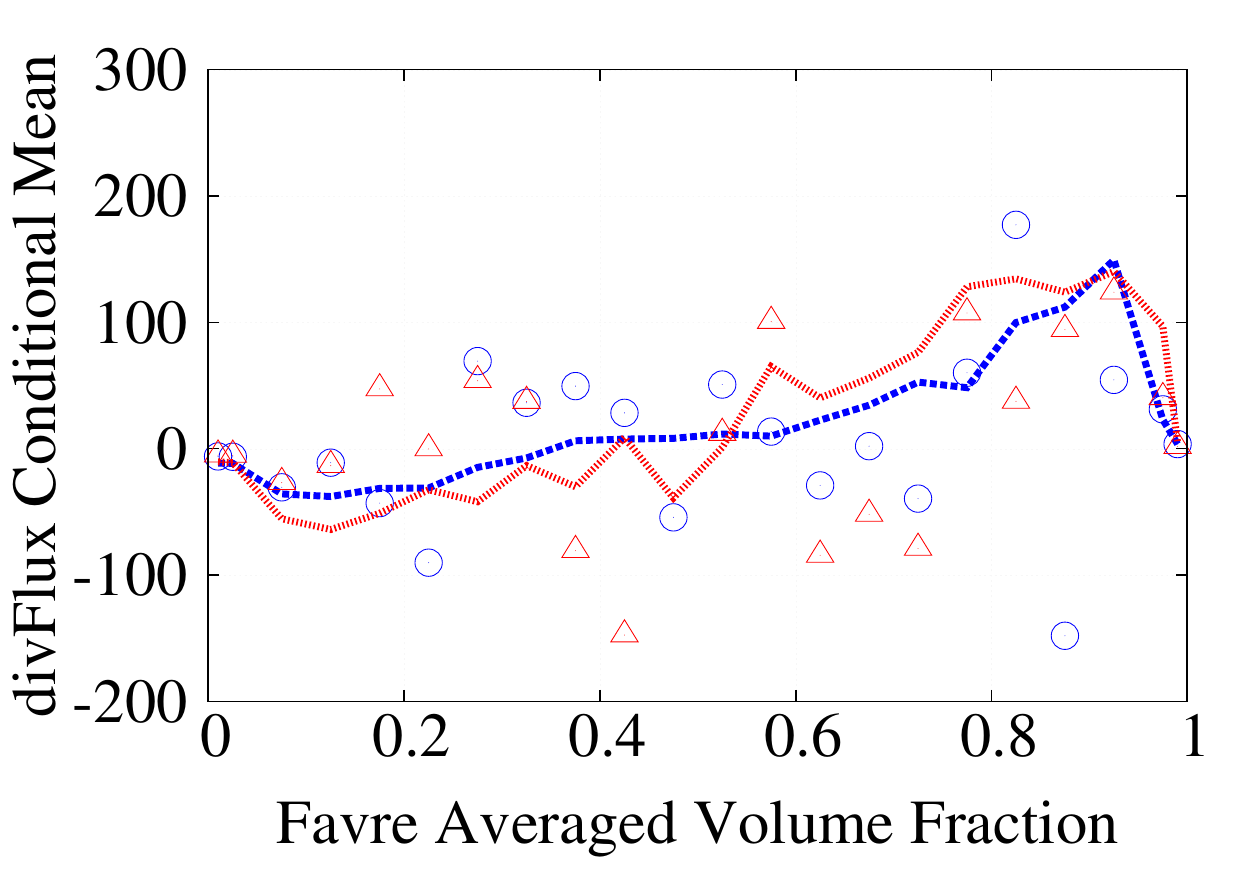}}

 \subfigure[]{\includegraphics[clip=true, trim=0 0 0 0, height=0.28\textwidth]{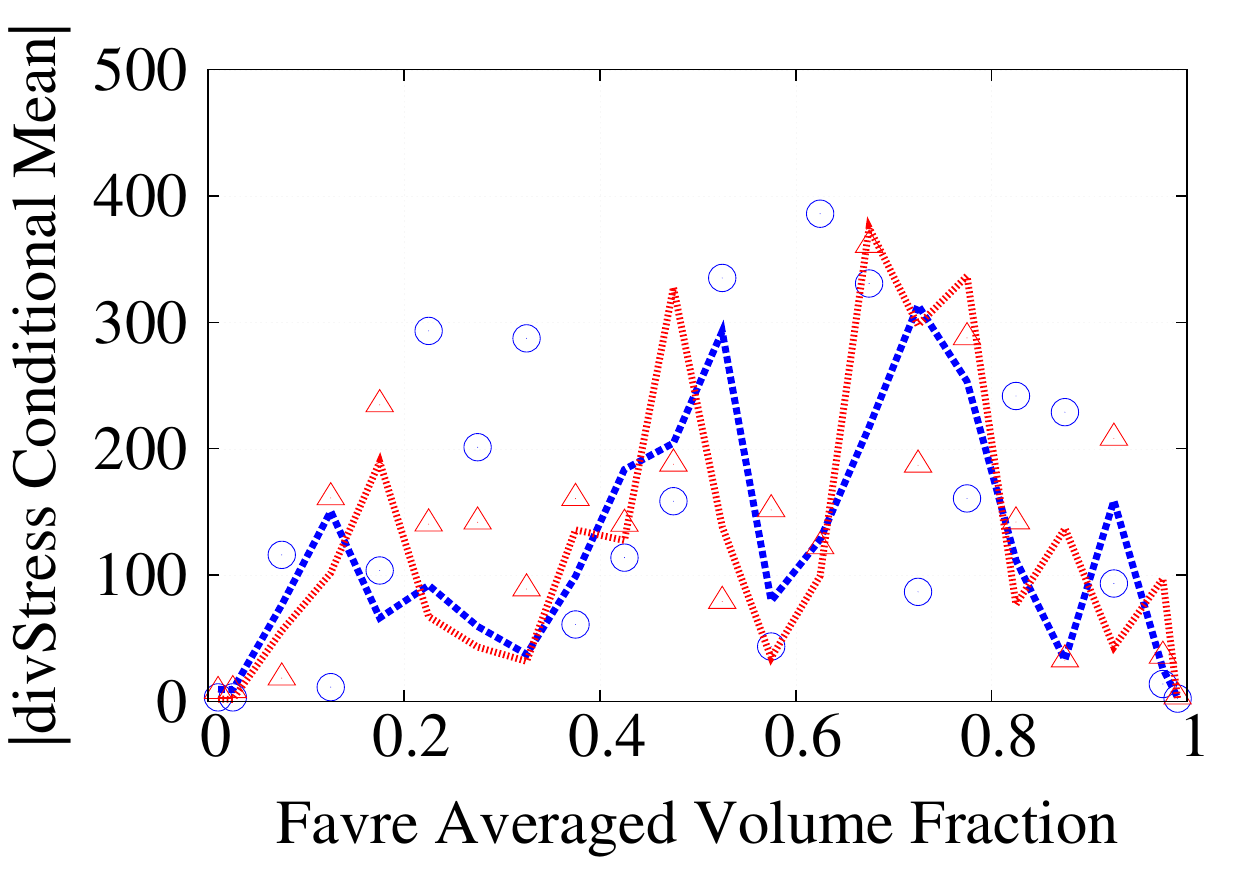}}
 \subfigure[]{\includegraphics[clip=true, trim=0 0 0 0, height=0.28\textwidth]{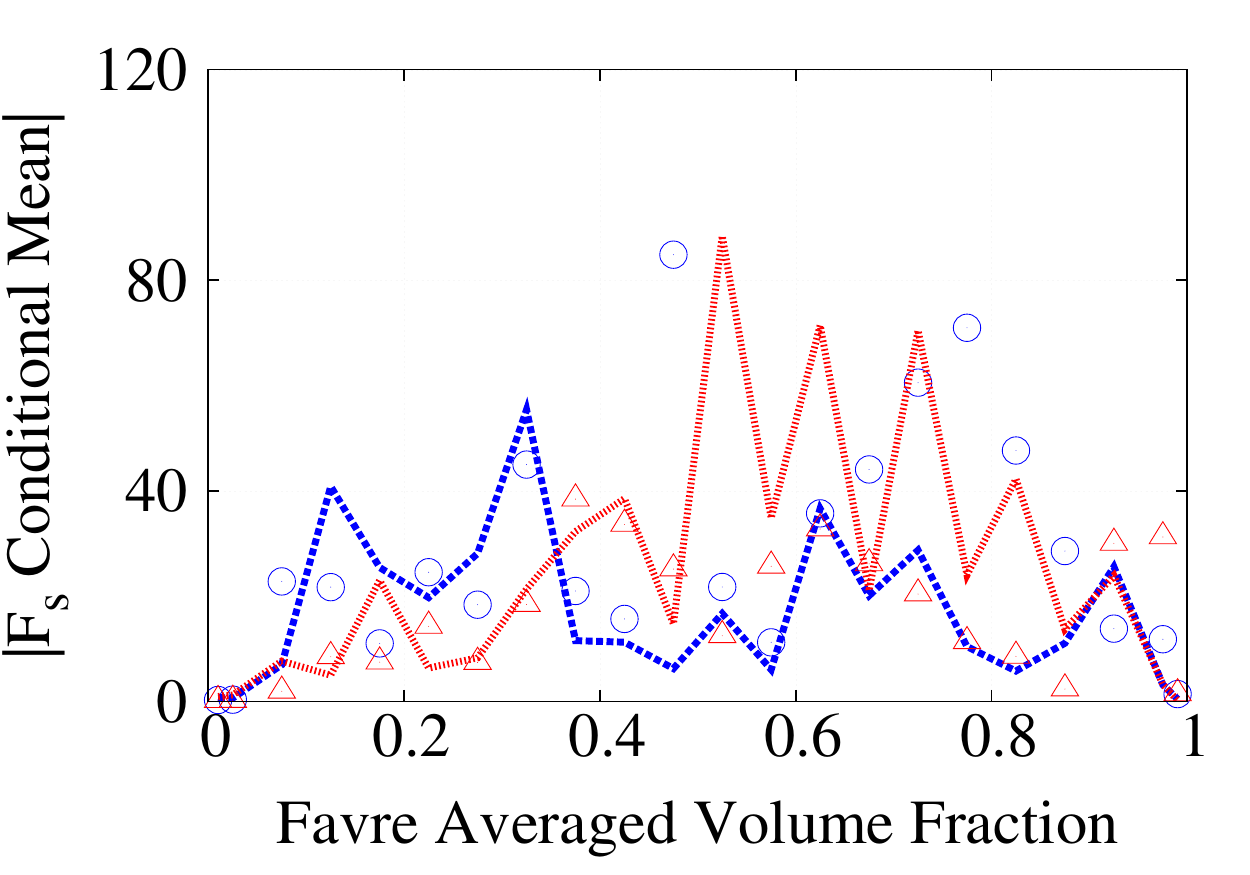}} 
\caption{Comparisons of sub-volume rms of velocity fluctuations (a), sub-volume flux (b), sub-volume stress (c), and volume averaged surface tension force (d), and their corresponding sub-volume closures in a conditionally averaged format for C1 at $\tau=1.6$ on $l_V=16\Delta_{RFS}$.}\label{fig:isoApriC1}
\end{figure}

\noindent This appendix compares RFS of two VoF-based solvers with different interface sharpening approaches for interfacial shear flow case C1. The first approach uses artificial compression (VoF-AC) that is described in Section~\ref{sec:aprioriSetup} and which was used as an EVD validation data set in Sections~\ref{sec:apriori} and \ref{sec:aposteriori}. The second approach, called \textit{isoAdvector}, uses an iso-surface model to reconstruct the interface at the sub-grid scales \cite{roenby2016computational}. Figure\,\ref{fig:picIsoAc} shows qualitative comparisons of volume fraction fields for the two approaches. Close similarity between the two sets of predictions can be observed at all three times, with the characteristic interfacial flows structures, including the unclosed vortexes sheltered by elongated ligaments extending into the upper stream, appearing in both. {For a quantitative comparison, Fig.\,\ref{fig:isoApriC1} shows conditionally averaged sub-volume velocity rms, divergences of sub-volume flux and sub-volume stresses, and volume averaged surface tension force extracted from both the VoF-AC and isoAd simulations of C1 with $l_V=16\Delta_{RFS}$ at $\tau=1.6$. \textit{A priori} model closures are also shown. The consistency of these statistical quantities produced by both models is clear, although there are some fairly small point to point differences.}

\bibliographystyle{unsrt}  
\bibliography{references}  

%
%
%
%
\end{document}